\documentclass[twocolumn,tighten,times]{aastex63}  % two-column new-style

\usepackage{
	amsmath,
	amssymb,
	amsthm,
	booktabs,
	epsfig,
	graphicx,
	grffile,
	hyperref,
	import,
	morefloats,
	microtype,
	natbib,
	subfigure,
	ulem,
	url,
	verbatim,
	wasysym,
	xcolor,
	xspace,
}
\usepackage[version=4]{mhchem}  % for molecule names and formulae
\usepackage{savesym}
\savesymbol{tablenum}
\usepackage{siunitx}
\restoresymbol{SIX}{tablenum}

\newcommand{\msun}{\ensuremath{
    \mathit{M}_\odot
}}
\newcommand{\lbol}{\ensuremath{
    \mathit{L}_\mathrm{bol}
}}
\newcommand{\lsun}{\ensuremath{
    \mathit{L}_\odot
}}
\newcommand{\eupper}{\ensuremath{
    E_\mathrm{u}
}}
\newcommand{\td}{\ensuremath{
    T_\mathrm{d}
}}
\newcommand{\mclump}{\ensuremath{
    M_\mathrm{cl}
}}
\newcommand{\sisrf}{\ensuremath{
    s_\mathrm{isrf}
}}
\newcommand{\nns}{\ensuremath{
    \delta_\mathrm{nns}
}}
\newcommand{\ljeans}{\ensuremath{
    \lambda_\mathrm{j,th}
}}
\newcommand{\ljeanstu}{\ensuremath{
    \lambda_\mathrm{j,tu}
}}
\newcommand{\jnns}{\ensuremath{
    \nns / \langle \ljeans \rangle
}}
\newcommand{\sigrms}{\ensuremath{
    \sigma_\mathrm{rms}
}}
\newcommand{\bfield}{\ensuremath{
    \mathbf{B}
}}
\newcommand{\spitzer}{ \textit{Spitzer} }
\newcommand{\uchii}{ UCH\textsc{ii} }
\newcommand{\radmc}{ \texttt{RADMC-3D} }
\newcommand{\casa}{ \texttt{CASA} }
\DeclareSIUnit{\jy}{Jy}
\DeclareSIUnit{\beam}{beam}
\DeclareSIUnit{\kms}{\kilo\meter\per\second}

\shorttitle{ALMA SCC Survey}
\shortauthors{Svoboda et al.}

\begin{document}
\defcitealias{svoboda16}{S16}
\defcitealias{ossenkopf94}{OH94}
\defcitealias{ellsworthbowers13}{E13}

\title{
ALMA observations of fragmentation, sub-structure, and protostars in high-mass starless clump candidates
}

\correspondingauthor{Brian E.\ Svoboda}
\email{bsvoboda@nrao.edu}

\author[0000-0002-8502-6431]{Brian E.\ Svoboda}
\altaffiliation{Jansky Fellow of the National Radio Astronomy Observatory}
\affiliation{
    National Radio Astronomy Observatory,
    1003 Lopezville Rd,
    Socorro,
    NM 87801 USA}
\affiliation{
    Steward Observatory,
	University of Arizona,
	933 North Cherry Avenue,
	Tucson,
	AZ 85721
	USA}

\author{Yancy L.\ Shirley}
\altaffiliation{Adjunct Astronomer of the National Radio Astronomy Observatory}
\affiliation{
    Steward Observatory,
	University of Arizona,
	933 North Cherry Avenue,
	Tucson,
	AZ 85721
	USA}

\author{Alessio Traficante}
\affiliation{
    IAPS-INAF,
	via Fosso del Cavaliere,
	100,
	I-00133,
	Roma
	Italy}
\affiliation{
    Jodrell Bank Centre for Astrophysics,
	The School of Physics and Astronomy,
	The University of Manchester,
	Oxford Road,
	Manchester M13 9PL
	UK}

\author{Cara Battersby}
\affiliation{
    Department of Physics,
    University of Connecticut,
    2152 Hillside Road,
    Storrs,
    CT 06269
    USA}
\affiliation{
    Harvard-Smithsonian Center for Astrophysics,
	60 Garden Street,
	Cambridge,
	MA 02138
	USA}

\author{Gary A.\ Fuller}
\affiliation{
    Jodrell Bank Centre for Astrophysics,
	The School of Physics and Astronomy,
	The University of Manchester,
	Oxford Road,
	Manchester M13 9PL
	UK}

\author{Qizhou Zhang}
\affiliation{
    Harvard-Smithsonian Center for Astrophysics,
	60 Garden Street,
	Cambridge,
	MA 02138
	USA}

\author{Henrik Beuther}
\affiliation{
	Max Planck Institute for Astronomy,
	K{\:o}nigstuhl 17,
	D-69117 Heidelberg
	Germany}

\author{Nicolas Peretto}
\affiliation{
    School of Physics \& Astronomy,
	Cardiff University,
	Queen's building,
	The parade,
	Cardiff,
	CF24 3AA
	UK}

\author{Crystal Brogan}
\affiliation{
    National Radio Astronomy Observatory,
	520 Edgemont Road,
	Charlottesville,
	VA 22903
	USA}

\author{Todd Hunter}
\affiliation{
    National Radio Astronomy Observatory,
	520 Edgemont Road,
	Charlottesville,
	VA 22903
	USA}

%\author{
%    Brian E.\ Svoboda$^{1}$,
%    Yancy L.\ Shirley$^{1,2}$,
%    Alessio Traficante$^{3,4}$,
%    Cara Battersby$^{5}$,
%    Gary A. Fuller$^{4}$,
%    Henrik Beuther$^{6}$,
%    Qizhou Zhang$^{5}$,
%    Nicolas Peretto$^{7}$,
%    Todd Hunter$^{8}$,
%    and
%    Crystal Brogan$^{8}$,
%}
%\affil{
%	$^1$Steward Observatory,
%	University of Arizona,
%	933 North Cherry Avenue,
%	Tucson,
%	AZ 85721,
%	USA;
%	\href{mailto:svobodb@email.arizona.edu}{svobodb@email.arizona.edu} \\
%	$^2$Adjunct Astronomer,
%	National Radio Astronomy Observatory,
%	USA \\
%	$^3$IAPS-INAF,
%	via Fosso del Cavaliere,
%	100,
%	I-00133,
%	Roma,
%	Italy \\
%	$^4$Jodrell Bank Centre for Astrophysics,
%	The School of Physics and Astronomy,
%	The University of Manchester,
%	Oxford Road,
%	Manchester M13 9PL,
%	UK \\
%	$^5$Harvard-Smithsonian Center for Astrophysics,
%	60 Garden Street,
%	Cambridge,
%	MA 02138,
%	USA \\
%	$^6$Max Planck Institute for Astronomy,
%	K{\:o}nigstuhl 17,
%	D-69117 Heidelberg,
%	Germany \\
%	$^7$School of Physics \& Astronomy,
%	Cardiff University,
%	Queen's building,
%	The parade,
%	Cardiff,
%	CF24 3AA,
%	UK \\
%	$^8$National Radio Astronomy Observatory,
%	520 Edgemont Road,
%	Charlottesville,
%	VA 22903,
%	USA \\
%	\vspace{-6mm}
%}

\begin{abstract}
The initial physical conditions of high-mass stars and protoclusters remain poorly characterized.
To this end we present the first targeted ALMA Band 6 \SI{1.3}{mm} continuum and spectral line survey towards high-mass starless clump candidates, selecting a sample of 12 of the most massive candidates ($\SI{4e2}{\msun} \lesssim M_\mathrm{cl} \lesssim \SI{4e3}{\msun}$) within $d_\odot < \SI{5}{kpc}$.
The joint $12+\SI{7}{m}$ array maps have a high spatial resolution of $\lesssim \SI{3000}{au}$ (\SI{0.015}{pc}, $\theta_\mathrm{syn} \approx 0\farcs8$) and have high point source mass-completeness down to $M \approx \SI{0.3}{\msun}$ at $6\sigrms$ (or $1\sigrms$ column density sensitivity of $N = \SI{1.1e22}{cm^{-2}}$).
We discover previously undetected signposts of low-luminosity star formation from \ce{CO} $J=2\rightarrow1$ and \ce{SiO} $J=5\rightarrow4$ bipolar outflows towards 11 out of 12 clumps, showing that current MIR/FIR Galactic Plane surveys are incomplete to low- and intermediate-mass protostars ($\lbol \lesssim \SI{50}{\lsun}$), and emphasizing the necessity of high-resolution followup.
We compare a subset of the observed cores with a suite of radiative transfer models of starless cores. We find a high-mass starless core candidate with a model-derived mass consistent with $29^{52}_{15} \, \msun$ when integrated over size scales of $R < \SI{2e4}{au}$.
Unresolved cores are poorly fit by radiative transfer models of externally heated Plummer density profiles,  supporting the interpretation they are protostellar even without detection of outflows.
A high degree of fragmentation with rich sub-structure is observed towards 10 out of 12 clumps.
We extract sources from the maps using a dendrogram to study the characteristic fragmentation length scale.
Nearest neighbor separations when corrected for projection with Monte Carlo random sampling are consistent with being equal to the clump average thermal Jeans length (\ljeans; i.e., separations equal to $0.4-1.6\times\ljeans$).
In context of previous observations that on larger scales see separations consistent with the turbulent Jeans length or the cylindrical thermal Jeans scale ($\approx 3-4\times\ljeans$), our findings support a hierarchical fragmentation process, where the highest density regions are not strongly supported against thermal gravitational fragmentation by turbulence or magnetic fields.
\end{abstract}


%%%%%%%%%%%%%%%%%%%%%%%%%%%%%%%%%%%%%%%%%%%%%%%%%%%%%%%%%%%%%%%%%%%%%%%%%%%%%%%%
%			       Begin Body Text
%%%%%%%%%%%%%%%%%%%%%%%%%%%%%%%%%%%%%%%%%%%%%%%%%%%%%%%%%%%%%%%%%%%%%%%%%%%%%%%%

\section{Introduction}\label{sec:Introduction}
High-mass stars ($M_* > \SI{8}{\msun}$) strongly influence the evolution of galaxies and the ISM, yet many fundamental questions remain to be answered concerning the incipient phases of high-mass star formation \citep[e.g.,][]{beuther07,tan14,motte17}.
Observational constraints on the initial physical conditions of protocluster evolution are a necessary prerequisite to improved understanding of high-mass star and cluster formation.
Of particular importance are observations of the quiescent environments before the initial conditions are disrupted by the extreme radiative and mechanical feedback of high-mass stars.
Thus our understanding of both how cluster formation is initiated and the ensuing protocluster evolution depend on identifying and constraining the physical properties of representative samples of starless molecular cloud ``clumps''\footnote{
In this paper we use the term ``core'' to refer to a dense gas structure that is $\sim\! \SI{0.1}{pc}$ in size and likely to form a single or bound multiple stellar system.
Such cores are embedded within larger scale ``clumps'' that are dense gas structures likely to form a stellar association or cluster, and are $\sim\! \SI{1}{pc}$ in size and $\sim\! \num{e2}-\SI{e4}{\msun}$ in mass \citep[c.f.][]{bergin07}.
}.

Recent blind surveys of dust continuum emission at (sub-)millimeter and far-infrared (FIR) wavelengths of the Galactic Plane have identified large statistical samples of clumps, enabling the discovery of those in the earliest evolutionary phases.
Such surveys include the Bolocam Galactic Plane Survey\footnote{Data products can be downloaded from \url{https://irsa.ipac.caltech.edu/data/BOLOCAM\_GPS/}} \citep[BGPS;][]{aguirre11,rosolowsky10,ginsburg13} at \SI{1.1}{mm}, ATLASGAL\footnote{\url{http://www3.mpifr-bonn.mpg.de/div/atlasgal/}} at \SI{870}{\um} \citep{schuller09,contreras13,csengeri14}, JCMT Galactic Plane Survey\footnote{\url{http://apps.canfar.net/storage/list/JPSPR1}} at \SI{850}{\um} \citep[JPS]{eden17}, and \emph{Herschel} Hi-GAL at 70, 160, 250, 350, and \SI{500}{\um} \citep{molinari10,molinari16a}. 
Starless clump candidates (SCCs) are identified by cross-matching clump catalogs to catalogs of star formation indicators and selecting clumps unassociated with any indicators.
These indicators include \SI{70}{\um} compact sources, color-selected young stellar objects (YSOs), \ce{H2O} and \ce{CH3OH} masers, and \uchii\ regions in \cite{svoboda16} for the BGPS and in \cite{yuan17} for ATLASGAL, in total identifying more than $\gtrsim \num{2e3}$ SCCs in the inner-Galaxy.
In addition, more than $\gtrsim \num{e4}$ clumps without \SI{70}{\um} sources have been identified from the Hi-GAL survey \citep{traficante15,elia17}.
In this study we aim to systematically study a representative sample of the highest mass SCCs within \SI{5}{kpc} in order to understand the fragmentation characteristics at high-spatial resolution, identify potential high-mass starless cores ($M \gtrsim \SI{30}{\msun}$, $R \lesssim \SI{0.1}{pc}$), and search for previously undetected low luminosity protostellar activity.

A variety of physics, including thermal gas pressure, turbulence, magnetic fields, and the geometry of filaments and density gradients, likely play a role in the fragmentation of molecular clouds and the resultant dense core populations.
Recent high-resolution observations with millimeter and sub-millimeter interferometers of high-mass clumps with little sign of star formation reveal significant fragmentation at the early stage of cluster formation \citep{zhang09,zhang11,wangk11,wangk12,wangk14,zhang15,lu15b,beuther15b,sanhueza17}.
These studies found that the most massive fragments in the clumps are at least ten times greater than the thermal Jeans mass, indicating that additional support from turbulence and/or magnetic fields are required.
Most of these studies focused on individual clumps and typically have not had the sensitivity to adequately detect fragments of a thermal Jeans mass (detections of $\gtrsim \! \SI{2}{\msun}$ at $4\sigrms$).
In contrast, the fragmentation scales in nearby molecular clouds have been studied extensively with recent notable analyses towards Serpens \citep{friesen17}, Orion Integral Shaped Filament \citep{kainulainen17a}, and Perseus \citep{pokhrel18}.
These studies find support for hierarchical, scale-dependent fragmentation with separations corresponding to a range between thermal Jeans fragmentation and thermal filamentary gravitational fragmentation.
It is not understood how these results extend towards earlier evolutionary stages in massive SCCs which are the focus of this work.

Publicly available millimeter and FIR Galactic Plane survey observations do not have sufficient angular resolution at $\sim \! 20-30\arcsec$ ($\sim \! \SI{0.5}{pc}$ at $\SI{4}{kpc}$) to study the sub-structure and dense core properties in distant SCCs.
The high-mass pre-stellar core candidate G028-C1S ($M_\mathrm{c} \sim \SI{60}{\msun}$) studied in \cite{tan13} for example was only identified as protostellar until interferometric followup of outflow tracers \citep{tan16,feng16}.
High-mass SCCs remain largely unstudied at high-spatial resolution owing to their historical difficulty in identification and typically large heliocentric distances, with only a handful of studies on individual objects to date \citep{beuther15a,sanhueza17}.
In particular the high-mass starless clump candidate ``MM1'' of IRDC G28.23--0.19 \citep{sanhueza13} has been studied in detail to determine that it is devoid of star formation indicators, including $3.6-\SI{70}{\um}$ point sources, $\mathrm{H_2O}$ and $\mathrm{CH_3OH}$ masers \citep{wangy06,chambers09}, and radio continuum \citep{battersby10,rosero16}.
The global physical properties of G28.23--0.19 MM1 (corresponding to BGPS catalog clump number 4649) are similar to the average properties of the SCCs presented in this work.
G28.23--0.19 MM1 is high-mass, cold, compact, and dense (i.e., $M_\mathrm{cl} \approx \SI{1.5e3}{\msun}$, $T_\mathrm{K} \approx \SI{12}{K}$, $R = \SI{0.6}{pc}$, $n \approx \SI{3e4}{cm^{-3}}$; \citeauthor{sanhueza17}\ \citeyear{sanhueza17}).
However the sensitivity and sample size of dense cores are not sufficient for a precise measurement of the fragmentation scale, and represents only a single clump.
In this survey we present observations on a sample of $12$ clumps that are selected from a blind Galactic Plane survey and are among the most massive SCCs.
%The sources detected with the SMA are of comparable integrated flux density, $S_\mathrm{1.3} \sim 5 - \SI{8}{mJy}$, to the brighter sources detected by ALMA but are only marginally resolved.
%The models presented in \S\ref{sec:Modeling} suggest that at spatial resolutions of $\sim \! \SI{2e4}{au}$ ($d_\odot = \SI{5.1}{kpc}$, \citeauthor{sanhueza12} \citeyear{sanhueza12}), compact, low- to intermediate-mass protostellar objects would be indistinguishable from a typical dense, starless core model, as the spatial resolution would be greater than or equal to the diameter of the flat-region (i.e.\ for $R_\mathrm{flat} \lesssim \SI{1e4}{au}$).

Existing large samples of SCCs have been primarily identified through the non-detection of coincident Hi-GAL \SI{70}{\um} sources \citep{veneziani13,traficante15,svoboda16,elia17} which is less affected than shorter wavelength \SI{8}{\um} or \SI{24}{\um} observations by both local extinction and from contamination of evolved stars (principally those on the asympototic giant branch).
The completeness of the \SI{70}{\um} maps to protostar bolometric luminosity, $L_\mathrm{bol}$, is affected by the survey depth and complex structure in the foreground and background emission which hinders the clear identification of compact sources.
\cite{svoboda16} calculate the $L_\mathrm{bol}$ completeness function for Hi-GAL \SI{70}{\um} compact sources associated to BGPS clumps and the respective distribution of heliocentric distances and find that for clumps with low \SI{70}{\um} backgrounds ($\sim \! 500 - \SI{1000}{MJy.sr^{-1}}$) the $90\%$ completeness limit is $L_\mathrm{bol} = \SI{50}{\lsun}$ \cite[see \S3.2.4 in][]{svoboda16}, which is greater than $\gtrsim 95\%$ of YSOs in the Gould's Belt \citep[n.b.~median \SI{1}{\lsun};][]{dunham14}.
Faint \SI{24}{\um} sources coincident with the clump column density peaks towards $9/18$ of \SI{70}{\um} dark SCCs suggest likely embedded intermediate-mass star formation that is undetected in the Hi-GAL observations \citep{traficante17}.
However, it is currently unknown what degree of star formation has initiated, if at all, in SCCs without sensitive and unambiguous tracers of protostellar activity such as bipolar molecular outflows.
Systematic observations of SCCs at high-resolution are necessary to determine what degree (if any) of low-luminosity star formation has begun in SCCs, with important implications for the protostellar accretion history.

The principal theories of high-mass star formation in dense Galactic molecular cloud clumps are the monolithic collapse of turbulent cores in virial equilibrium \citep{mckee02,mckee03,hosokawa09} and the accretion of sub-virial cores through gravitionally-driven cloud inflow \citep{smithr09,hartmann12}.
The latter replace the competitive Bondi-Hoyle accretion of cores \citep{bonnell01,wangp10} with cores fed the gas reservoir through inflowing streams.
The turbulent core model predicts monolithic high-mass starless cores whereas the competitive model predicts a fragmentation of cores near the thermal Jeans mass.
The existence of high-mass starless cores is a key distinction between these models, yet few, if any, observational candidates are known \citep{kong17}, and some promising candidates have revealed embedded protostellar activity upon further observational investigation \citep[i.e.~G028-C1S][]{tan13,tan16,feng16b}.
Irrespective of the specific theoretical model, measurements of the fragmentation properties at early evolutionary phases provide valuable observational constraints on the initial physical conditions of high-mass star and cluster formation.
To this end, we perform a systematic search for high-mass starless cores towards massive SCCs with the Atacama Large Millimeter/submillimeter Array.

In this paper we present a systematic survey of $12$ high-mass SCCs at sub-arcsecond resolution.
We present our sample selection, observational setup, and data reduction methodology in Section \ref{sec:Obs}.
We describe detections of previously unknown, low-luminosity protostellar activity in Section \ref{sec:Activity} and the modeling of continuum sources in Section \ref{sec:Modeling}.
We measure and analyze the fragmentation scale between sources in Section \ref{sec:FragmentationScale}, discuss the implications in Section \ref{sec:Discussion}, and report our conclusions in Section \ref{sec:Conclusions}.
In Appendix \ref{apx:Models} we include a detailed description of the setup and computation of radiative transfer models analyzed in Section \ref{sec:Modeling}.

%How do high-mass starless clumps fragment into dense cores that form individual or bound multiple stars?
%Monolithic starless cores capable of forming a high-mass star are of particular interest as potential progenitors of high-mass stars.
%Measurements of the star formation efficiency in dense cores \citep{lada10} and the offset between the core mass function and stellar initial mass function of $\mathrm{IMF/CMF} \approx 0.3$ (Motte) suggest high-mass starless cores require $M \gtrsim \SI{30}{\msun}$.
%Do such very massive starless cores exist?
%What is the importance of sub-virial competitive accretion versus a virialized turbulent core accretion in forming intermediate and high-mass stars?

%\begin{itemize}
%    \item Are high-mass starless cores found? Detail theoretical considerations for turbulent core and competitive core accretion, namely in the context of whether high-mass turbulently supported starless cores should, or are expected to, exist or not.
%    \item What physical processes set the fragmentation scale before high-mass star formation begins? Detail how LMSF that precedes HMSF or coeval LMSF \& HMSF are significant in interpreting current theories (such as the Rowan Smith competitive accretion-like simulations).
%\end{itemize}

\section{Observations}\label{sec:Obs}
\subsection{Sample Selection}\label{ssec:ObsSample}
\begin{figure*}
    \centering
    \includegraphics[width=0.99\textwidth]{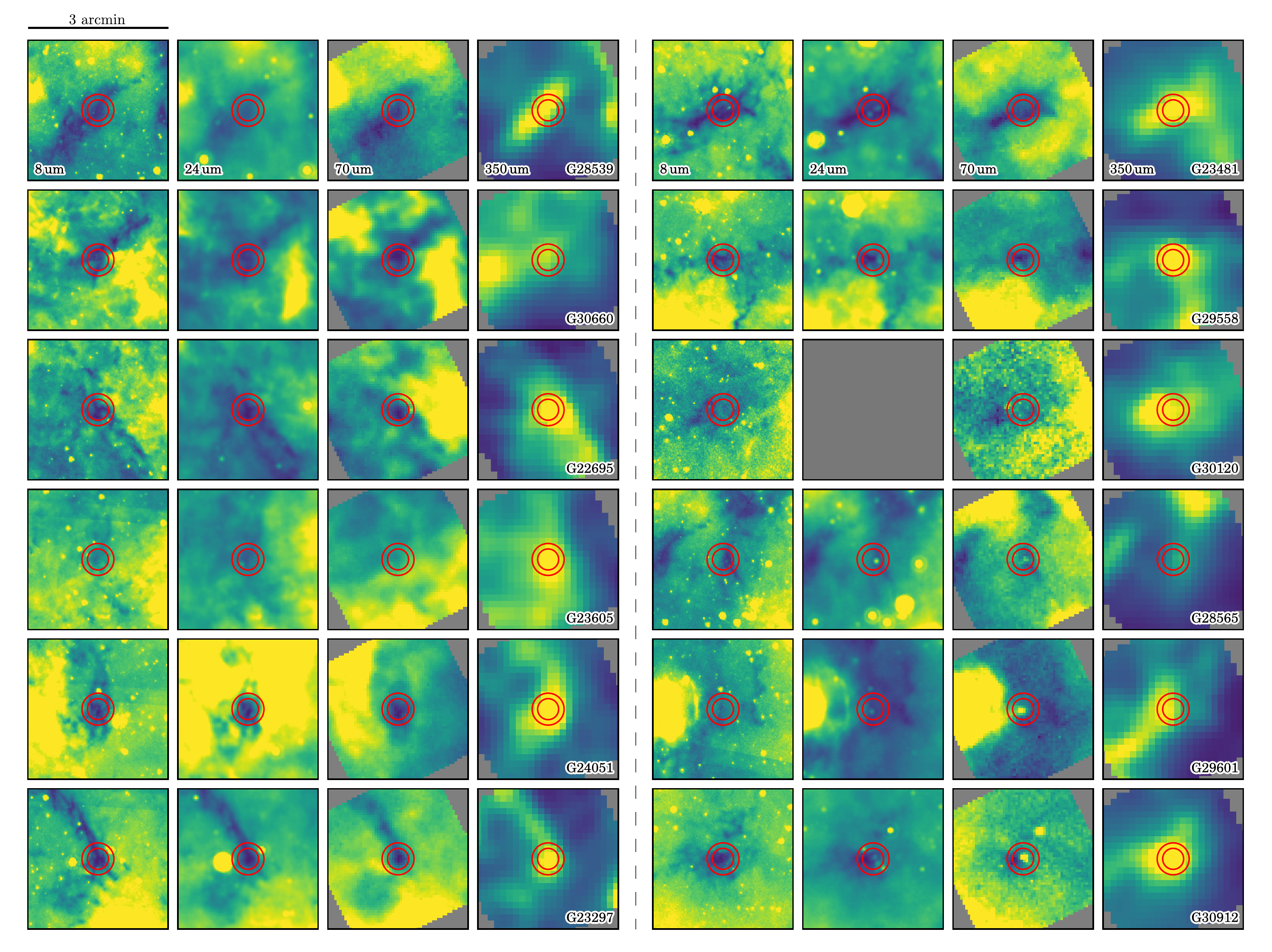}
\caption{
Mid- and far-infrared $3\arcmin \times 3\arcmin$ maps of the clumps in the survey sample, showing GLIMPSE \SI{8}{\um}, MIPSGAL \SI{24}{\um}, and Hi-GAL \SI{70}{\um} and \SI{350}{\um}.
The ALMA Band 6 single pointings target the peak flux positions derived from the BGPS \SI{1.1}{mm} observations.
The inner and outer red circles show the 50\% (27\arcsec) and 20\% (40\arcsec) power points of the primary beam for the ALMA \SI{12}{m} array images.
Clumps from \cite{svoboda16} were selected to have no detected indicators of star formation activity such as embedded \SI{70}{\um} sources, \ce{H2O} and \ce{CH3OH} masers, and \uchii regions.
Clumps from \cite{traficante15} were selected to be \SI{70}{\um} dark using an automated extraction, one of which shows a marginal detection and two of which show clear detections upon visual inspection.
Note that G30120 at $b \approx \ang{-1.1;;}$ is outside the MIPSGAL survey and does not contain \textit{Spitzer} \SI{24}{\um} data.
}
\label{fig:IrMosaic}
\end{figure*}

We have identified SCCs through the combined catalogs and images of primarily two dust continuum Galactic Plane surveys: (1) an evolutionary analysis of BGPS \SI{1.1}{mm} \citep[][hereafter \citetalias{svoboda16}]{svoboda16}, and (2) comparison of the \cite{peretto09} infrared dark cloud (IRDC) catalog with Hi-GAL images \citep{traficante15}.
The BGPS observed between $\ang{-10} < \ell < \ang{90}$ with $|b| < \ang{0.5}$ (expanding to $|b| < \ang{1.5}$ for selected $\ell$) at $\lambda_{\rm c} = \SI{1.12}{mm}$ with a $\theta_\mathrm{hpbw} = 33\arcsec$ synthesized angular resolution.
In the region $\ang{10} < \ell < \ang{65}$ the BGPS has been compared to a diverse set of a observational indicators for star formation activity (\citetalias{svoboda16}).
These indicators include compact \SI{70}{\um} sources, mid-IR color-selected YSOs, \ce{H2O} masers, Class II \ce{CH3OH} masers, extended \SI{4.5}{\um} outflows, and \uchii regions.
From the sample of more than $2500$ SCCs in the combined samples of \citetalias{svoboda16} and \cite{traficante15}, we target the $12$ highest mass SCCs within $d_\odot < \SI{5}{kpc}$.
Point sources at \SI{70}{\um} were identified by visual inspection in \citetalias{svoboda16} and by an automated extraction in \cite{traficante15}.
Three clumps (G28565, G29601, and G309120 which were initially determined from the automated extraction to be dark at \SI{70}{\um}, upon closer scrutiny by visual inspection show association to weak sources.
Among the 12 ALMA targets, nine have no detectable point source emission from Hi-GAL \SI{70}{\um} (flag 0 in \citetalias{svoboda16}), two have low-confidence or marginal detections (flag 4, G28565 \& G29601), and one has bright, compact detection (flag 1, source G30912).
We emphasize that starless clump candidates are designated based on the observational data sets and identification techniques used, and that these factors are reflected in the completeness and purity of the resulting catalogs of SCCs.
Table \ref{tab:TargetPositions} details the target positions and velocities.
Table \ref{tab:PhysicalProperties} details the physical properties of the sample and Figure \ref{fig:IrMosaic} shows images of the clumps at wavelengths from \SI{8}{\um} to \SI{350}{\um}. 

\begin{figure}
    \centering
    \includegraphics[width=0.47\textwidth]{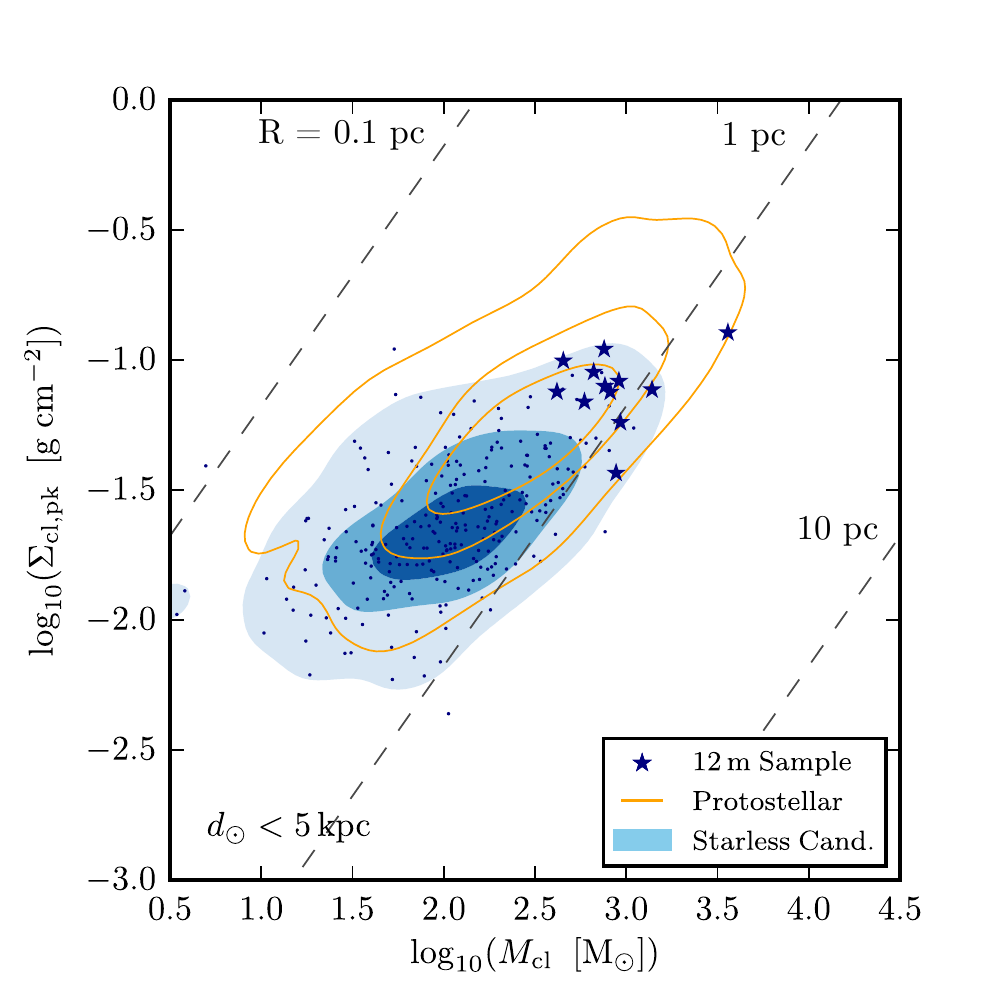}
\caption{
Peak mass surface density $\Sigma_\mathrm{cl,pk}$ versus total mass $\mclump$.
Values are derived from the BGPS at \SI{1.1}{mm} ($\theta_\mathrm{hpbw} = 33\arcsec$) for clumps with well-constrained distances $d_\odot < \SI{5}{kpc}$.
Starless clump candidates (blue points, contours), protostellar clumps (orange contours), and the ALMA sample (blue stars).
Total masses of the sample range between $\mclump \approx 400-\SI{3000}{\msun}$ and $\Sigma_\mathrm{cl,pk} \sim \SI{0.1}{g.cm^{-2}}$.
The dashed lines show $\Sigma_\mathrm{cl,pk}$ as a function of $\mclump$ for constant radii at \SI{0.1}{pc}, \SI{1.0}{pc}, and \SI{10}{pc}.
}
\label{fig:Tanogram}
\end{figure}

The clump average physical properties in \citetalias{svoboda16} are shown in Figure \ref{fig:Tanogram}, plotting peak mass surface density $\Sigma_\mathrm{cl,pk}$ (at $\theta_\mathrm{hpbw} = 33\arcsec$) and total mass \mclump, for sources with well-constrained distances less than $d_\odot < \SI{5}{kpc}$ and $\ang{10} < \ell < \ang{65}$.
Protostellar clumps and SCCs are plotted, where SCCs have quiescent background emission and no detected compact sources from the Hi-GAL \SI{70}{\um} images (flag 0, see \citetalias{svoboda16} \S3.2.4).
Protostellar clumps are typically higher in both mass and mass surface density compared to SCCs.
The ALMA targets are shown, occupying the highest $\Sigma_\mathrm{cl,pk}$ and $\mclump$ portions of the SCC distribution where typical values for the sample are $\mclump \sim \SI{800}{\msun}$ and $\Sigma_\mathrm{cl,pk} \sim \SI{0.1}{g.cm^{-2}}$ (measured over $\sim \! \SI{0.6}{pc}$ scales).
G28539 in particular stands out as the most massive clump in the sample at $\mclump \sim \SI{3e3}{\msun}$ and also the highest peak mass surface density.

Assuming a star-formation efficiency of $\epsilon_\mathrm{sf} = 0.3$ and a standard stellar initial mass function \citep[IMF;][]{kroupa01}, a \SI{320}{\msun} clump meets the criteria of forming a \SI{8}{\msun} star (see \citetalias{svoboda16} \S6.1).
All of the clumps that comprise this sample are above this mass threshold, and are similarly above the mass-radius relationship for high-mass star formation proposed by \citep{kauffmann10}\footnote{\cite{kauffmann10} define the prescription $M \leq \SI{580}{\msun} (R / \si{pc})^{1.33}$ for a clump to form high-mass stars. The pre-factor has been scaled for consistency with the dust opacity used in this work. For radius $R = \SI{0.8}{pc}$, this relation yields a mass threshold of $M \approx \SI{430}{\msun}$.}.
However in practice it is difficult to assess the high-mass star formation potential of clumps beyond these simple heuristics.
It should be kept in mind that if the star formation efficiency of the targets is substantially lower than the typical assumed value of $\epsilon_\mathrm{sf} = 0.3$ then they are unlikely to form high-mass stars.

\subsection{ALMA Band 6}\label{ssec:ObsAlma}
As part of Atacama Large Millimeter/submillimeter Array\footnote{ALMA is operated by the National Radio Astronomy Observatory, which is a facility of the National Science Foundation, operated under cooperative agreement by Associated Universities, Inc.} (ALMA) Cycle 3 program 2015.1.00959.S, we observed 12 clumps in Band 6 in a compact configuration (C36-2; joint $12+\SI{7}{m}$ array baselines range from $\sim \! 9 - \SI{450}{m}$).
Data were taken between $3-20$ March, 2016, for the \SI{12}{m} array and between 30 April to 19 August, 2016, for the \SI{7}{m} array.
Including time for calibration and overheads, the \SI{12}{m} array observations lasted for approximately \SI{12}{hr}, with typical precipitable water vapor of \SI{1.5}{mm}.
Titan and J1733--1304 were used as flux calibrators, J1751+0939 to calibrate the bandpass, and J1743--0350 and J1830+0619 to calibrate the time-dependent gains.
Identical \SI{1}{hr} scheduling blocks were configured to interleave and observe all 12 targets within the same block, and because sources are within a \ang{5} radius on the sky ($22\fdg7 < \ell < 30\fdg9$), the same calibrators can be used.
Thus due to nearly identical observing conditions, the individual maps have similar $uv$-coverage, atmospheric noise, and beam size.

Positions for the sample were chosen from the BGPS \SI{1.1}{mm} continuum peak flux density position, and compared for consistency with the ATLASGAL \SI{870}{\um} peak emission and Hi-GAL \SI{70}{\um} peak absorption (when present) positions.
The Band 6 receiver was configured in dual-polarization mode with lower and upper sidebands centered near 215 and \SI{230}{GHz}, respectively.
The observations targeted each clump peak with a single pointing with half-power beam width (HPBW) of the measured primary beam $26\farcs6$ ($\sim \! \SI{0.5}{pc}$ at $d_\odot = \SI{4}{kpc}$) and 20\%-power beam width of $40\arcsec$ ($\sim \! \SI{0.8}{pc}$), the effective limit of the \SI{12}{m} array field of view.

\subsubsection{ALMA \SI{1.3}{mm} continuum reduction}\label{sssec:AlmaContinuumReduction}
\begin{figure*}
    \centering
    \includegraphics[width=0.99\textwidth,trim={12mm 3mm 19mm 7mm},clip]{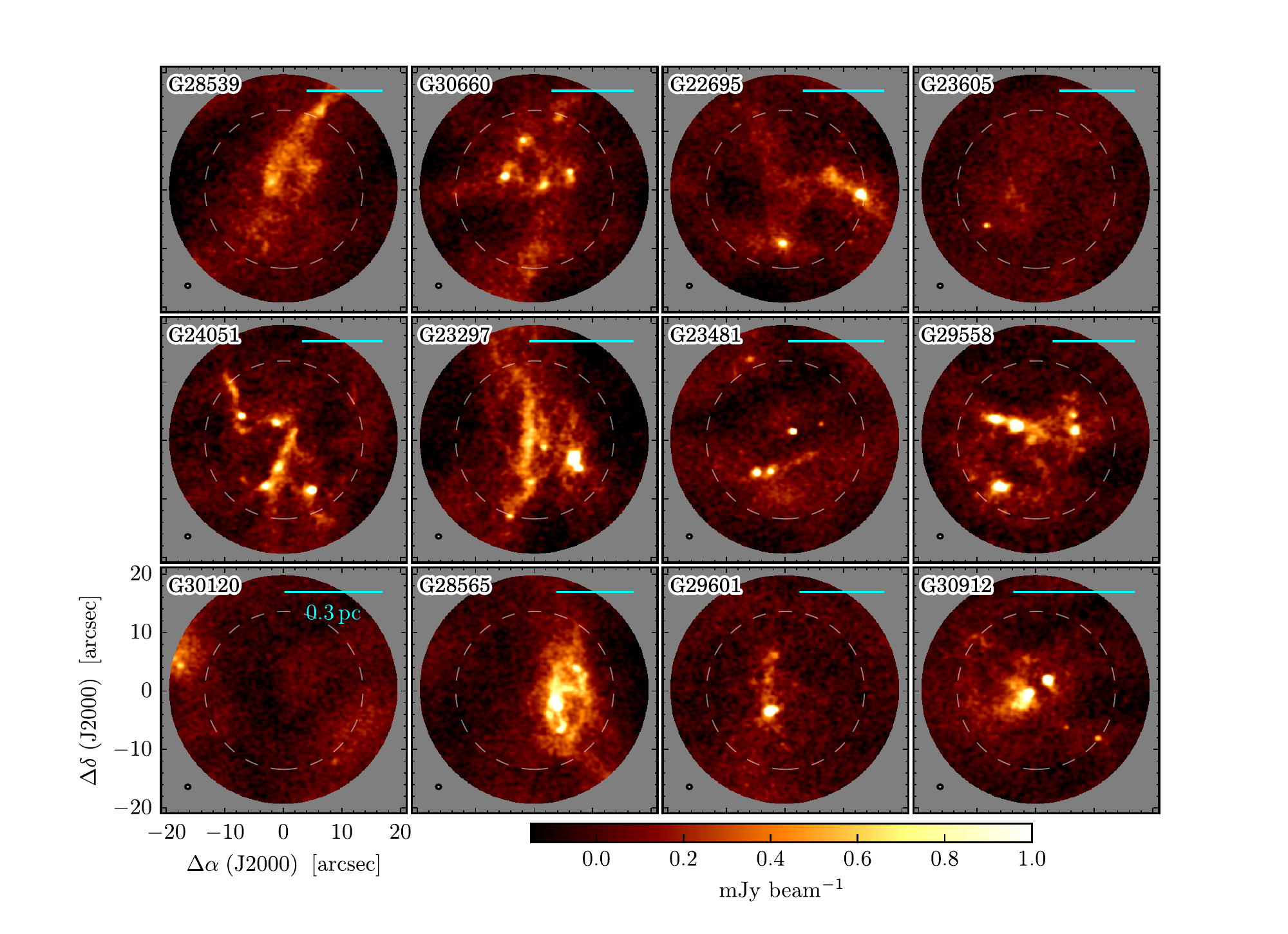}
\caption{
ALMA $12+\SI{7}{m}$ array jointly deconvolved \SI{230}{GHz} line-free continuum images.
The clumps show a rich degree of fragmentation with multiple condensations connected by filamentary structures, although sources G30120 and G23605 are largely devoid of detected emission on the scale of the synthesized beam ($0.85\arcsec \times 0.75\arcsec$, visualized at lower-left).
The images are uncorrected for primary beam attenuation for visual display purposes.
The color scale ranges from $-0.15$ to \SI{1.0}{\milli\jy\per\beam} on a linear scaling.
The scalebar (cyan) visualizes \SI{0.3}{pc} at the clump heliocentric distance.
The dashed circle shows the half-power beam width ($27\arcsec$) and the image extends down to the 20\% power point ($40\arcsec$).
}
\label{fig:ContinuumTwelve}
\end{figure*}

Data reduction was performed using \casa\ (version 4.7.134-DEV, r38011, for consistency with QA2 delivered products).
Line-free continuum visibilities were created by flagging channels contaminated by spectral lines, where the input spectral windows were further visually inspected to check for emission at unexpected velocity ranges, partitioned out into a new measurement set with the \texttt{split} task, and channel averaged to \SI{25}{MHz} to avoid bandwidth smearing.
Together, this yields $\approx \SI{3.5}{GHz}$ of dual polarization continuum bandwidth. The continuum image root mean square (RMS, $\sigrms = \sqrt{\sum_n I_n^2 / n}$) is measured for each \textsc{clean}ed image within a region that excludes identified emission using the \texttt{casaviewer} tool.
None of the images are dynamic range limited with peak image intensity divided by the RMS less than 200.
We estimate the fiducial mass sensitivity given $T_\mathrm{K}(\ce{NH3}) \approx \SI{12}{K}$, thermally coupled gas and dust ($T_\mathrm{d}=T_\mathrm{K}$), and OH5 dust opacity $\kappa(\lambda = \SI{1.3}{mm}) = \SI{0.899}{cm^2.g^{-1}}$).
The methods for deriving dust mass values from the continuum emission are discussed in more detail in \S\ref{sec:Modeling}.
The joint $12+\SI{7}{m}$ continuum was then iteratively \textsc{clean}ed with manual masking using the \texttt{tclean} task using the multiscale deconvolver and a robust weighting of 1, down to a brightness threshold of $2-3\sigrms$.
An image cell size of $0\farcs1$ was used for all continuum and spectral line maps.
Self-calibration was not applied because the brightest sources in the image are only a few mJy, and not sufficiently bright such that a conservative self-cal produces a noticeable improvement without also increasing the image noise.
The resultant images have a synthesized beam size of $\theta_\mathrm{maj} \approx 0\farcs85$ by $\theta_\mathrm{min} \approx 0\farcs75$ ($0\farcs8$ angular diameter yields $2800-\SI{3800}{au}$ at $d_\odot=3.5-\SI{4.8}{kpc}$).
The continuum images are shown in Figure~\ref{fig:ContinuumTwelve}.

\subsubsection{ALMA spectral line reduction}\label{sssec:AlmaLineReduction}
The flexibility of the ALMA correlator enabled simultaneous observation of several molecular line transitions.
Table \ref{tab:Correlator} reports the details of the correlator configuration.
We observed nine spectral windows (SPWs) with one wide-band, low-spectral resolution window centered at \SI{233.8}{GHz} and eight high-spectral resolution windows centered on lines of interest.
Table \ref{tab:LineList} reports the transition quantum numbers, rest frequencies, and upper energy levels ($E_\mathrm{u}/k$).
The SPWs containing the \ce{H2CO} and \ce{CH3OH} transitions have a spectral resolution of \SI{0.34}{\kms} and the other line SPWs have \SI{0.68}{\kms} resolution.
Line rest frequencies were taken from a combination of the SLAIM\footnote{\href{http://splatalogue.net}{http://splatalogue.net}} (F.~J.~Lovas, private communication, \citeauthor{slaim}~\citeyear{slaim}) and the CDMS \citep{cdms} online spectroscopic databases.
The line SPWs from the $12+\SI{7}{m}$ arrays were jointly imaged using the CASA task \texttt{tclean} with a Briggs robust parameter of $1.0$, cell size of $0\farcs1$, and re-gridded to common spectral resolutions listed in Table \ref{tab:LineList}.
We find a typical RMS noise levels in the image cubes of \SI{1.8}{mJy/(\kms)} (i.e., \SI{2.2}{mJy} per \SI{0.68}{\kms} channel or \SI{3.0}{mJy} per \SI{0.34}{\kms} channel) or \SI{71}{mK/(\kms)} when converted to brightness temperature units (HPBW beam size of $0\farcs85\times0\farcs75$).

In this work we inspect the line image cubes for detection of emission, and for the presence of outflows traced by \ce{CO} and \ce{SiO}, but do not \textsc{clean} the data cubes.
Due to the lack of full $uv$-coverage, the CO maps in particular show strong effects of spatial filtering near the systematic velocities that make the deconvolution process complex and error prone.
Detailed analysis of the spectral line data is left to a future work.
Table \ref{tab:Detections} lists the detection flags per target for the continuum, molecular lines, and outflows.
Features are considered detections if they have peak intensities $\geq 7\sigrms$ (``D''), weak detections if between $5 - 7\sigrms$ (``W''), and non-detections if $\leq 5\sigrms$.
Targets that exhibit bipolar outflows in CO or SiO are flagged ``B'' (discussed below in \S\ref{ssec:OutflowDetections}).

\subsection{Image fidelity and MIR comparison}\label{ssec:MirComparison}
\begin{figure}
    \centering
    \includegraphics[width=0.47\textwidth]{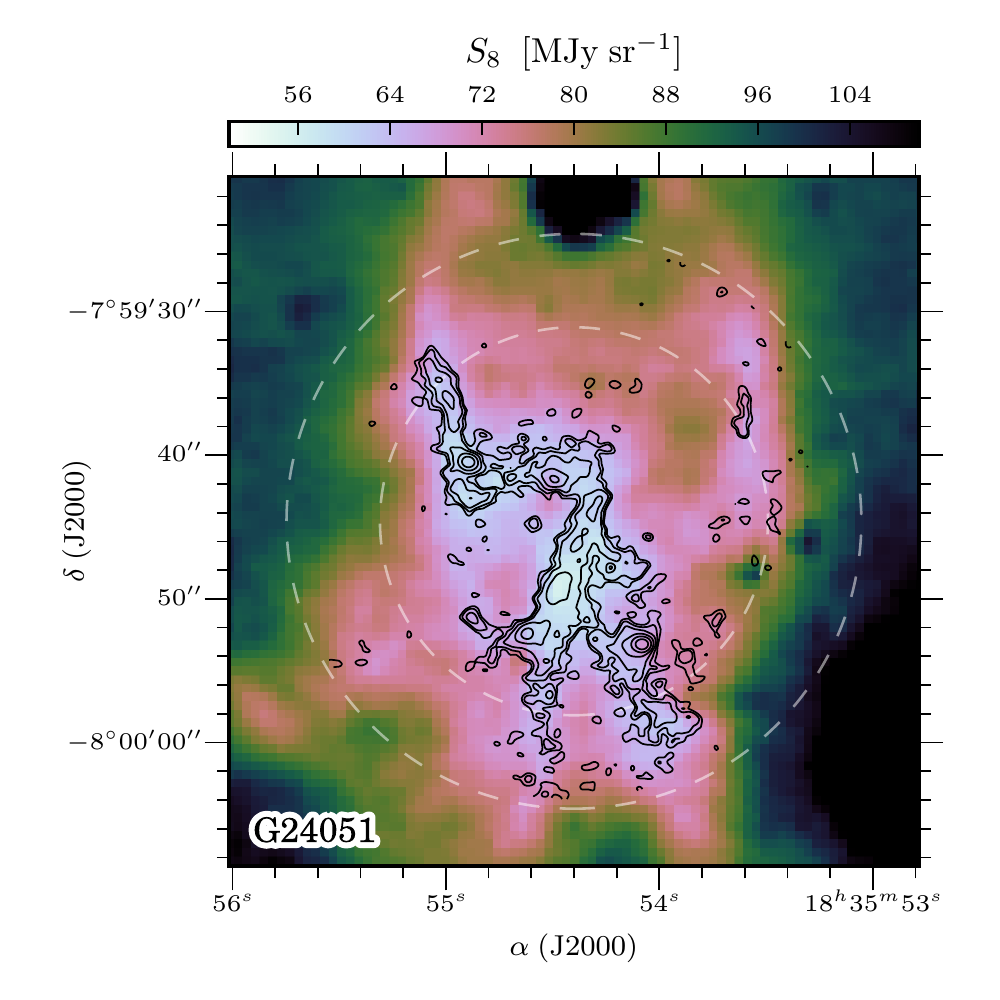}
\caption{
Comparison between the ALMA 230 GHz continuum (black lines) and IRAC \SI{8}{\um} intensity $S_8$ (color map, inverted) for clump G24051.
Good spatial correspondence is observed between the sub-structure in the ALMA continuum and the highest extinction features in the GLIMPSE \SI{8}{\um} map ($\theta_\mathrm{fwhm} \approx 2\arcsec$).
The continuum images are shown without correction for primary beam attenuation for visual display purposes, and the contours are at steps of 2, 3, 5, 10, 20, and $40\sigrms$.
The dotted lines show the 50\% and 20\% power points of the ALMA primary beam.
}
\label{fig:EightMicronTau}
\end{figure}

The dense gas features revealed in the continuum maps clearly show hierarchical structure, with bright ridges, filaments, and cores contained within larger, lower surface brightness features.
Given the complexity within the maps and the systematic uncertainties of imaging, we compare the continuum images to an additional measure of gas column density at comparable resolution, MIR extinction.
For appropriate configurations of distance and the MIR radiation field, clumps can appear associated with \SI{8}{\um} absorption features (EMAFs), where high column densities at close distances typically yield the strong MIR shadows that identify infrared dark clouds (IRDCs).
MIR extinction mapping has the dual advantages of comparatively high-resolution, insensitivity to dust temperature, and lack of spatial filtering.
We use the \spitzer\ GLIMPSE \citep{benjamin03,churchwell09} IRAC Band 4 ($\lambda_{\rm c} = \SI{7.9}{\um}$, $2\arcsec$ FWHM) mosaic to show the EMAF contrast.
Figure \ref{fig:EightMicronTau} presents a map of the flux density $S_8$ with the ALMA \SI{230}{GHz} continuum for source G24051 overlaid.
The dense gas structures observed in the millimeter continuum show a remarkable consistency when compared with the column density features inferred from the of MIR contrast.
This holds similarly true for the other clumps in the sample, as all show at least some MIR extinction.
Qualitatively this good correspondence supports the fidelity of the emission structure detected in the ALMA maps.

\subsection{Core identification and dendrogram}\label{ssec:CoreIdentification}
\begin{figure*}
    \centering
    \includegraphics[width=0.99\textwidth,trim={12mm 7mm 19mm 7mm},clip]{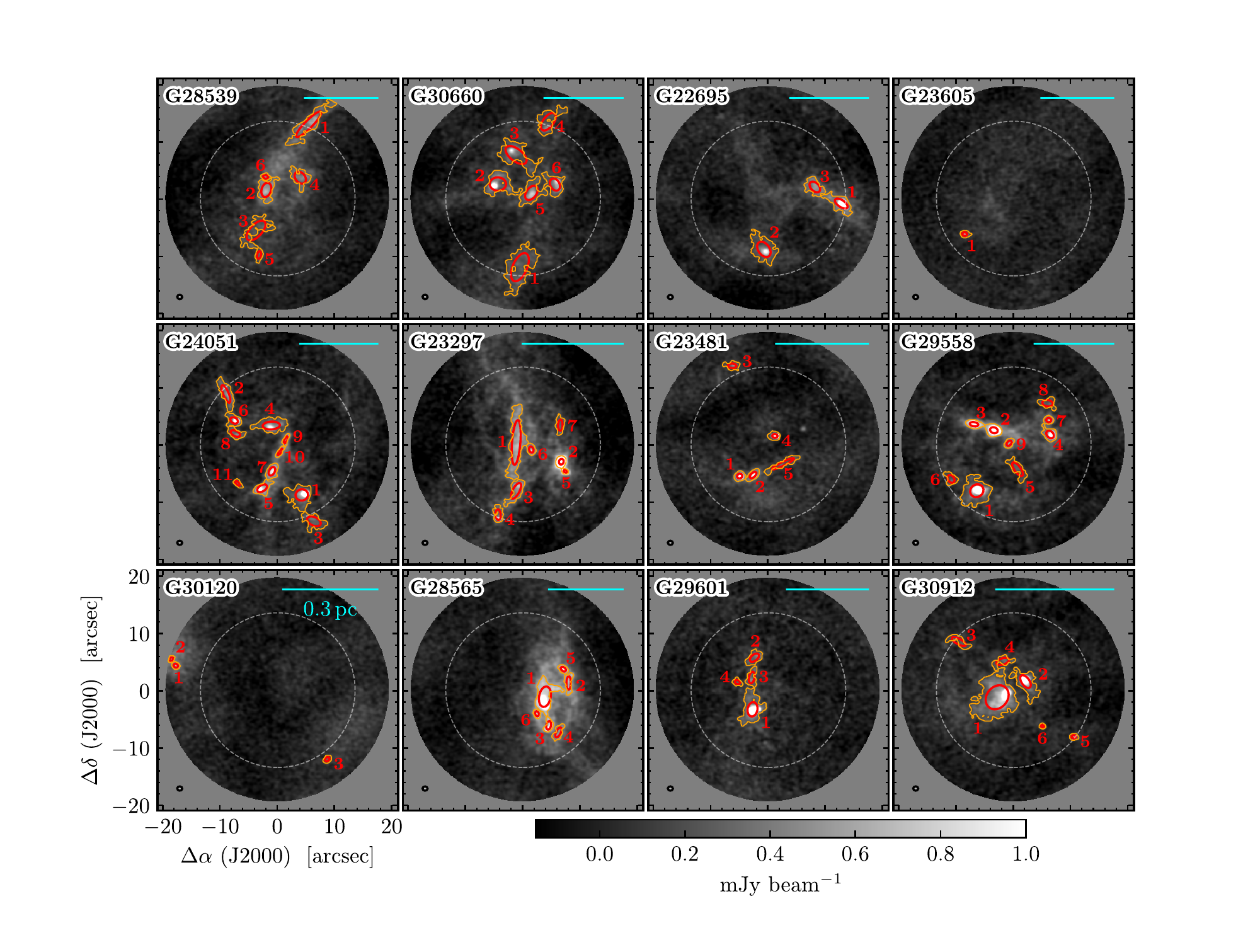}
\caption{
Dendrogram extracted dense gas sub-structures (orange contour) over-plotted on the ALMA $12+\SI{7}{m}$ array jointly deconvolved \SI{230}{GHz} line-free continuum images.
Elliptical sources are visualized (red ellipses).
Sub-structures are labeled by their catalog number from Table \ref{tab:CoreProperties}.
The maps are uncorrected for primary beam attenuation for visual display purposes.
The color scale ranges from $-0.15$ to \SI{1.0}{\milli\jy\per\beam} on a linear scaling.
The scalebar (cyan) visualizes \SI{0.3}{pc} at the clump heliocentric distance.
The dashed circle shows the half-power beam width ($27\arcsec$) and the image extends down to the 20\% power point ($40\arcsec$).
}
\label{fig:CorePositions}
\end{figure*}

\begin{figure}
    \centering
    \includegraphics[width=0.47\textwidth]{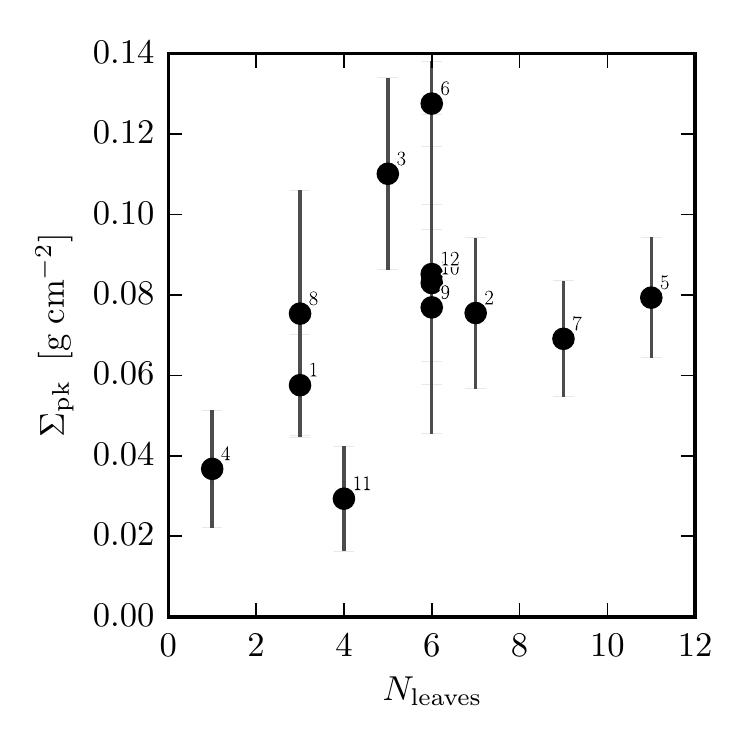}
\caption{
Peak clump mass surface density from the BGPS \SI{1.1}{mm} data versus the number of leaves (i.e.\ dendrogram leaves) per clump from the ALMA observations.
The data hints at an increasing trend of higher mass surface density clumps associated with a higher degree of fragmentation.
}
\label{fig:ClumpScatterSigma}
\end{figure}

In order to analyze the fragmentation scale we first identify dense gas sub-structures using a segmentation algorithm.
The nature of the tree data-structure in the dendrogram algorithm makes it well-suited to identifying and categorizing structure in images with hierarchical structure \citep[see][]{rosolowsky08b}, as opposed to a simpler segmentation algorithms, such as that done with a seeded watershed algorithm \citep[e.g., \textsc{clumpfind};][]{williams94}.
We use the open source Python software library \texttt{astrodendro} to create the dendrogram and catalog of cores.
The dendrogram has three principal tunable parameters, defining a minimum threshold value $v_\mathrm{min}$ setting the floor or outer boundary of each tree, the minimum contrast or step size $\delta_\mathrm{step}$ between nodes, and the minimum area $\Omega_\mathrm{min}$.
Because the noise varies considerably across the primary beam of each image, we apply the dendrogram to maps that have \textit{not} been corrected for the weight of the primary beam.
This effectively works to identify features with outer contours of constant statistical significance across the field of view, rather than outer contours of constant flux.
Sources are extracted out to the limit of the maps, set to the 20\% power point of the primary beam.
We choose conservative values for each parameter, using $v_\mathrm{min} = 3\sigrms$, $\delta_\mathrm{step} = 3\sigrms$, and $\Omega_\mathrm{min} = \Omega_\mathrm{bm}$, applied to the unmasked images.
Sources (i.e.\ leaves, or nodes without children) are then sub-selected to meet the criteria that the peak flux is $>5\sigrms$.
In total, we identify 67 sub-structures for the sample of 12 clumps.
Figure \ref{fig:CorePositions} shows the dendrogram extracted dense gas sub-structures in each clump.
Table \ref{tab:CoreProperties} catalogs the measured positions, sizes, and flux densities of the sub-structures.
We find an average number of sub-structures per clump of $N_\mathrm{src}=5.6$ (median 6), with the maximum $N_\mathrm{src}=11$ in G24051 and minimum $N_\mathrm{src}=1$ in G23605.
G23605 is the only clump with $N_\mathrm{src} < 3$ and is thus not included in the source nearest neighbor distance analysis.
Figure \ref{fig:ClumpScatterSigma} presents $N_\mathrm{src}$ per clump versus $\Sigma_\mathrm{cl,pk}$, where a tentative increasing trend is observed, where clumps that have high $\Sigma_\mathrm{cl,pk}$ are more fragmented than lower $\Sigma_\mathrm{cl,pk}$.

While in theory the distribution of integrated flux densities can be analyzed to measure a CMF, large observational uncertainties exist in practice that complicate its interpretation.
The principal contributor arises from at least a factor of three uncertainty in $T_\mathrm{d} \sim 6-\SI{35}{K}$ ($\sim \! 10\times$ uncertainty in $M$), due to uncertainty in the ISRF, local extinction, and uncertainty in the protostellar activity of each source.
From single wavelength observations we do not have enough information to construct spectral energy distributions (SEDs) and measure average line-of-sight dust temperatures.
Other significant systematics also arise from uncertainty in the missing flux density due to spatial filtering by the interferometer, dust opacity ($\delta \kappa / \kappa \approx 50\%$), kinematic-derived heliocentric distance ($\delta d_\odot / d_\odot \approx 15\%$), and the aperture or source boundary used to extract $S_\nu$.
For these reasons, we shall leave the study of the characteristic fragmentation mass and the CMF in SCCs to a future work utilizing complementary JVLA \ce{NH3} observations that will provide both gas kinetic temperature and kinematic information (Svoboda et al.\ \textit{in prep.}).
The characteristic fragmentation length scale, on the other hand, can be inferred directly from the distribution of angular separations between sources with assumptions on how to correct for geometric projection.

\section{Protostellar Activity}\label{sec:Activity}
In this section we describe new evidence for protostellar activity, and in Sect.~\ref{sec:FragmentationScale} we perform an analysis of the fragmentation scale from the sub-structure detected in the continuum.
With the improved sensitivity and resolution of ALMA, multiple indicators of protostellar activity are observed for the first time.
In particular bipolar molecular outflows detected in \ce{CO} $J=2\rightarrow1$ and \ce{SiO} $J=5\rightarrow4$ provide unambiguous evidence of embedded protostellar activity.
The detection of molecular transitions with comparatively high upper excitation temperatures ($E$-\ce{CH3OH} $4_{2,2} \rightarrow 3_{1,2}$, $\eupper / k = \SI{45.5}{K}$; p-\ce{H2CO} $3_{2,2} \rightarrow 2_{2,1}$, $\eupper / k = \SI{68.1}{K}$) and detection of bright, compact continuum emission (unresolved on scales smaller than $\lesssim \SI{3000}{au}$) are also suggestive of embedded, low-$\lbol$ protostellar activity.
Together, these data provide a clear indication of embedded protostars towards 11 out of 12 clumps.

\subsection{Compact continuum sources}\label{ssec:CompactSources}
Numerous high SNR (signal-to-noise ratio, $S_\nu/\sigrms$), point-like sources are observed in the continuum images (Figure~\ref{fig:ContinuumTwelve}).
We speculate such sources originate from the dense, centrally heated inner-envelopes of embedded protostars.
In \S\ref{sec:Modeling} we investigate whether the compact continuum sources are inconsistent with radiative transfer models of dense, starless cores.

We designate continuum sources as ``compact`` if they are unresolved or are marginally resolved on the scale of the ALMA synthesized beam $\theta_\mathrm{syn} \approx 0\farcs8$.
Continuum sources are determined to be unresolved if a Gaussian fit to the image plane data using the CASA task \texttt{imfit} reports an deconvolved angular sizes $\theta_\mathrm{dec} \lesssim \theta_\mathrm{syn}$.
The deconvolved Gaussian FWHM are determined from subtracting the synthesized HPBW in quadrature from the fitted width, i.e.\ $\theta_\mathrm{dec} = \sqrt{\theta_\mathrm{fit}^2 - \theta_\mathrm{syn}^2}$.
These angular widths correspond to physical sizes of $\lesssim\! \SI{1500}{au}$ at heliocentric distances of $d_\odot \approx \SI{4}{kpc}$.
The brightest compact sources have typical peak flux densities between $S_\mathrm{1.3,pk} \approx 1-\SI{7}{\milli\jy\per\beam}$.
All clumps aside from G28539 host a compact source with $S_\mathrm{1.3,pk} > \SI{1}{\milli\jy\per\beam}$.
Indeed, sources G23605 and G30120 host compact sources, even though they show limited fragmentation otherwise.
While lacking extended continuum emission, the compact source G23605 S1 in has clear association with emission from multiple molecular species (\ce{C^18O}, \ce{H2CO}, \ce{CH3OH}) at the LSR velocity of the clump, determined from the \ce{NH3} emission (\ang{;;32}, \SI{0.7}{pc} resolution; \citetalias{svoboda16}).
G30120 S1 is a compact source near the eastern edge of the field with a strong \ce{CO} outflow and other molecular detections.

For comparison to nearby low-mass star forming regions, \cite{enoch11} carried out a survey of Class 0 YSOs in Serpens at \SI{230}{GHz} with CARMA.
The envelope masses range between $M_\mathrm{env} = 0.5 - \SI{20}{\msun}$ (median $M_\mathrm{env} = \SI{3.7}{\msun}$) and with integrated flux densities between $S_{1.3} = \num{1.4e1} - \SI{4.0e3}{mJy}$ (median $S_{1.3} = \SI{120}{mJy}$) and deconvolved size scales between $D = 400-\SI{3000}{au}$ (median $D \approx \SI{700}{au}$).
With a heliocentric distance of $d_\odot = 415 \pm \SI{25}{pc}$ to Serpens \citep{dzib10}, the \SI{120}{mJy} median source flux density and \SI{700}{au} size measured by \cite{enoch11} correspond to \SI{1.2}{mJy} and $0\farcs18$ when scaled to a fiducial distance of \SI{4}{kpc}.
If there are low- to intermediate-mass Class 0 YSOs with similar physical properties in these SCCs as in Serpens, then they would be consistent with the observed bright ($\gtrsim 20\sigrms$) unresolved point continuum sources.
This is further supported by the frequent coincidence of outflows towards such sources, discussed in section \S\ref{ssec:OutflowDetections}.
To determine whether the observed compact continuum sources are consistent with starless cores ($\sim\! \SI{0.1}{pc}$) embedded within the mapped clumps ($\sim\! \SI{1}{pc}$), in \S\ref{sec:Modeling} we compare a subset of the observations to radiative transfer models of starless cores.

Some continuum sources without molecular line detections may be background galaxies.
Deep surveys performed with ALMA \citep{hatsukade13,carniani15} have determined source counts of background galaxies at \SI{1.3}{mm}.
The number of sources expected in the images with flux densities greater than $N(S_{1.3} > \SI{0.3}{mJy}) \lesssim 3$ over the 12 fields, measured as HPBW area for each pointing, outside of which the degraded sensitivity yields negligible background sources.
This represents approximately $\lesssim \! 5\%$ of the detected sources, and thus does not have a significant effect on the calculation of the nearest neighbor separations or other estimated distributions of core properties.

\subsection{CO \& SiO Outflows}\label{ssec:OutflowDetections}
Ordered, bipolar molecular outflows driven by protostellar accretion provide a sensitive and unambiguous detection of embedded protostellar activity \citep[see reviews by][]{arce07,frank14}.
In this section we describe the properties of outflows detected with ALMA in \ce{CO} $J=2\rightarrow1$ and \ce{SiO} $J=5\rightarrow4$.

Outflows are identified through visual inspection of the \ce{CO} datacubes in conjunction with the \SI{1.3}{mm} maps overplotted.
While the emission structures in the \ce{CO} cubes are complex, bipolar outflows are clearly apparent as paired linear emission structures.
These features are identified as linear features radiating from the same location with highly ordered red and blue velocity components that are detected over many velocity channels ($\gtrsim\! \SI{10}{\kms}$, $\gtrsim\! 15$ channels).
Outflow candidate features with only a single red or blue component are also observed, but due to the greater ambiguity in identification these are not regarded as clear signatures of star formation activity.
The \ce{CO} outflows are generally highly ordered in position and velocity, but spatial filtering of bright, extended emission and self-absorption near the source systemic velocity complicate the identification of low-velocity ($|v| \lesssim \SI{1.5}{\kms}$) outflow components.
Higher velocity components of the spectra also suffer both self-absorption from foreground \ce{CO} clouds and confusion with bright Galactic emission, which can bias measurements of the maximum outflow velocity to lower values.
Analysis of an example outflow in G24051 is presented in Appendix \ref{apx:PvDiagram}.

\begin{figure*}
    \centering
    \includegraphics[width=0.99\textwidth,trim={12mm 3mm 19mm 7mm},clip]{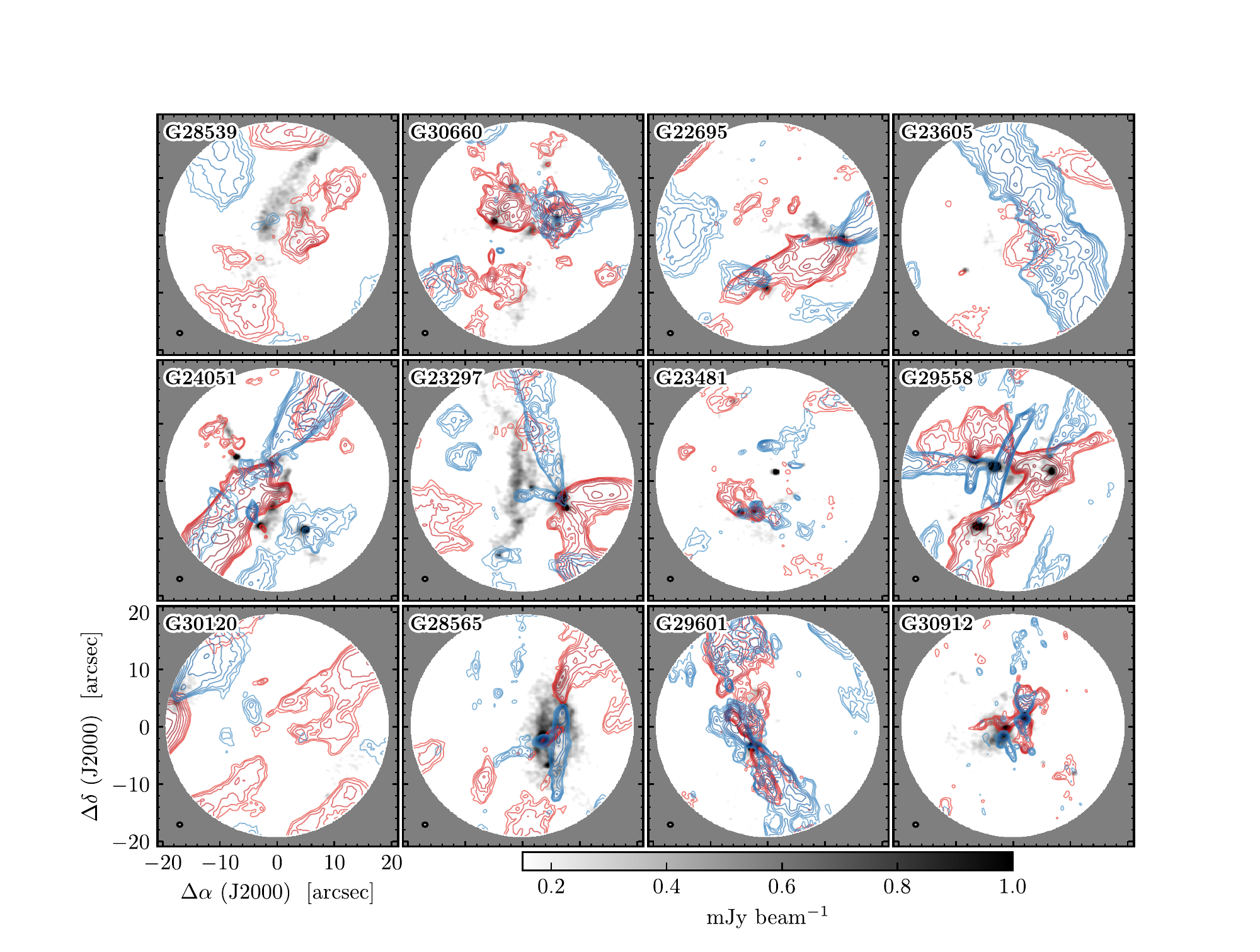}
\caption{
ALMA joint $12+\SI{7}{m}$ array \ce{CO} $J=2\rightarrow1$ intensity of velocity components integrated between offsets \SIrange{+5}{+15}{\kms} (red contours) and between offsets \SIrange{-5}{-15}{\kms} (blue contours).
Bipolar outflows are observed towards $9/12$ clumps.
Contours are shown at logarithmically spaced steps of 0.16, 0.22, 0.29, 0.40, 0.54, 0.74, 1.00, 1.36, 1.85, and \SI{2.6}{\jy\kilo\meter\per\second}.
The inverted grayscale image shows the \SI{230}{GHz} continuum.
The image extends down to the 20\% power point ($40\arcsec$).
The maps are made from the dirty image cubes and have not been deconvolved with \textsc{clean}.
}
\label{fig:CoOutflowMosaic}
\end{figure*}

We find that 9 out of 12 clumps are associated with bipolar \ce{CO} outflows and 16 outflows in total are observed.
We also find that 3 out of 12 clumps are associated with bipolar \ce{SiO} outflows and 4 outflows in total are observed.
The clumps with outflows are reported in Table \ref{tab:Detections}.
Pairs of \ce{CO} outflows originating from the same continuum source are also observed, as seen in G23297 S2 and G29601 S1, which point to unresolved protostellar multiple systems.
Figure \ref{fig:CoOutflowMosaic} presents the ALMA joint $12+\SI{7}{m}$ array \ce{CO} $J=2\rightarrow1$ integrated intensity maps for blue- and red-shifted velocity components.

We also detect \ce{SiO} emission towards several more continuum sources and positions without clear signs of ordered bipolar outflows.
\ce{SiO} emission detection is a strong indicator of protostellar activity because of its origin in high-velocity shocks driven by protostellar outflows \citep{schilke97}.
However, recent work has shown that low-velocity shocks ($\lesssim\! \SI{10}{\kms}$) created by colliding flows may produce substantial distributed \ce{SiO} emission \citep{jimenezserra10,luong13,louvet16}.
Thus, considered by itself, a detection of relatively narrow linewidth ($\Delta v \lesssim \SI{10}{\kms}$) \ce{SiO} $J=5\rightarrow4$ emission is not an unambiguous indicator of star formation activity.
Maps of \ce{SiO} integrated intensities are presented in Appendix \ref{apx:SiOOutflows}.

\subsection{G28539, a true starless clump?}\label{ssec:G28539RadioContinuum}
The \SI{70}{\um} dark clump G28539 (upper-left corner of Fig.~\ref{fig:ContinuumTwelve} \& \ref{fig:CoOutflowMosaic}) shows no clear sign of \ce{CO} or \ce{SiO} outflows, and thus remains a starless clump candidate at the improved sensitivity of ALMA.
Several indirect tracers of star formation are observed towards G28539 however and we discuss these in turn.

Moderately high-excitation molecular lines ($E_\mathrm{u}/K \gtrsim \SI{50}{K}$) are unlikely to be excited in the cold \SI{10}{K} gas expected to be found in starless cores and quiescent clump gas.
Detection of such lines in our observations are thus indirect evidence of embedded protostars, although as discussed in \S\ref{ssec:OutflowDetections} it is possible some of these lines are excited from low-velocity shocks originating from colliding flows.
In G28539, a compact source of weak emission \ce{CH3OH} and \ce{H2CO} $3_{2,2}\rightarrow2_{2,1}$ is detected.
These features are not coincident with continuum emission and may originate from non-protostellar shocks, shocks of undetected protostellar outflows, or of embedded protostellar cores that are below our detection limit.
Similarly, a compact source of \ce{SiO} is also detected does not coincide with any continuum emission feature (it may be seen on the west side of the field in Fig.~\ref{fig:SiOOutflowMosaic}).

There exist weak \SI{24}{\um} sources in the vicinity of the clump boundaries as defined by the \SI{350}{\um} and \SI{500}{\um} emission, notably a faint source within the extinction feature $\sim \! 1\arcmin$ east of the ALMA field (see \SI{24}{\um} panel in Fig.~\ref{fig:IrMosaic}), a brighter source on the south-eastern outskirt of the clump, and a marginal feature coincident with the continuum source in the NW edge of the ALMA field of view.
Because of the substantial contamination from evolved stars, \SI{24}{\um} emission alone is not a robust indicator of protostellar activity.
If these sources are indeed protostars associated with the clump then they would be evidence that star formation has begun in G28539.

Deep radio continuum observations when available also provide a diagnostic of star formation activity because they are sensitive to the ionized gas in ultra- and hyper-compact \textsc{Hii} regions, ionized winds, and jets from low- to intermediate-mass protostars.
\cite{rosero16} carried out deep Jansky Very Large Array (JVLA) C \& K-band observations towards a sample of high-mass clumps which contains source G28539 in the field ``G28.53--00.25''.
The HPBW of the primary beam for the JVLA at C-band is $9.2\arcmin$ at \SI{4.2}{GHz} (LSB) and $4.2\arcmin$ at \SI{7.4}{GHz} (USB), with synthesized HPBW resolution of approximately $\sim \! 0.4\arcsec$ in the A-configuration.
Using the radio-continuum to bolometric luminosity scaling relations for protostars in \cite{shirley07} (Eq.~3), the measured $\sigrms = \SI{3}{\micro\jy\per\beam}$ sensitivity at $d_\odot = \SI{4.7}{kpc}$ can be converted to a $1\sigma$ bolometric luminosity sensitivity of $\sim \! 30 \, L_\odot$, that is reasonably comparable to the PACS \SI{70}{\um} sensitivity from Hi-GAL.
Here a faint point source is detected near the center of the ALMA pointing, detected in both side-bands at moderate significance (8 and $5\sigma$ in LSB, USB respectively).
The measured in-band spectral index ($S \propto \nu^{\mathbf{+}\alpha}$) $\alpha = -0.65 \pm 0.46$ favors a non-thermal synchrotron dominated source, but the weak constraint is consistent with thermal free-free emission $\alpha = -0.1$ at $1.2\sigrms$.
The location $18\degr 44\arcmin 22\farcs621$~$-4^\mathrm{h} 02^\mathrm{m} 00\fs380$ (J2000) is not coincident with millimeter continuum or spectral line emission in the ALMA data.
Given the lack of a clear association, we conclude that this radio continuum source is likely an extra-galactic contaminant and not an indicator of protostellar activity.

In summary, indirect evidence for star formation exists from two different tracers: (1) \SI{24}{\um} sources at the edge or outside of the ALMA field of view, and (2) ALMA detections of \ce{CH3OH} and \ce{SiO} that are not clearly associated with continuum sources.
G28539 is the most massive clump in the sample ($M_\mathrm{cl} \approx \SI{3600}{\msun}$) and shows fairly limited signs of fragmentation.
After the ALMA observations G28539 is the only starless clump candidate remaining in our sample.
It is thus a target of great interest for studying the initial conditions of high-mass star formation.

\section{Modeling Continuum Sources}\label{sec:Modeling}
\subsection{Starless core models}\label{ssec:StarlessCoreModels}
A diverse range of continuum sub-structures are found to be present in SCCs, from unresolved compact sources, filaments, to lower surface brightness extended emission.
In this section we analyze whether cores with bright, unresolved continuum emission on scales $< \! \SI{1500}{au}$ ($\sim \! \theta_\mathrm{syn} / 2$) are necessarily protostellar even without detections of outflows or strong high-excitation molecular lines.
We also model whether low- to intermediate-mass starless cores are accurately recovered in the observations and perform detailed modeling of high-mass starless core candidates in clump G28539.

To characterize the continuum features in our images we apply the radiative transfer code \radmc\ \citep{dullemond12} to self-consistently calculate the equilibrium dust temperature distributions of externally heated starless cores and to produce synthetic images.
We follow a similar approach to modeling starless cores as found in \cite{shirley05} and \cite{lippok16}.
We apply conventional assumptions for the dust properties \citep{ossenkopf94,weingartner01,young05} and interstellar radiation field \citep[ISRF,][]{draine78,black94}.
A detailed description of the computed models may be found in Appendix \ref{apx:Models}.

We apply a spherically symmetric Plummer-like function to parametrize the model radial density profile \citep{plummer1911,whitworth01,lippok16}.
The gas density profile $n_\mathrm{H}$ can be expressed as:
\begin{equation}
    n_\mathrm{H}(r) = \left( n_\mathrm{in} - n_\mathrm{out} \right) \left[ 1 + \left( \frac{r}{R_\mathrm{flat}} \right)^2 \right]^{-\eta/2} + n_\mathrm{out}
\end{equation}\label{eq:Plummer}
for radius $r$, inner gas density $n_\mathrm{in}$, outer gas density $n_\mathrm{out}$, flat radius $R_\mathrm{flat}$, and power law exponent $\eta$ \citep[n.b.\ an isothermal Bonnor-Ebert sphere may be approximate with $\eta = 2$;][]{ebert55,bonnor56}.
The strength of the interstellar radiation field (ISRF) is varied from the local value by a multiplicative scale factor \sisrf.
We compute \num{e4} models, randomly sampling the parameter space by drawing values from a uniform distribution in log-space within the ranges for the parameters 
$n_\mathrm{in} = \num{1e4} - \SI{1e7}{cm^{-3}}$,
$n_\mathrm{out} = \num{1e1} - \SI{1e3}{cm^{-3}}$,
$R_\mathrm{flat} = \num{1e3} - \SI{2e4}{au}$,
and $\sisrf = 1 - 100$,
while $\eta = 2.5 - 5.5$ is drawn uniformly in linear space.
Models are evaluated on a logarithmic radial grid from \SI{2.5e2}{au} to \SI{6.0e4}{au}.
These values are chosen to cover the range of values from the sample of low- and intermediate mass cores in \cite{lippok16} but extended to higher $n_\mathrm{in}$ and smaller $R_\mathrm{flat}$.
After computing the radiative transfer, the models are ray-traced by \radmc\ and projected to a fiducial distance of $d_\odot = \SI{4}{kpc}$.

\subsection{Model recovery}\label{ssec:ModelRecovery}
We find that $53\%$ of the computed models ($5268/\num{e4}$) meet the detection threshold of $S_\mathrm{1.3mm,pk} > 5\sigrms$ when convolved with a $\theta = 0\farcs8$ Gaussian beam.
The cut in peak flux density has no effect on the recovered distributions of $\eta$ and $n_\mathrm{out}$, and minimal effects on $R_\mathrm{flat}$ and $\sisrf$, with an increase in the median values by a factor of $1.5$ and $1.2$, respectively, over the distribution of model cores.

\begin{figure*}
    \centering
    \includegraphics[width=0.49\textwidth]{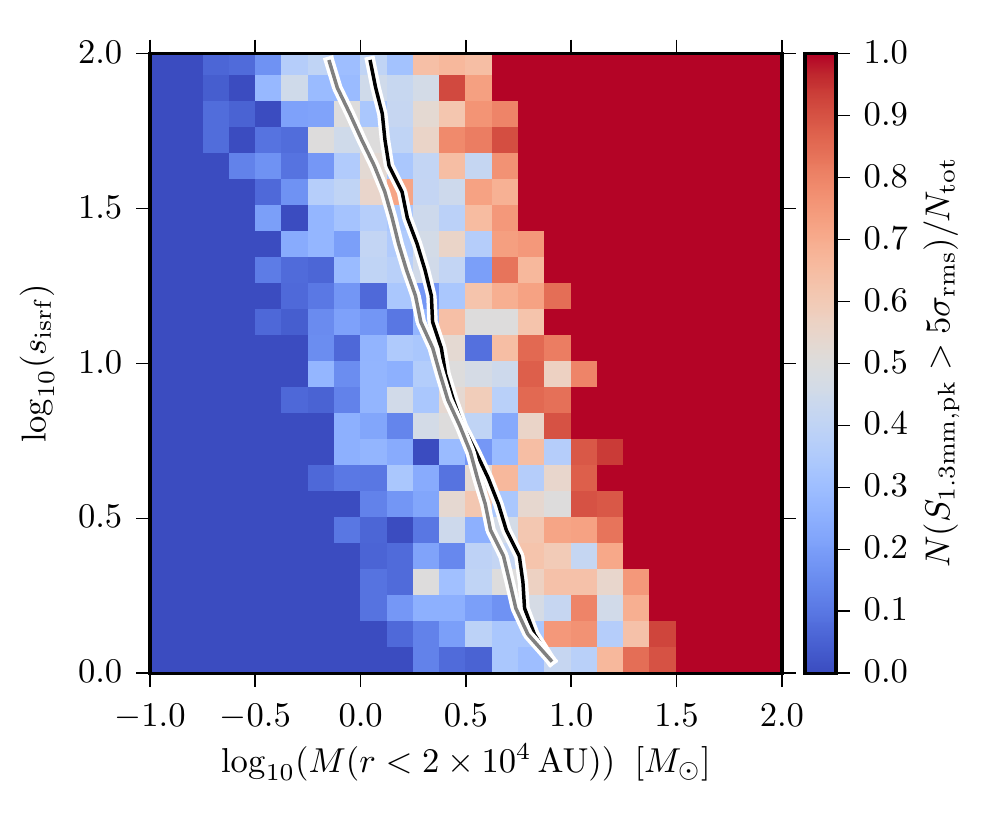}
    \includegraphics[width=0.423\textwidth]{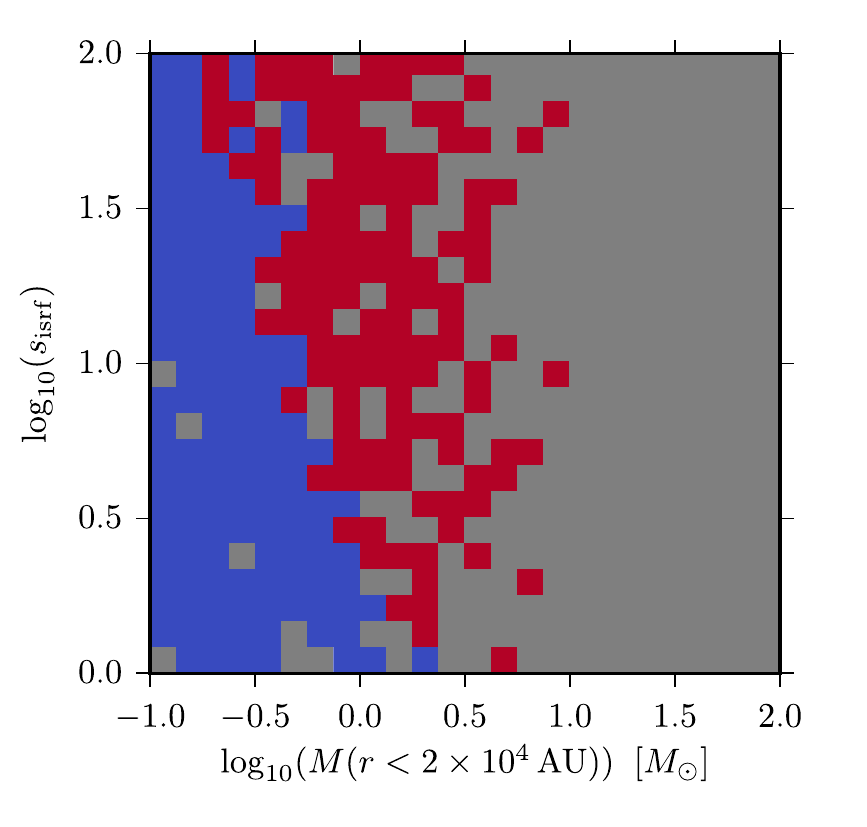}
\caption{
\textit{Left:}
Fraction of the computed models with peak flux densities $S_\mathrm{1.3mm,pk}$ meeting the source detection criteria $> 5 \sigrms$ in the images as a function of $M(r < \SI{2e4}{au})$ and $\sisrf$.
Lower mass cores meet the criteria for larger values of $\sisrf$.
The $50\%$ detection threshold for all models (black line) and $50\%$ detection threshold for models with $n_\mathrm{in} > \SI{e5}{cm^{-3}}$ (gray line) are shown.
Cores in this range with $M \sim 1 - 6$ lie above this threshold, depending on \sisrf, and are relatively insensitive to the choice in model parameters.
Note that for $\log_{10}(\sisrf) \sim 0.5$ (or $\sisrf \sim 3)$ the distribution above $M \gtrsim \SI{4}{\msun}$ meet the detection criteria.
\textit{Right:} Detection criteria for ``compact'' models ($R_\mathrm{flat} < \SI{3e3}{au}$, $\eta > 4.5$) with high central densities ($n_\mathrm{in} > \SI{e5}{cm^{-3}}$), for cases where any model meets the criteria (red) and cases where none do (blue).
Compact starless core models with $M \sim \SI{1}{\msun}$ are detectable at $\sisrf \sim 3$.
Regions that are not sampled by the compact subset of models are shown in gray, note that because the maximum $n_\mathrm{in} = \SI{e7}{cm^{-3}}$, models with $M \gtrsim \SI{10}{\msun}$ are more extended.
}
\label{fig:DetectionFraction}
\end{figure*}

It is important to keep in mind that the suite of model cores is constructed to span the parameter space of relevant values, not to represent an observed or predicted core mass function.
We do not use the fractions of detectable cores to infer completeness, but to show the expected range of physical parameters for which cores can be recovered.
To estimate this, we sort the models by $M$ and \sisrf, counting both the fraction of detectable cores, and regions of parameter space with at least one detectable model (Fig.\ \ref{fig:DetectionFraction}).
Computing the detection fraction in this way has the effect of marginalizing over our uncertainty in $R_\mathrm{flat}$, $\eta$, $n_\mathrm{in}$, and $n_\mathrm{out}$, that are poorly constrained with our single wavelength maps.
Figure \ref{fig:DetectionFraction} (left) shows that at 50\% completeness, the cores at $\sisrf \sim 3$ are recovered for $M \gtrsim \SI{4}{\msun}$, and that this extends down to $M \sim \SI{1}{\msun}$ for the extreme value $\sisrf = 100$.
Figure \ref{fig:DetectionFraction} (right) shows that it is possible however to recover lower-mass cores if the ranges of models is restricted to those that are the most compact (where $R_\mathrm{flat} < \SI{3e3}{au}$, $\eta > 4.5$) and have high central densities ($n_\mathrm{in} > \SI{1e5}{cm^{-3}}$).
For these compact sources, $M \sim \SI{1}{\msun}$ models may be recovered at $\sisrf \sim 3$ and down to $M \sim \SI{0.2}{\msun}$ for $\sisrf = 100$.

From these models we can infer that the completeness expected from our point-source sensitivity, $\sim \! \SI{0.3}{\msun}$ at $6\sigrms$, is an underestimate if the majority of cores are resolved \citep[see also Appendix A in][]{beuther18}.
Low-mass cores with extended profiles will thus go undetected with a criteria based on peak intensity, leading to our seemingly shallow limit of $\sim \! 1 - \SI{4}{\msun}$.
Observationally, we must approach extended emission at low-SNR with caution because there is extended structure in the maps on scales larger than the \SI{12}{m} primary beam that cannot be adequately \textsc{clean}ed.
For this reason we do not attempt to identify or catalog sources down to the limits of statistical significance for extended and spatially-integrated flux densities but maintain a conservative detection limit based on source peak flux density.
The typical integrated flux density of a source is $S_{1.1} \sim \! 1-\SI{10}{mJy}$ ($M \sim \! 1-\SI{10}{\msun}$ assuming $T_\mathrm{d} = \SI{12}{K}$ and $d_\odot = \SI{4}{kpc}$), and generally consistent with the thermal Jeans mass $M_\mathrm{j,th}$ for a uniform medium at the density of the clump, $M_\mathrm{j,th} \sim \SI{2}{\msun}$ where $M_\mathrm{j,th} \equiv (4 \pi / 3)(\lambda_\mathrm{j,th} / 2)^3 \rho_0$ for thermal Jeans length $\lambda_\mathrm{j,th}$ and average density $\rho_0$ (\citeauthor{mckee07} \citeyear{mckee07}; see \S\ref{sec:FragmentationScale} for an analysis of the Jeans length \ljeans).

\subsection{Synthetic observations with CASA}\label{ssec:SyntheticObservations}
We now investigate whether the models of starless cores provide adequate fits to the brightness profiles present in the SCCs of this survey.
We find that compact sources of continuum emission that are unresolved (i.e.\ deconvolved sizes  $\lesssim \! \SI{1500}{au}$, $\approx \theta_\mathrm{syn}/2$) are poorly fit by models of starless cores.
Without multiple wavelength observations or gas kinetic temperature information, the radial dust temperature profiles of the cores are poorly constrained.
Because of the substantial systematic uncertainties presented in single wavelength observations and potentially undetected embedded protostars, we do not perform a fit to every continuum source, but select a few characteristic examples for quantitative comparison.
We create synthetic observations from the models using the \casa\ \texttt{sm} module by predicting onto the observed visibilities (gridded beforehand for computational efficiency) and imaged without noise using the same \texttt{tclean} configuration as the observations.
This does not introduce a significant effect on the models however because nearly all the flux is concentrated on radii $r < \SI{2e4}{au}$ or angular diameters of $\lesssim \! 8\arcsec$, appreciably less than half of the \SI{12}{m} array $27\arcsec$ HPBW, and substantially less than the maximum recoverable scale of $33\arcsec$ from the \SI{7}{m} array.
A subset of models were further tested for consistency, because the aim of this comparison is for an understanding of a few representative sources and not detailed parameter estimation, we do not image the full suite of models; rather, we convolve the models with the angular size of the synthesized beam ($\theta_\mathrm{syn} = 0\farcs8$) and convert to radial brightness profiles.

\subsection{Comparison to observations}\label{ssec:ComparisonToObservations}
We compare the observations and models using a method based on the $\chi^2$-statistic, where the reduced $\chi_\mathrm{r}^2$ may be expressed as
\begin{equation}
    \chi^2_\mathrm{r} = \frac{1}{\nu} \sum_i \frac{\left(o_i - m_i\right)^2}{\sigma^2}
\end{equation}\label{eq:ChiSquare}
for degrees of freedom $\nu$, independent measurements $i$, measurements $o_i$, model values $m_i$, and variances $\sigma^2$.
We discriminate between models based on the goodness of fit metric $\Delta \chi^2_\mathrm{r} \equiv \chi^2_\mathrm{r} - \chi^2_\mathrm{r,best}$ 
from \cite{robitaille07} and \cite{robitaille17}.
Robitaille et al.\ apply the heuristic that models with $\Delta \chi^2_\mathrm{r} < 3$ are considered good fits and rejected as poor fits otherwise.
Robitaille et al.\ further note that the Bayesian likelihood under the assumption of normal errors (i.e., $P(D|\theta_j,M) \propto \exp\left[ - \chi^2 / 2\right]$ for data $D$, parameters $\theta_j$, and model $M$) yields too stringent a definition of probability given systematic sources of error in the measurements and poor physical correspondence of the model to nature.
This ultimately provides a more conservative criteria for rejecting poor fits as the $\Delta \chi^2_\mathrm{r}$ heuristic likely overestimates uncertainties.

We consider two example starless core candidates, G28539 S2 and S4 (see Fig.~\ref{fig:CorePositions} and Table~\ref{tab:CoreProperties}, because they (1) lack unresolved continuum emission at their center, (2) host no outflows or other indicators of star formation activity, and (3) are relatively isolated such that radial brightness profiles can be adequately extracted.
G28539 S2 and S4 are also of interest because they are among the brightest such sources, and thus are good high-mass starless core candidates.

\begin{figure*}
    \centering
    \includegraphics[width=0.8\textwidth]{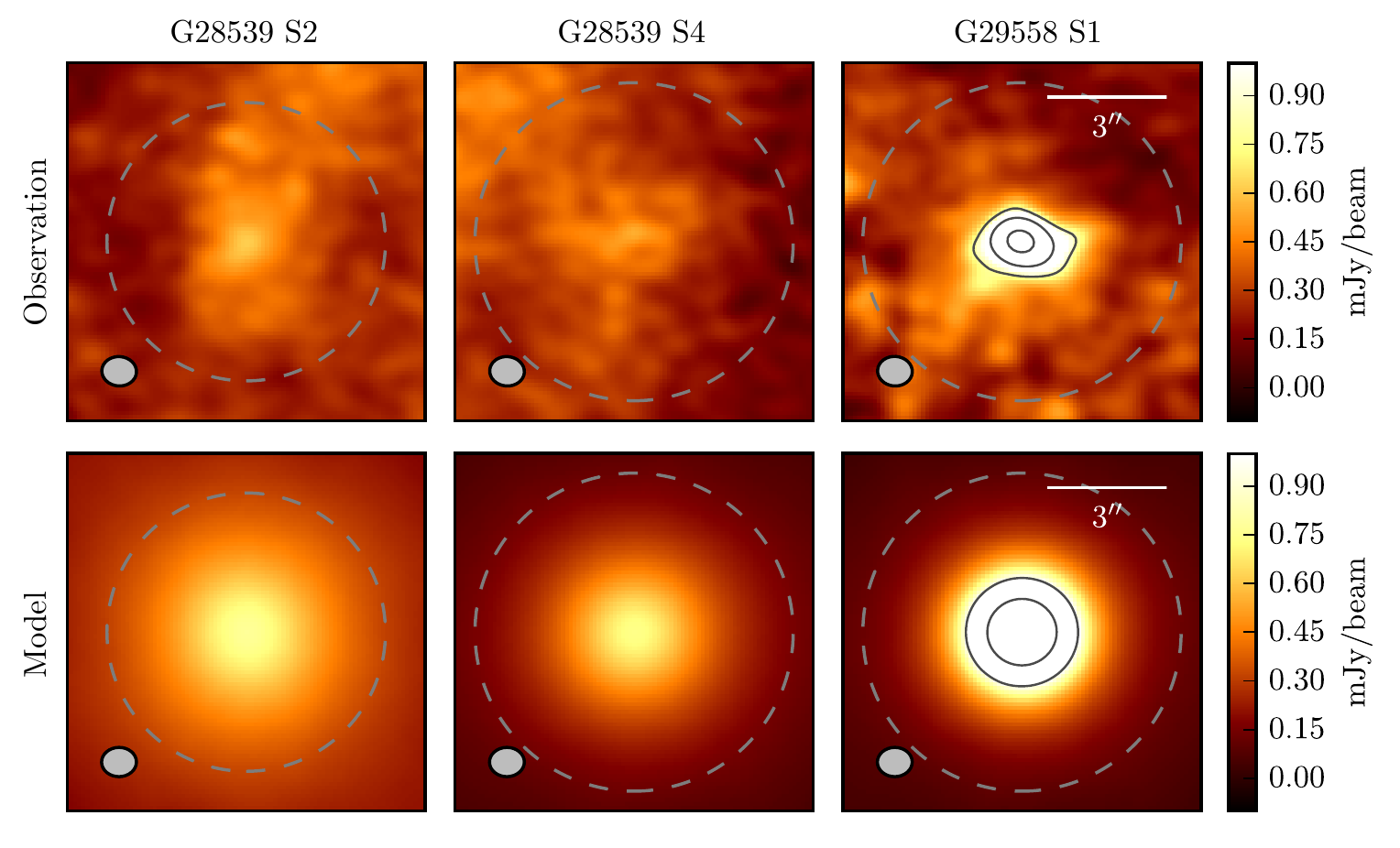}
\caption{
\textit{Top row:}
230 GHz continuum images of example sources G28539 S2, G28539 S4, and G29558 S1.
Contours (black solid) show 10, 20, and $50\sigrms$, and the $3\farcs5$ and $4\arcsec$ radius apertures (gray dashed) show the region the radial brightness profiles used for the model comparison were extracted over.
The beam ($\theta_\mathrm{syn} \sim 0\farcs8$), scalebar ($3\arcsec$), and colorbar ($-0.1$ to \SI{1}{mJy.beam^{-1}}) are visualized.
\textit{Bottom row:}
Best-fit models when run through the \casa\ simulator (bottom row, same color-scale as above).
The models for the resolved sources G28539 S2 and S4 are well fit by models of starless cores ($\chi^2_\mathrm{r} \sim 0.1-1$), while the unresolved source G29558 S1 is poorly fit ($\chi^2_\mathrm{r} > 20$).
}
\label{fig:ExampleCores}
\end{figure*}

\begin{figure}
    \centering
    \includegraphics[width=0.49\textwidth]{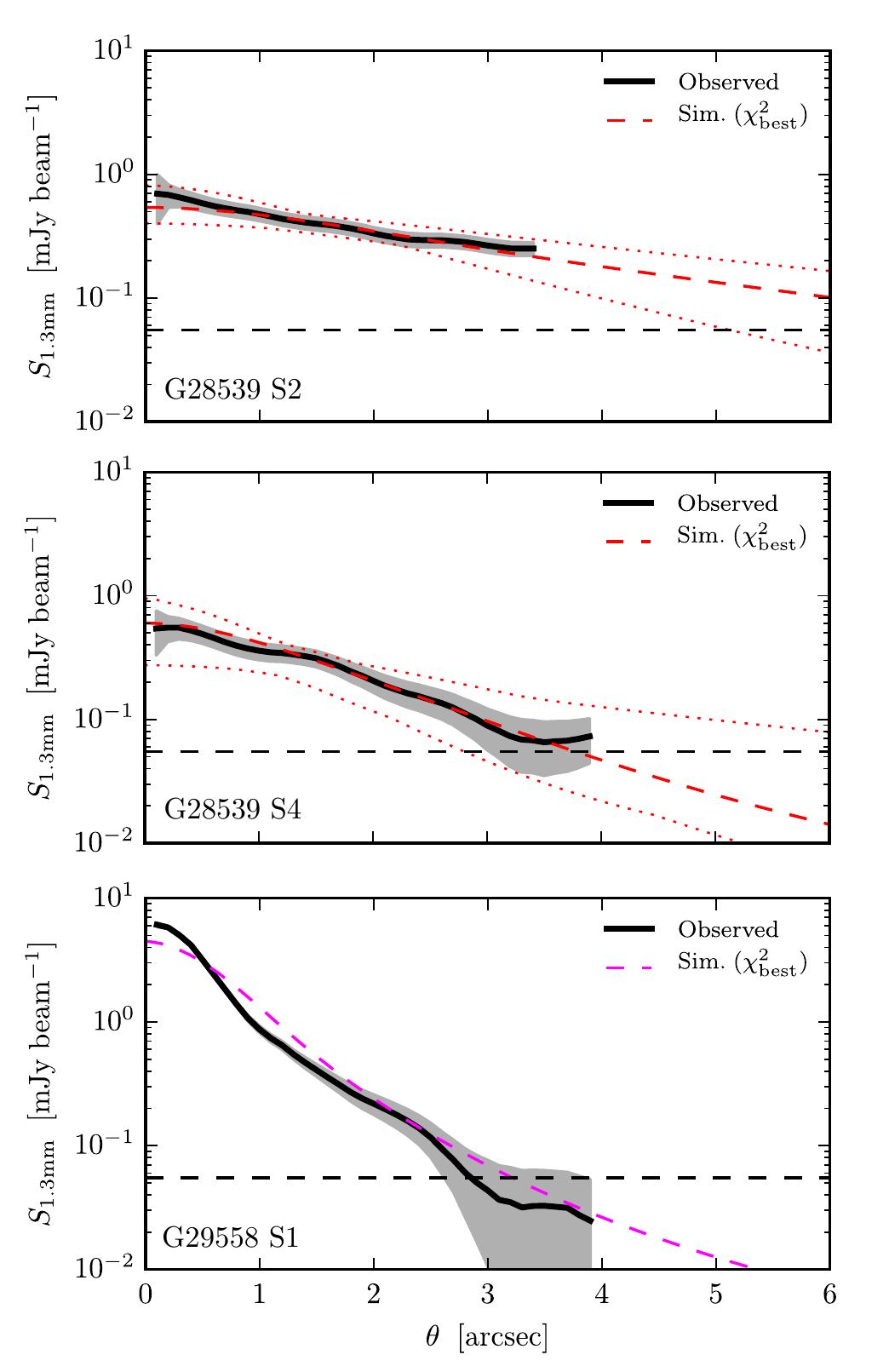}
\caption{
Example ALMA observed sources fit with the suite of starless core models.
The observed radial brightness profiles (black) and the image $1\sigrms$ (gray region) are shown with the best fit model (red dashed) and envelope of all models that satisfy $\chi^2_\mathrm{r} - \chi^2_\mathrm{r,best} < 3$ (red dotted).
The error envelope is calculated as the $\pm1\sigma$ uncertainty of the integrated intensity within the annular aperture at the angular radius $\theta$.
The profiles are truncated to where the source is mostly symmetric.
The map RMS is visualized (gray dashed line).
{\it Top:} $\chi^2_\mathrm{r,best}=1.1$.
{\it Middle:} $\chi^2_\mathrm{r,best}=0.12$.
{\it Bottom:} $\chi^2_\mathrm{r,best}=23.9$ (magenta dashed), no models for $\chi^2_\mathrm{r} - \chi^2_\mathrm{r,best} < 9$.
The resolved sources G28539 S2 and S4 are well fit by starless core models, while the models fail to fit the high-SNR, unresolved inner component in G29558 S1.
}
\label{fig:RadialModelProfiles}
\end{figure}

We extract radial brightness profiles for the cores by extracting the integrated flux density within $0\farcs2$ diameter annuli about the central position.
Uncertainties in the integrated flux densities are calculated as the $\delta S_\nu = \sigrms \sqrt{\Omega_\mathrm{ann}/\Omega_\mathrm{bm}}$ for the solid angle of the annulus $\Omega_\mathrm{ann}$ and the synthesized beam solid angle $\Omega_\mathrm{bm}$.
The radial brightness profiles of the models are then compared by Equation \ref{eq:ChiSquare} for degrees of freedom $\nu = r_\mathrm{max}/0\farcs2 - 5 \approx 15$ (maximum radius $r_\mathrm{max}=3\farcs5-4\farcs0$).
Well-fit models are then selected where $\Delta \chi^2_\mathrm{r} < 3$.
Figure \ref{fig:ExampleCores} shows the best fit models compared to the observations, and Figure \ref{fig:RadialModelProfiles} shows the radial brightness profiles with the range of fits.
We find that the extended brightness profiles are well fit by the starless core models ($\chi^2_\mathrm{r,best} = 1.1$ and $0.12$ for S2 and S4 respectively).
If the range of models are limited to those that resulted in $T_\mathrm{d}(r = \SI{1e3}{au}) = 7-\SI{13}{K}$ in order to be broadly consistent with the clump-average temperature derived from the Hi-GAL SED and GBT \ce{NH3} fits \citep[see also the detailed considerations in][]{tan13}, then $M_\mathrm{S2} = 29^{52}_{15} \, \msun$ and $M_\mathrm{S4} = 14^{34}_{6.0} \, \msun$, for the median, maximum, and minimum model mass.
With a core star-formation efficiency of $30\%$ it is possible that these cores may form high-mass stars ($M_* > \SI{8}{\msun}$).
Assuming a $50\%$ formation efficiency from models regulated by outflows \citep{zhangy14} the maximum expected stellar mass for S2 could be $M_* \approx \SI{26}{\msun}$.
Given the fact that these cores are not associated with outflows in the ALMA data or other high-excitation molecular lines, they are excellent candidates for high-mass starless cores.

G29558 S1 represents the class of compact continuum sources in our data set.
Analysis of this source is then a test of whether the compact sources are well described by starless core models, or alternatively, likely to host embedded protostars.
This continuum source has some surrounding extended continuum emission, does not show clear outflows traced by \ce{CO} or \ce{SiO}, but is associated with weak \ce{CH3OH} and p-\ce{H2CO} emission.
It is bright with peak flux density \SI{6.6}{mJy} and is similar to other continuum sources with associated outflows.
We find that the models poorly fit the observations, with $\chi^2_\mathrm{r,best} = 23.9$ and no models for $\Delta \chi^2_\mathrm{r} < 9$.
The properties are pushed to the extremes of parameter space: $\sisrf \sim 100$, $\eta \sim 5.5$, and $n_\mathrm{in} \gtrsim \SI{1e7}{cm^{-3}}$.
The moderate $R_\mathrm{flat} \sim \SI{5e3}{au}$ is a compromise between the compact and extended components of the brightness profile.
The poor model fits to G29558 S1 do not strictly require that it or any other individual source is protostellar (models with $n_\mathrm{in} \gtrsim \SI{e8}{cm^{-3}}$ and $R_\mathrm{flat} < \SI{e3}{au}$ would likely fit the observations).
However, such extreme starless cores are unlikely to be observed in significant numbers in our sample, where $\sim \! 40\%$ of fragments are compact continuum sources.
The free-fall timescale of a core with $n_\mathrm{in} = \SI{e8}{cm^{-3}}$ would be $t_\mathrm{ff} \approx \SI{3e3}{yr}$, and for $n_\mathrm{in} = \SI{e7}{cm^{-3}}$ would be $t_\mathrm{ff} \approx \SI{1e4}{yr}$.
These are shorter than the inferred ages from the extent and velocity of the observed outflows, although these have substantial uncertainties.
Together, the observed properties of these compact continuum sources are more favorably explained as embedded low- to intermediate-mass YSOs, which at $\sim \SI{4}{kpc}$ would be both of comparable brightness and unresolved (see \S\ref{ssec:CompactSources}).
A detailed analysis of the starless and protostellar core properties and dynamics will follow in a future work incorporating \ce{NH3} data from the VLA observations.

\section{Fragmentation Scale}\label{sec:FragmentationScale}
\subsection{Nearest neighbor separations and Monte Carlo simulations}\label{ssec:NearestNeighbor}
We characterize the linear fragmentation scale in terms of the nearest neighbor separation $\nns^\prime$ between dendrogram leaves in each clump.
Geometric projection of sources in the plane of the sky will systematically decrease $\nns^\prime$ from the true value, $\nns$.
In this work we employ Monte Carlo random sampling to de-project $\nns^\prime$ statistically.
Thus while the uncertainty in $\nns$ may make constraints for any individual pair of sources quite weak, with prior assumptions on the relative positions of sources, the posterior distribution from the ensemble of all $\nns$ measurements in our sample of SCCs can be readily constrained.

Monte Carlo sampling is used to draw realizations of relative line-of-sight distances $z$, computing \nns\ for each source from the Cartesian coordinates $(x,y,z)$.
We use the hierarchical classification of sources in the dendrogram to discriminate between two methods of drawing $z$ values: (i) isolated sources and (ii) sources with common surrounding emission.
If sources are isolated (Case i), forming a tree with a single branch, then for each trial we draw line-of-sight distances from a Gaussian distribution with standard deviation $\sigma_\mathrm{z} = \SI{0.15}{pc}$ (FWHM \SI{0.35}{pc}), chosen such that the double-sided $2\sigma_\mathrm{z}$ interval is \SI{0.6}{pc}, which is the approximate diameter inferred from the \SI{8}{\um} maps (cf.\ Fig.\ \ref{fig:IrMosaic} \& \ref{fig:EightMicronTau}).
If sources are associated within the same branch of the dendrogram (Case ii; ie.\ they are within a common base iso-contour of emission) then we assume that those sources are connected in a filamentary gas structure with unknown inclination with respect to the observer.
For each trial, we draw a common inclination $\phi$ for the group, pivoting along the major axis, with the pivot axis fixed to $z=0$ at the projected geometric center.
Inclinations are drawn such that the length between the two components with the maximum separation $\delta_\mathrm{max}$ is less than $D = \SI{0.6}{pc}$, thus where $\phi$ is drawn uniformly within the interval $(-\arccos(\delta_\mathrm{max}/D), \, +\arccos(\delta_\mathrm{max}/D))$.
If $\delta_\mathrm{max} > D$, then $\phi$ is drawn uniformly within $(\ang{-65}, \, \ang{+65})$, such that $\nns \lesssim \SI{1.4}{pc}$ to extend out to a typical clump effective radius of $R \approx \SI{0.7}{pc}$.
In total there are $17$ (26\%) isolated sources and $49$ (74\%) grouped sources.
Without more detailed knowledge available, informed from either additional observational data or theoretical simulations, we consider this scheme a conservative way to correct the data for geometric projection.
While the assumptions in the correction are simple and imperfect, for brevity we refer to the distributions of MC trials as ``projection-corrected'' below to distinguish it from the projected data.
Extensions of this method may opt to use more sophisticated schemes to group sources beyond common millimeter continuum emission, such as grouping sources through a lower density kinematic tracer or a source-density based clustering algorithm.

With no correction applied, the distribution of projected separations has a median value of $\mu_{1/2}(\nns^\prime) = \SI{0.083}{pc}$ with $(16,84)$ percentile interval of $(0.051,0.140) \, \mathrm{pc}$.
To calculate the projection-corrected separations, we compute \SI{1e4} realizations for each clump, and find $\mu_{1/2}(\nns) = \SI{0.118}{pc}$ with $\mu_{1/2}(\nns)/\mu_{1/2}(\nns^\prime) = 1.42$ and a $(16,84)$ percentile interval of $(0.065,0.232) \, \mathrm{pc}$.
For comparison, if we assume that all sources are uniformly distributed within a spherical volume of radius $R_\mathrm{s}$ the following projection correction may be applied:
\begin{equation}
    \nns = \nns^\prime \left( \frac{4 R_\mathrm{s}}{3 \nns^\prime} \right)^{1/3} \quad ,
\end{equation}
as is done in \citet{myers17}.
If we assume $R_\mathrm{s} = \SI{0.38}{pc}$ from the radius of the 20\%-power point of the ALMA primary beam at \SI{4}{kpc}, then this correction factor would be $\nns/\nns^\prime \approx 1.84$ and $\nns = \SI{0.153}{pc}$, which is larger than the median value computed above from the MC trials by $29\%$.

\subsection{Jeans length comparison}\label{ssec:JeansLengthComparison}
To consistently compare $\nns$ values between clumps with different physical conditions, we scale the values by the clump average thermal Jeans length, the minimum wavelength for gravitational fragmentation in an isothermal, uniform medium.
The thermal Jeans length \ljeans\ can be expressed as \citep{mckee07}:
\begin{equation}
    \ljeans = \left( \frac{\pi c^2_\mathrm{s}}{G \rho_0} \right)^{1/2} \quad ,
\end{equation}\label{eq:JeansLength}
where $c_\mathrm{s} = \sqrt{k T / \mu m_\mathrm{p}}$ is the isothermal sound speed (\SI{0.21}{\kms} for $T_\mathrm{d} = \SI{12}{K}$), $G$ is the gravitational constant, and $\rho_0$ is the average volume density.
For the accurate propagation of uncertainties in the calculation of \ljeans, we perform MC random sampling of the relevant observational uncertainties in $\rho_0$ from the dust mass surface density ($\rho_0 = 3 \Sigma / 4 R$) and heliocentric distance.
The total (i.e., gas) mass surface density are calculated with
\begin{equation}
    \Sigma = \frac{S_\mathrm{\nu,int}}{B_\nu(\td) f_\mathrm{d} \kappa \mu m_\mathrm{p} \Omega} \quad ,
\end{equation}\label{eq:MassSurfaceDensity}
for source integrated flux density $S_\mathrm{\nu,int}$, source solid angle $\Omega$, Planck function $B_\nu(\td)$ evaluated at dust temperature \td, opacity per mass of dust $\kappa \left( \lambda = \SI{1.3}{mm} \right) = \SI{0.90}{cm^2.g^{-1}}$ \cite{ossenkopf94}, mean molecular weight $\mu=2.33$, and dust-to-gas mass ratio $f_\mathrm{d} \equiv \left( m_\mathrm{d} / m_\mathrm{g} \right) = 1/110$ (values are further described in Appendix \ref{apx:Models}).

\begin{figure}
    \centering
    \includegraphics[width=0.47\textwidth]{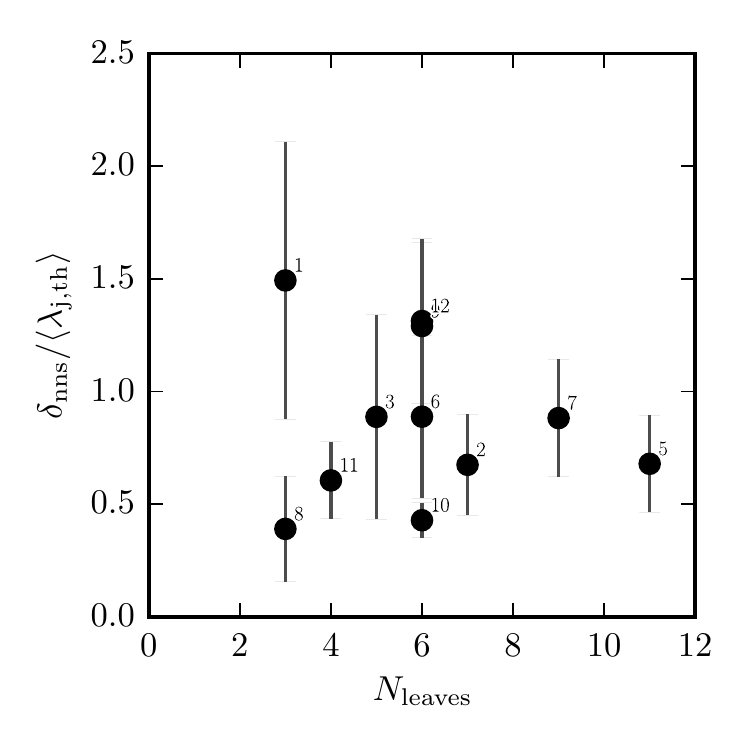}
\caption{
Nearest neighbor separation scaled by the clump thermal Jeans length (\jnns) versus number of leaves.
}
\label{fig:ClumpScatterJnns}
\end{figure}

The fragmentation measured within the ALMA maps is most sensitive within the HPBW ($27\arcsec$) of the primary beam, so an estimate of $\rho_0$ within this volume we consider to be the most representative density for the computation of \ljeans.
Clump average densities on angular scales ($\sim \! 1-2\arcmin$) larger than the HPBW likely underestimate $\rho_0$.
Likewise, image-integrated flux densities from the $12+\SI{7}{m}$ array data possibly underestimate the $\Sigma$ from spatial filtering.
Due to the unfavorable match in resolution compared to the Hi-GAL \SI{500}{\um} ($\theta_\mathrm{hpbw} \approx 35\arcsec$) or BGPS \SI{1.1}{mm} ($\theta_\mathrm{hpbw} \approx 33\arcsec$), we extract flux densities from the ATLASGAL \SI{870}{\um} maps ($\theta_\mathrm{hpbw} \approx 19\arcsec$) at the position of the ALMA pointing for each clump within a beam-sized $27\arcsec$ diameter circular aperture to measure $\Sigma_\mathrm{cl}$.
Use of the single millimeter flux mitigates one systematic uncertainty in choosing between Hi-GAL SED fits with or without the \SI{160}{\um} band included, or using Hi-GAL SED fits that are over the emission for the full clump rather than the peak at consistent angular resolution.
The clump average dust temperatures from SED fits to the Hi-GAL data range from $\td=10-\SI{14}{K}$, but some systematic uncertainty exists with averaging over larger volumes than the ALMA field of view and choices in including the \SI{160}{\um} band.
We choose a conservative dust temperature distribution by assuming a Gaussian dust temperature distribution $\langle \td \rangle = 12 \pm \SI{2}{K}$ ($1\sigma$ interval).
For consistency this temperature is also used for the gas kinetic temperature in $c_\mathrm{s}$. 
We propagate the uncertainty in heliocentric distance based on the distance probability density function (DPDF) from \cite{ellsworthbowers15a} for each clump.
All sources are well-resolved to the near kinematic distance, and have a $\delta d_\odot / d_\odot \approx 0.15$ fractional uncertainty.
We sample the distributions for $S_{870}$, \td, and $d_\odot$ for each MC trial of $\rho_0$ in the calculation of \ljeans\ to combine with a trial of \nns\ to compute the quotient \jnns\ for each clump.
The computed median volume densities for the clumps in the sample range between $n(\mathrm{H_2}) = (2 - 6)\times\SI{e4}{cm^{-3}}$ and with associated values of the thermal Jeans length between $\ljeans = 0.10 - \SI{0.17}{pc}$ ($2.1-\SI{3.5e4}{au}$).
The median of samples from all clumps is $\ljeans = \SI{0.135}{pc}$ (\SI{2.77e4}{au}).
No correlation is observed between $\jnns$ and the number of cores/leaves in each clump (Fig.~\ref{fig:ClumpScatterJnns}).

\begin{figure*}
    \centering
    \includegraphics[width=0.97\textwidth]{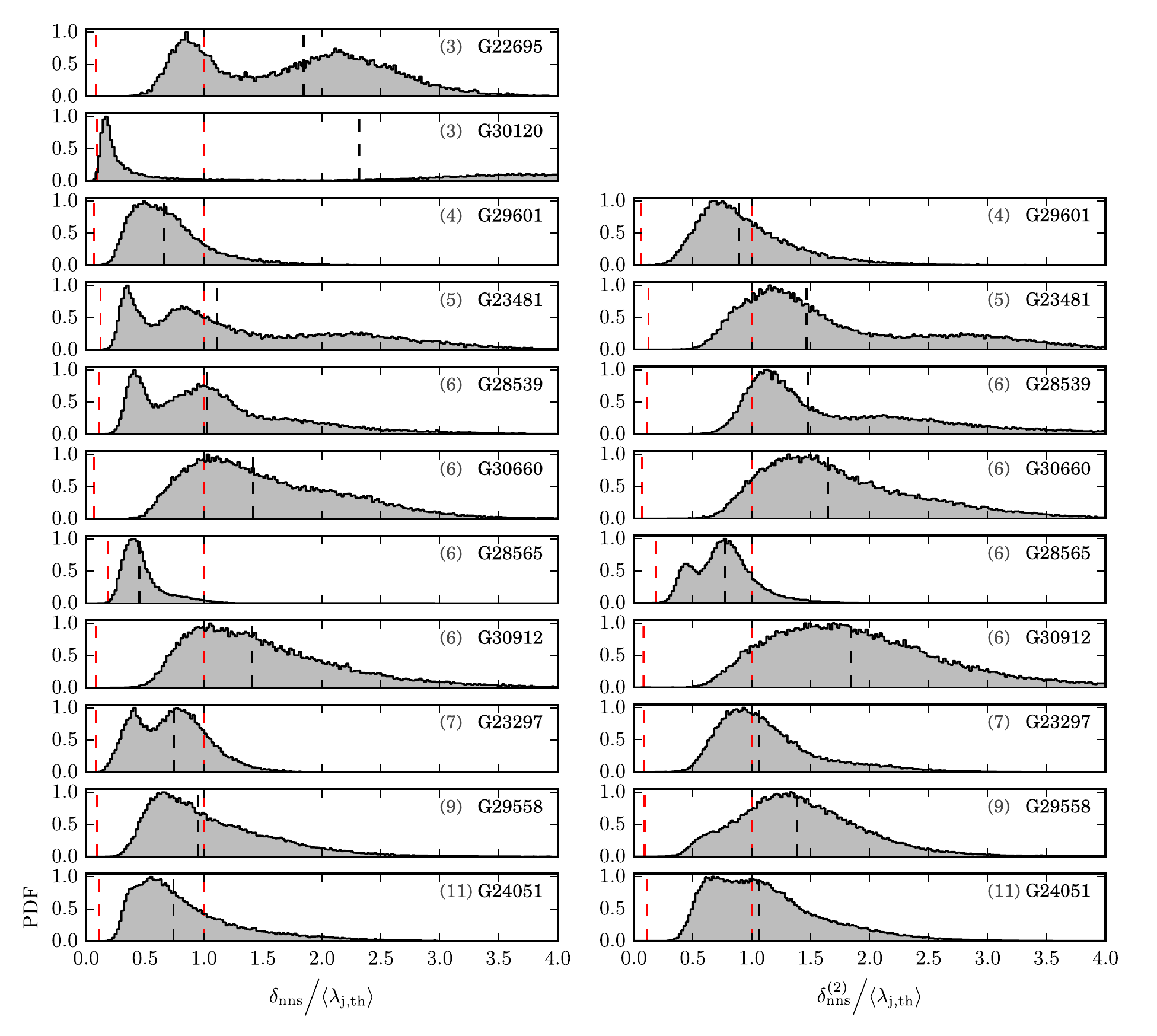}
\caption{
\textit{Left:}
Probability density functions (PDFs) of the projection-corrected nearest neighbor separations between sources in each clump, scaled by the clump average thermal Jeans length.
PDFs are scaled such that the peak probability equals $1$.
The thermal Jeans length is shown with a dashed red line at 1 ($\sim \! \SI{0.1}{pc}$), dashed red line near zero shows scale of the synthesized beam ($\sim \! \SI{0.015}{pc}$), and the black dashed line shows the 50th percentile of the distribution.
The source names are shown in the upper right and the number of sources are shown in parentheses.
\textit{Right:}
PDFs for the second nearest neighbor separations ($\nns^{(2)}$).
The distributions are more uni-modal near $1$ and show moderately larger median separations than $\nns$.
}
\label{fig:NnsBySource}
\end{figure*}

Probability density functions (PDFs) of $\jnns$ are computed for each clump by performing Monte Carlo random sampling of the observational uncertainties in \ljeans\ as described above and sampling the de-projected source separations (see \S\ref{ssec:NearestNeighbor}).
Figure \ref{fig:NnsBySource} (left) shows the distributions of \jnns\ for each clump sorted in descending order by the number of continuum sources.
The separation distributions show a bi-modal tendency with peaks at $\jnns \sim 0.3$ and $\jnns \sim 1$, and with long-tails extending to high values $\gtrsim 1.5$.
The distinct peaks at small values of \jnns\ (all well-resolved) likely result from closely spaced, connected sources where \nns\ is not strongly effected from sampling the inclination distribution.
Median values of the distributions range between $\jnns = 0.4 - 1.5$.
The values are generally consistent with the thermal Jeans length, but the high frequency of sources with sub-Jeans separations may indicate hierarchical fragmentation at multiple scales.
With the initial fragmentation on the clump scale, a further fragmentation on the ``core scale'' would proceed on sizes $\lesssim \SI{2e4}{au}$ and densities $\gtrsim \SI{3e5}{cm^{-3}}$.
If such hierarchical fragmentation proceeds principally with two resultant fragments on the core scale, then the second nearest neighbor distance would measure the above level in the hierarchy and recover the spacing of the clump scale.
This is supported by a plot of the second nearest neighbor distance $\delta^{(2)}_\mathrm{nns}$ distributions, shown in Figure \ref{fig:NnsBySource} (right), that shows clumps with more uni-modal distributions, with modes and median values at or slightly above the thermal Jeans length.
Median values of the $\delta^{(2)}_\mathrm{nns}$ distributions are greater than those for $\nns$ but generally fall within a similar range between $\delta^{(2)}_\mathrm{nns} / \langle \lambda_\mathrm{j,th} \rangle = 0.75 - 1.7$.

\begin{figure*}
    \centering
    \includegraphics[width=0.95\textwidth,trim={0mm 3mm 0mm 0mm},clip]{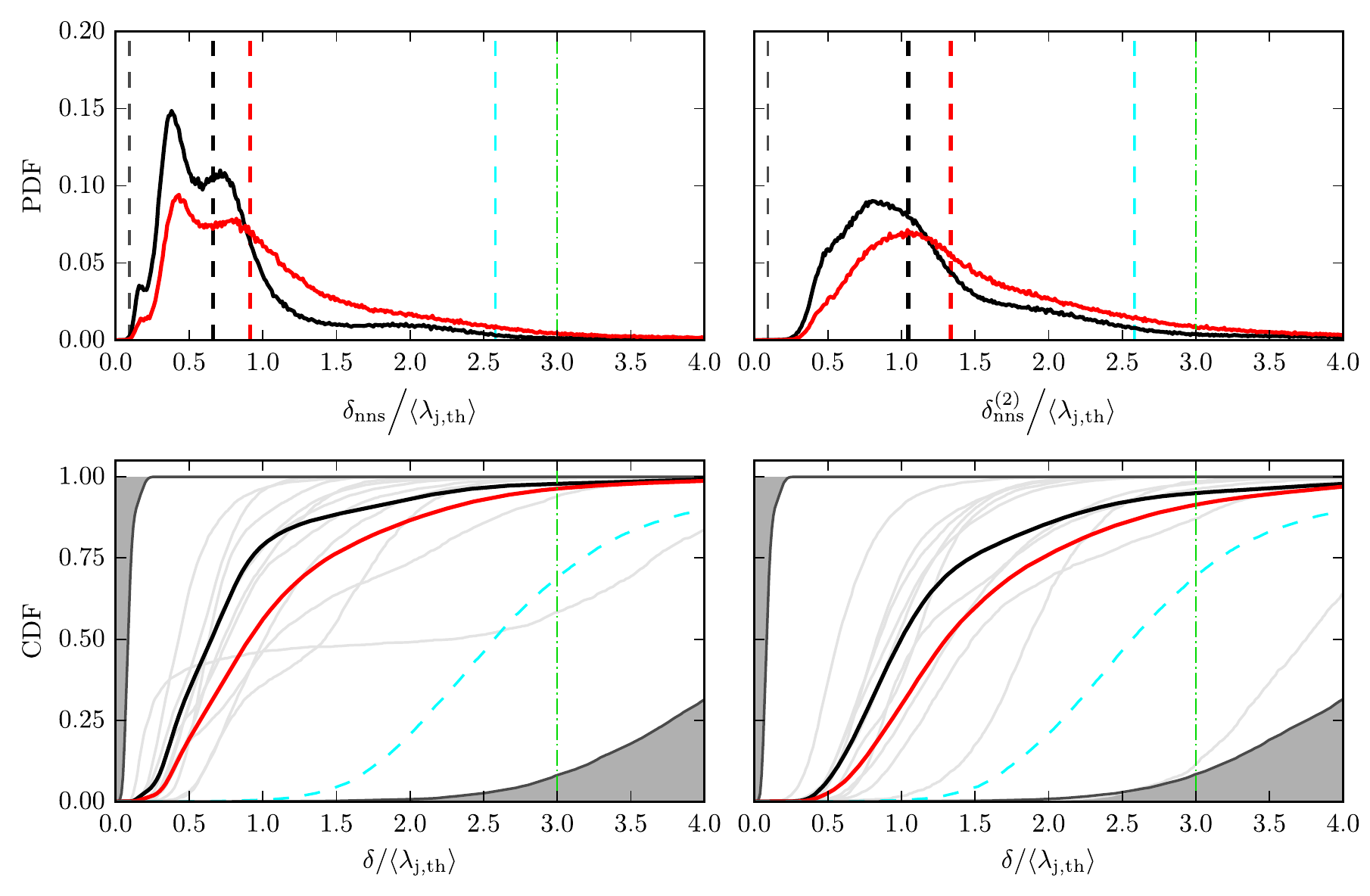}
\caption{
\textit{Lower left:}
CDFs of different lengths $\delta$ when scaled as multiples of the thermal Jeans length computed with Monte Carlo random sampling.
CDFs of the projection-corrected nearest neighbor separations for sources in all clumps (red), similarly for sources of individual clumps (thin gray), and projected nearest neighbor separations for all sources (black).
The median value of $\jnns = 0.82$ with a $(25,75)$ percentile interval from $0.52-1.25$, consistent with fragmentation primarily occurring at the thermal Jeans length on the clump scale.
The value $\jnns = 2$ occurs at the $90.3$ percentile.
The dash-dotted green line visualizes the median value turbulent jeans length of $\ljeanstu / \ljeans \approx 3$.}
The inner and outer grey areas show the scaled synthesized beam ($0.8\arcsec$) and scaled $20\%$-point of the primary beam ($40\arcsec$).
The scaled HPBW ($27\arcsec$) of the primary is also shown (dashed cyan).
\textit{Upper left:}
PDFs of different length scales $\delta$, with the same color coding.
The dashed lines show the values of the 50th percentiles.
\textit{Lower right:}
CDF for the second nearest neighbor separations, $\nns^{(2)}$.
\textit{Upper right:}
PDF for the second nearest neighbor separations, $\nns^{(2)}$.
\label{fig:CumulativeNns}
\end{figure*}

We compute PDFs for each clump (see above) and the ensemble distribution composed of all separation measurements from each clump aggregated together (Fig.~\ref{fig:CumulativeNns}).
The ensemble separation distribution is used to define a representative fragmentation scale from the SCCs in this survey.
As these clumps are at similar distances and blindly selected from Galactic Plane dust continuum surveys, the measured ensemble sample properties may be used to cautiously infer the properties of the Galactic high-mass SCC population ($\mclump \gtrsim \SI{e3}{\msun}$).
Additional observations are required to directly constrain the properties of SCCs with $\mclump \gtrsim \SI{e4}{\msun}$ (if they exist outside of the Central Molecular Zone) or SCCs below the mass range of this sample, $\mclump \lesssim \SI{400}{\msun}$.
Figure \ref{fig:CumulativeNns} shows the cumulative distribution function (CDF) for the ensemble of \jnns\ measurements as drawn from the MC sampling for the projected separations, projection-corrected separations, and relevant scales such as the resolution and primary beam HPBW.
The projection-corrected ensemble distribution has a median value of $\jnns = 0.82$ with a $(25,75)$ percentile interval of $0.52-1.25$.
The percentiles for $\jnns = 0.5$, $1$, $2$, and $3$ are, respectively, $23.6$, $63.3$, $90.3$, and $97.4$.
Overall, the sample of SCCs show a fragmentation scale that is well characterized by the thermal Jeans length.

A relatively small fraction of the separation distribution is inconsistent with the thermal Jeans length, $<10\%$ for $> 2 \times \ljeans$.
The large separations do not result from a single or small number of clumps with consistently large separations, but from isolated individual sources within clumps that show fragmentation near the thermal Jeans length.
G30660 and G30912, for example, have a significant proportion of the distribution at large separations (see Fig.~\ref{fig:NnsBySource} left), but do not have peculiar dust temperatures, between $\td = 11 - \SI{12}{K}$ from Hi-GAL SED fits.
This portion of the separation distribution may indicate an additional scale for hierarchical fragmentation where a source of non-thermal support prevents fragmentation at the thermal Jeans scale.

The Jeans length can further take into account sources of non-thermal support, such as turbulence or magnetic fields, by using an effective sound speed 
\begin{equation}
   c_\mathrm{s,eff} = \left( c_\mathrm{s}^2 + \sigma_\mathrm{nt}^2 \right)^{1/2} 
\end{equation}
through the contribution of a non-thermal velocity dispersion $\sigma_\mathrm{nt}$.
From \citetalias{svoboda16}, 9 out of 12 clumps have $T_\mathrm{K}$ measured from \ce{NH3} (at $32\arcsec$ resolution).
The measured velocity dispersions (i.e.\ $c_\mathrm{s,eff}$) determined from the spectral line model fit range between $\sigma(\ce{NH3}) = 0.50 - \SI{0.95}{\kms}$ with a median value of \SI{0.65}{\kms}, corresponding to $\sigma_\mathrm{nt} \approx \SI{0.62}{\kms}$ for $c_\mathrm{s} = \SI{0.21}{\kms}$ at $T_\mathrm{K} = \SI{12}{K}$ (where $T_\mathrm{K} = 11-\SI{14}{K}$).
Replacing $c_\mathrm{s}$ with $c_\mathrm{s,eff}$ in Equation \ref{eq:JeansLength} yields the effective Jeans length, or when turbulence is the dominant source of non-thermal support, the turbulent Jeans length $\ljeanstu$.
Because $\lambda_\mathrm{j} \propto c_\mathrm{s}$, the increase $c_\mathrm{s,eff} / c_\mathrm{s} \sim 2.4 - 4.5$ (median $3.1$) yields a similar scaling for $\ljeanstu / \ljeans$.
In comparison, $\jnns = 3$ ($\delta_\mathrm{nns} / \langle \ljeanstu \rangle \approx 0.32$) occurs at the $97.4$ percentile, and thus while such separations are not absent from the data, they are also not representative of the fragmentation measured within the ALMA maps.
The length scale distribution is incomplete beyond $\jnns > 3.1$, where $10\%$ of the MC trials would have 3D separations greater than or equal to the FOV ($40\arcsec$).

\section{Discussion}\label{sec:Discussion}
The physical processes regulating fragmentation in molecular clouds remain an open problem in star formation.
How much are SCCs supported against gravitationally induced fragmentation from non-thermal forms of pressure, such as magnetic fields (\bfield-fields) and/or turbulence?
Individual SCCs have been studied at high-resolution \citep{beuther15a,sanhueza17}, but we shall discuss a systematic set of observations on a representative sample of high-mass SCCs.
Here we describe the fragmentation characteristics of SCCs in the context of theoretical models of star and cluster formation and compare to existing high-resolution observations of clumps and IRDCs.

\subsection{Cylindrical Fragmentation in SCCs}\label{ssec:HierarchicalFragmentation}
As shown in \S\ref{sec:FragmentationScale}, we find that clumps fragment at scales consistent with the thermal Jeans length in SCCs.
It is known however that geometry and non-thermal support affect the predicted fragmentation scale, producing deviations from that expected for an isothermal, uniform medium.
In this section we discuss how the fragmentation scale observed with ALMA compares to different characteristic length scales.

Filaments are ubiquitous in both observed molecular clouds and simulations \citep[e.g.,][]{barnard07,andre14,smithr16}, and thus cylindrical geometry is of special significance to dense molecular regions.
On larger spatial scales observable in MIR extinction, it is clear that the clump peaks are embedded in filamentary gas structures (see Fig.\ \ref{fig:IrMosaic} and G23297 for a good example).
An infinite, self-gravitating cylinder is unstable to axisymmetric perturbations or ``sausage'' instability, where the cylinder fragments at the scale of the fastest growing mode of the fluid instability \citep{chandrasekhar53,ostriker64,larson85,nagasawa87,inutsuka92}.
For a pressure confined isothermal gas cylinder of radius $R$ and scale height $H = c_\mathrm{s} \left( 4 \pi G \rho_\mathrm{c} \right)^{-1/2}$ (where $\rho_\mathrm{c}$ is the central density of the cylinder) then the fastest growing mode depends on the ratio of $R$ and $H$ \citep{nagasawa87}.
In the case where $R \ll H$ then $\lambda_\mathrm{cyl} \approx 10.8 R$; and, alternatively where $R \gg H$, then $\lambda_\mathrm{cyl} \approx 22.4 H$ \citep{nagasawa87,jackson10}.

Is the isothermal, cylindrical fragmentation scale representative in SCCs?
The approximation of SCCs as isothermal is imperfect due to shielding that decreases the temperatures of inner regions, but the assumption is generally more valid than clumps with active HMSF and substantial internal protostellar heating and feedback.
Observed aspect ratios of $\gtrsim \! 5$ over the full clump extent support the approximation of an infinite cylindrical geometry.
The typical radial extent of the SCCs as observed in the MIR extinction maps suggests $R \sim \SI{0.4}{pc}$.
Assuming that the cylinder central density is equal to the observed clump peak density (i.e.\ $\rho_\mathrm{c} = \rho_0 \approx \SI{3e4}{cm^{-3}}$) then $H \sim \SI{0.02}{pc}$, and thus $R / H \sim 20$ roughly satisfies the condition $R \gg H$.
Note that for $\rho_\mathrm{c}$ equal to the clump peak density then this simplifies to  $\lambda_\mathrm{cyl} / \ljeans \approx 3.50$, or for the median $\ljeans = \SI{0.137}{pc}$, $\lambda_\mathrm{cyl} = \SI{0.480}{pc}$.
We find that $\lambda_\mathrm{cyl}$ is not representative of the $\nns$ distribution in SCCs, with the observed $\jnns \sim 1$.

Observational studies carried out on larger spatial scales than this work support $\lambda_\mathrm{cyl}$ as a characteristic scale in filaments \citep{beuther15b,friesen16}.
While these studies did not have sufficient spatial resolution to adequately resolve the thermal Jeans length, they probe separations on the clump scale and larger than $\sim \! \SI{1}{pc}$, as observed with ALMA.
This work complements the larger-scale studies by identifying fragmentation on the clump Jeans length at an early evolutionary phase.
This is supported by the results of \cite{kainulainen13b} who with \textit{Spitzer} MIR extinction mapping find that the molecular filament G11.11--0.12 is well described by filament fragmentation and turbulent $\lambda_\mathrm{cyl}$ on $\delta \gtrsim \SI{0.5}{pc}$ and $\ljeans$ on smaller scales.
\cite{beuther15b} in an analysis of the fragmentation in the star-forming filament IRDC 18223 find a mean fragment separation of $\delta = 0.40 \pm \SI{0.18}{pc}$, consistent with a thermal $\lambda_\mathrm{cyl} = \SI{0.44}{pc}$ of the filament, approximately twice that of $\ljeans = 0.07 - \SI{0.23}{pc}$, however the authors note that measures of $\delta$ should be considered an upper limit due to the sensitivity and resolution of the data.
\cite{friesen16} in a survey of the entire Serpens South molecular cloud \citep[as part of the GBT Ammonia Survey, GAS;][]{friesen17} find that the nearest neighbor separations of dense gas structures within the same filament are significantly larger than \ljeans\ and are well represented by $\lambda_\mathrm{cyl}$.
The spatial resolution is limited however to approximately $\ljeans \sim \SI{0.07}{pc}$, and thus does not properly resolve \ljeans\ in sources with $\langle n \rangle \gtrsim \SI{2e3}{cm^{-3}}$.
The above surveys support the view of hierarchical fragmentation by gravitationally unstable filaments, but lack the resolution to test what fragmentation process dominates on the scales of individual cores embedded within the clumps.
The measurements of the fragmentation scale presented in \S\ref{sec:FragmentationScale} complement the above studies at resolutions down to $\sim \! \SI{3000}{au}$ and provide further support to the view that filaments initially fragment at $\lambda_\mathrm{cyl}$ and then further fragment at $\ljeans$.

\subsection{Comparison to more active regions}\label{ssec:ActiveCompare}
Direct observations of star-forming IRDCs and embedded protoclusters have found fragmentation consistent with the thermal Jeans length \citep{palau15,beuther15a,teixeira16,busquet16,beuther18} but it is unknown if these systems represent the initial state of fragmentation.
Because high-mass SCCs may represent an initial stage of protocluster evolution before the formation of a high-mass star, they offer unique insight into the physical processes regulating fragmentation when compared to more evolved systems.
From a survey of dense star-forming cores \cite{palau15} find that the fragmentation on $\sim \! \SI{0.1}{pc}$ scales is best explained through thermal fragmentation.
Similar results are found at sub-core spatial scales of $\lesssim \! \SI{1000}{au}$ towards the Orion Molecular Cloud 1S \citep[OMC-1S;][]{palau17} and also consistent with the fragmentation measured in OMC-1N \citep{teixeira16}.
While the measured median nearest neighbor separation in SCCs is consistent with the thermal Jeans length of the clump gas, the distribution also shows a distinct peak at approximately an order of magnitude higher gas density near $\jnns \approx 0.3$ (see Figs. \ref{fig:NnsBySource} \& \ref{fig:CumulativeNns}).
These results may indicate continued thermal Jeans fragmentation such as in OMC-1S and OMC-1N.
\cite{beuther15a} find results approximately consistent with thermal Jeans fragmentation towards the $\sim \! \SI{800}{\msun}$ IRDC 18310--4, and while showing faint \SI{70}{\um} emission, has similar physical properties to the SCCs in this sample.
Similarly an analysis of the star-forming IRDC G14.225--0.506 favors thermal Jeans fragmentation \citep{busquet16}.
\cite{beuther18} present a minimal spanning tree analysis of the separations in the CORE survey of 20 luminous ($L_\mathrm{bol} > \SI{e4}{\lsun}$) high-mass star forming regions and find fragmentation at scales on the order of the thermal Jeans length or smaller.
As a possible explanation for the sub-Jeans length scales, \cite{beuther18} suggest that bulk motions from ongoing global collapse may have brought the fragments within closer proximity after having initially fragmented on the thermal Jeans scale.
All of the sources in the CORE survey are high-mass protostellar objects (HMPOs) and more evolved than this sample.
Thus our finding of fragmentation on the thermal Jeans length at an earlier evolutionary stage supports the interpretation of the COREs results and conclusion that the measured fragmentation scale may be impacted by the dynamical evolution of the protocluster.

The agreement between the nearest neighbor separations and the thermal Jeans length appears to favor a Jeans fragmentation process for stellar cluster formation. 
Indeed the thermal Jeans mass in typical star forming clumps is approximately \SI{1}{\msun}, which corresponds well with the stellar mass at the peak of the Initial Mass Function.
Therefore, \cite{larson05} argued that the thermal Jeans process is responsible for the formation of lower mass stars in a cluster.
\cite{zhang09} found that cores forming massive stars often $\gtrsim \! \SI{10}{\msun}$, an order of magnitude greater than the thermal Jeans mass of its parental clump.
These cores require additional support from turbulence to account for their formation.
Furthermore, the observed measurements imply that thermal physics provide the dominant form of support, but additional models exist to describe the thermal fragmentation process that differ in geometry and density profile.
For example, \cite{myers17} present 2D axisymmetric models of filamentary structure that fragment through the thermal instability of Bonnor-Ebert spheres above a threshold minimum density.
Because the Bonnor-Ebert radius and Jeans length have the same dependence on temperature and density with only slight differences in numerical coefficients, this leads to a fragmentation approximately equal to \ljeans.
When compared to the observed median $\jnns = 0.91$ from \S\ref{sec:FragmentationScale}, the spacings between cores predicted by \cite{myers17} is $\jnns = 0.71$ (for a concentration factor $q_\mathrm{Z} \equiv \langle n \rangle / n_\mathrm{min} = 2$) are broadly consistent.

\subsection{Coeval formation of low- and high-mass protostars?}\label{ssec:CoevalFormation}
It is not clear if SCCs are the progenitor environments of high-mass star formation.
Their high total masses ($M_\mathrm{cl} \sim \SI{1000}{\msun}$), high central densities ($\langle n \rangle \sim \SI{5e4}{cm^{-3}}$), cold gas kinetic temperatures ($\langle T_\mathrm{K}(\ce{NH3}) \rangle \sim \SI{11}{K}$), and low virial parameters ($\alpha_\mathrm{vir} \sim 0.1 - 1$) \citep{wienen12,svoboda16} all point to persistent, bound clumps with the likely necessary physical conditions for high-mass star formation
\citep{mckee07}.
However, no high-mass protostars are observed.
These observational facts are consistent with a scenario where high-mass stars form in SCCs through thermal fragmentation, and then accrete clump gas as initially low-mass protostars.
Thus, SCCs may represent a very early and unique stage in protocluster evolution preceding the formation of high-mass protostars.
This view is supported by cluster-scale theoretical simulations that incorporate protostellar and stellar feedback \citep{smithr09,wangp10,peters10a,peters10b,peters11}.
\cite{smithr09} find that no high-mass starless cores are formed in their models, and that massive stars originate from low-to-intermediate mass cores that become high-mass protostars via accretion.
The mass accreted comes primarily from the surrounding clump at scales $> \SI{0.1}{pc}$ \citep{smithr09,wangp10}.

\cite{cyganowski17} in a study towards the deeply embedded protocluster G11.92--0.61 discover low-mass cores in the accretion reservoir of the accreting high-mass protostellar object ``MM1'' with mass $M_* \sim 30-\SI{60}{\msun}$ \citep{ilee16}.
The detection of coeval low- and high-mass protostars is consistent with competitive accretion-type models of star formation (see \S\ref{sec:Introduction}).
At a comparable distance of $d_\odot = 3.37^{+0.39}_{-0.32}$ kpc \citep[derived from maser parallax][]{sato14} and total mass to the SCCs in this study, G11.92-0.61 is more evolved, coincident with several indicators of high-mass star formation, such as Class I \& II \ce{CH3OH} masers, \ce{H2O} masers, a GLIMPSE Extended Green Object \citep{cyganowski08}, numerous ``hot core'' molecular lines, and high-velocity collimated outflows.
The sample of SCCs in this study complement the study of G11.92--0.61  in \cite{cyganowski17} through ALMA observations at similar resolution and sensitivity for clumps in a less active evolutionary state.
In contrast, we find no clear high-mass protostellar cores or high-mass protostars in our sample of SCCs, while numerous accreting low-mass protostars are observed, as evidenced by bipolar outflows in \ce{CO}/\ce{SiO}.
If a few of the protostars in SCCs will accrete up to high-mass stars, for which the accretion reservoir of the clump is sufficient, then these observations support a coeval mode of protocluster formation at earlier phases.
When initially only low- to intermediate-mass protostars are present, this coeval formation may also be termed ``low-mass first'' to lie in contrast to the monolithic collapse of turbulently supported high-mass cores.
The competitive accretion-type simulations performed by \cite{smithr09} find that high-mass stars form initially from intermediate mass pre-stellar cores near the center of the gravitational potential which accrete principally from collapsing clump gas up to high-mass condensations.
An important feature of the \cite{smithr09} model is that low-mass protostars form within the accretion reservoir of the central protostar, at separations $< \! \SI{0.15}{pc}$.
This is well matched to the distribution of nearest neighbor separations found in this work of $\mu_{1/2}(\nns) = \SI{0.118}{pc}$ (see \S\ref{sec:FragmentationScale}).
As \cite{smithr09} point out, this signature is likely the most detectable at the early evolutionary phases of the clump where sources are less centrally concentrated in the potential and bright sources of emission are not present.

In contrast to the results of \cite{cyganowski17}, \cite{zhang15} in a study of the protocluster G28.24+0.06 P1 failed to detect a distributed population of low-mass cores with Cycle~0 ALMA observations.
Based on this \cite{zhang15} draw the conclusion that the distributed population of low-mass cores forms at a later evolutionary stage and that they are not, at least for the initial generation of protostars, coeval.
Because G11.91--0.61 is at a later evolutionary stage, the distributed population of low-mass protostars observed in it may have developed after the massive cores formed.
The SCCs in this study are in a similar early evolutionary phase as G28.24+0.06 and also similarly lack high-mass protostars (the maximum core mass in G28.4+0.06 is $M_\mathrm{core} \sim \SI{16}{\msun}$).
Accurate core masses are required for a quantitative analysis of the mass segregation and related length scales, but the diversity in morphologies shown within the sample, from distributed (e.g.~G30660, G29558) to weakly fragmented (e.g.~G28539, G29601), supports the presence of a distributed low-mass core population at the initial evolutionary phase for some systems.
It is possible that depending on the initial level of support provided against fragmentation individual systems develop with varying degrees of hierarchy and segregation, and that the conclusions of \cite{zhang15} and \cite{cyganowski17} may both be correct for sources of different initial physical conditions.

The short evolutionary timescales of high-mass starless clumps, $\tau_\mathrm{SCC} \sim 0.5-\SI{0.1}{Myr}$ for $\mclump = 1-\SI{3e3}{\msun}$ \citetalias{svoboda16}, is also consistent with the simulations of \cite{smithr09} that show that the central, resultant high-mass protostar accretes in $0.25 \times t_\mathrm{dyn} \sim \SI{0.12}{Myr}$ the clump dynamical time, over a diameter of $\sim \! \SI{0.4}{pc}$ (equivalent to the ALMA HPBW) \citep[see also][]{wangp10}.
Similarly, \citep{battersby17} perform a lifetime analysis of dense, molecular gas ($N(\ce{H2}) \gtrsim \SI{e22}{cm^-2}$) analyzed on a per-pixel basis from a Hi-GAL $2\deg \times 2\deg$ field near $\ell = 30\deg$.
They find a timescale that is consistent for starless regions of $0.2-\SI{1.7}{Myr}$, although with substantial uncertainty.
The similarity in timescales is reasonable, as once a high-mass protostar forms, it would be accompanied by observational star formation indicators that identify it as a protostellar clump and remove it from the SCC category, as determined in \citetalias{svoboda16}.
Further, we also observe hierarchical fragmentation as evidenced by the multi-modal distribution of nearest neighbor separations (see Fig.\ \ref{fig:CumulativeNns}), as seen in G11.92--0.61.
The ubiquity of filamentary structures observed (see Fig.\ \ref{fig:ContinuumTwelve}) may also point to accretion mediated by sub-sonically gravitationally contracting filaments \citep{smithr16}.
This may suggest that while self-gravitating, turbulent clumps are not globally collapsing, accretion may yet be mediated through locally collapsing filaments.
This latter point will be the topic of further research investigated with ALMA observations of $\mathrm{N_2H^+}$ $J=1\rightarrow0$ to study the kinematics of the filaments observed in this sample SCCs.

\section{Conclusions}\label{sec:Conclusions}
We present the first systematic observations of a large sample of well-vetted starless clump candidates with ALMA at high-resolution ($\sim \! \SI{3000}{au}$) capable of resolving the thermal Jeans length and sensitivity (\SI{50}{\micro\jy\per\beam}) sufficient for detecting point sources down to $\sim \! \SI{0.3}{\msun}$ and moderately compact starless cores down to $\sim \! \SI{1.0}{\msun}$).
The targets are selected from a complete sample of clumps identified from large Galactic Plane surveys.
The sample is composed of 12 high-mass SCCs within \SI{5}{kpc} from \cite{svoboda16} and \cite{traficante15} which did not show detected emission at \SI{70}{\um} or other star formation indicators.
Because these systems have not been affected by the extreme (proto-)stellar feedback of high-mass stars they are ideal environments to study the initial conditions of protocluster evolution.
Our main findings are:
\begin{enumerate}
    \item The newly sensitive ALMA Band 6 $12+\SI{7}{m}$ ($\nu_\mathrm{c} \approx \SI{230}{GHz}$) data show multiple indicators of low-/intermediate-mass star formation activity present in 11 out of 12 formerly starless clump candidates.
    This is determined through the presence of bipolar outflows detected in \ce{CO} $J=2\rightarrow1$ and \ce{SiO} $J=5\rightarrow4$ emission, and high-excitation p-\ce{H2CO} $3_{2,2} \rightarrow 2_{2,1}$ emission ($\eupper / k = \SI{68.1}{K}$).
    These observations caution the interpretation of infrared dark clouds and SCCs identified from Galactic Plane surveys as quiescent, and unless shown otherwise are, given the findings towards this sample, likely to host low-/intermediate-mass star formation activity below the luminosity completeness of current surveys.
    \item We compare representative examples of resolved and unresolved continuum sources with radiative transfer models of starless cores computed with \radmc.
    Unresolved sources are poorly fit by starless core models with typical physical properties.
    The range of models does not encompass the most compact and dense cores ($R_\mathrm{flat} < \SI{1e3}{au}$, $n_\mathrm{in} \gtrsim \SI{1e7}{cm^{-3}}$), but the short core free-fall times ($t_\mathrm{ff} \lesssim \SI{1e4}{yr}$) and the observed similar flux-density to Gould's Belt low-/intermediate-mass protostars, support the conclusion that these cores are protostellar even without identified outflows in \ce{CO} or \ce{SiO}.
    \item Two high-mass starless core candidates in G28539 are identified and well fit by starless core models, with $M_\mathrm{S2} = 29^{52}_{15} \, \msun$ and $M_\mathrm{S4} = 14^{34}_{6.0}$.
    Without supplementary measurements to infer the dust temperature profile, the masses are highly uncertain, and are consistent within the uncertainties of only forming an intermediate mass star ($M_* < \SI{8}{\msun}$).
    \item G28539 is the sole remaining starless clump candidate without any definitive indications of protostellar activity from the ALMA observations.
    It is the most massive SCC in the sample ($M_\mathrm{cl} \approx 3600^{+600}_{-500} \, \msun$, $d_\odot =  4.8^{+0.3}{-0.3}\, \mathrm{kpc}$), and stands as an excellent target to study the initial conditions of protocluster evolution.
    A marginal \SI{24}{\um} source, however, is observed coincident with \SI{1.3}{mm} continuum source (G28539 S1) near the NW edge of the ALMA field, which may be evidence or protostellar activity.
    Further indirect evidence for star formation exists from compact \ce{SiO} and \ce{CH3OH} emission, although the source of emission is not associated with a continuum source.
    If these signatures are indeed associated with protostellar activity there would be no true high-mass starless clumps in this sample.
    \item A high degree of fragmentation is observed, with nearest neighbor separations consistent with the clump scale thermal Jeans length ($\sim \! \SI{0.1}{pc}$).
    In context of previous observations that on larger scales see separations consistent with the turbulent Jeans length or cylindrical thermal Jeans length, our findings support a hierarchical fragmentation process, where the highest density regions of SCCs are not strongly supported against fragmentation by turbulence or magnetic fields.
    \item Observed embedded low- to intermediate-mass star formation and thermal Jeans fragmentation in high-mass SCCs are consistent with models of star formation that form high-mass stars through gravitationally driven cloud inflow, in which low- and high-mass stars form coevally.
    However, further observations and followup study are necessary to properly characterize the clump star formation efficiency, protostellar accretion rates, and presence of dynamical flows in molecular tracers to validate this conclusion.
\end{enumerate}

\section{Acknowledgments}\label{sec:Acknowledgements}
We are grateful to the anonymous referee for many helpful comments and suggestions.
BES would like to thank Al Wootten and the staff of the NRAO North American ALMA Science Center, Kimberly Ward-Duong for helpful discussions, and Viviana Rosero for access to the JVLA C Band image towards G28539.
BES was supported in part by the NSF Graduate Research Fellowship under grant No.\ DGE-114395.
YLS and BES acknowledge the support from the NSF AAG grant No.\ AST-1410190.
HB acknowledges support from the European Research Council under the European Community's Horizon 2020 framework program (2014--2020) via the ERC Consolidator Grant ``From Cloud to Star Formation (CSF)'' (project number 648505).

\software{
This research has made use of the following software projects:
    \href{https://astropy.org/}{Astropy} \citep{astropy18},
    \href{https://matplotlib.org/}{Matplotlib} \citep{matplotlib07},
    \href{http://www.numpy.org/}{NumPy} and \href{https://scipy.org/}{SciPy} \citep{numpy07},
    \href{https://pandas.pydata.org/}{Pandas} \citep{pandas10},
    \href{https://ipython.org/}{IPython} \citep{ipython07},
    \href{https://casa.nrao.edu/}{CASA} \citep{casa07},
    and
    the NASA's Astrophysics Data System.
}

\facilities{
    ALMA
}

%%%%%%%%%%%%%%%%%%%%%%%%%%%%%%%%%%%%%%%%%%%%%%%%%%%%%%%%%%%%%%%%%%%%%%%%%%%%%%%%
%				Bibliography
%%%%%%%%%%%%%%%%%%%%%%%%%%%%%%%%%%%%%%%%%%%%%%%%%%%%%%%%%%%%%%%%%%%%%%%%%%%%%%%%

\bibliographystyle{aasjournal}
\bibliography{mybib}

\begin{thebibliography}{}
\expandafter\ifx\csname natexlab\endcsname\relax\def\natexlab#1{#1}\fi
\providecommand{\url}[1]{\href{#1}{#1}}
\providecommand{\dodoi}[1]{doi:~\href{https://doi.org/#1}{\nolinkurl{#1}}}
\providecommand{\doeprint}[1]{\href{https://ascl.net/#1}{\nolinkurl{https://ascl.net/#1}}}
\providecommand{\doarXiv}[1]{\href{https://arxiv.org/abs/#1}{\nolinkurl{https://arxiv.org/abs/#1}}}

\bibitem[{{Aguirre} {et~al.}(2011){Aguirre}, {Ginsburg}, {Dunham}, {Drosback},
  {Bally}, {Battersby}, {Bradley}, {Cyganowski}, {Dowell}, {Evans}, {Glenn},
  {Harvey}, {Rosolowsky}, {Stringfellow}, {Walawender}, \&
  {Williams}}]{aguirre11}
{Aguirre}, J.~E., {Ginsburg}, A.~G., {Dunham}, M.~K., {et~al.} 2011, \apjs,
  192, 4, \dodoi{10.1088/0067-0049/192/1/4}

\bibitem[{{Andr{\'e}} {et~al.}(2014){Andr{\'e}}, {Di Francesco},
  {Ward-Thompson}, {Inutsuka}, {Pudritz}, \& {Pineda}}]{andre14}
{Andr{\'e}}, P., {Di Francesco}, J., {Ward-Thompson}, D., {et~al.} 2014,
  Protostars and Planets VI, 27,
  \dodoi{10.2458/azu_uapress_9780816531240-ch002}

\bibitem[{{Arce} {et~al.}(2007){Arce}, {Shepherd}, {Gueth}, {Lee}, {Bachiller},
  {Rosen}, \& {Beuther}}]{arce07}
{Arce}, H.~G., {Shepherd}, D., {Gueth}, F., {et~al.} 2007, Protostars and
  Planets V, 245

\bibitem[{{Barnard}(1907)}]{barnard07}
{Barnard}, E.~E. 1907, \apj, 25, \dodoi{10.1086/141434}

\bibitem[{{Battersby} {et~al.}(2010){Battersby}, {Bally}, {Jackson},
  {Ginsburg}, {Shirley}, {Schlingman}, \& {Glenn}}]{battersby10}
{Battersby}, C., {Bally}, J., {Jackson}, J.~M., {et~al.} 2010, \apj, 721, 222,
  \dodoi{10.1088/0004-637X/721/1/222}

\bibitem[{{Battersby} {et~al.}(2017){Battersby}, {Bally}, \&
  {Svoboda}}]{battersby17}
{Battersby}, C., {Bally}, J., \& {Svoboda}, B. 2017, \apj, 835, 263,
  \dodoi{10.3847/1538-4357/835/2/263}

\bibitem[{{Benjamin} {et~al.}(2003){Benjamin}, {Churchwell}, {Babler}, {Bania},
  {Clemens}, {Cohen}, {Dickey}, {Indebetouw}, {Jackson}, {Kobulnicky},
  {Lazarian}, {Marston}, {Mathis}, {Meade}, {Seager}, {Stolovy}, {Watson},
  {Whitney}, {Wolff}, \& {Wolfire}}]{benjamin03}
{Benjamin}, R.~A., {Churchwell}, E., {Babler}, B.~L., {et~al.} 2003, \pasp,
  115, 953, \dodoi{10.1086/376696}

\bibitem[{{Bergin} \& {Tafalla}(2007)}]{bergin07}
{Bergin}, E.~A., \& {Tafalla}, M. 2007, \araa, 45, 339,
  \dodoi{10.1146/annurev.astro.45.071206.100404}

\bibitem[{{Beuther} {et~al.}(2007){Beuther}, {Churchwell}, {McKee}, \&
  {Tan}}]{beuther07}
{Beuther}, H., {Churchwell}, E.~B., {McKee}, C.~F., \& {Tan}, J.~C. 2007,
  Protostars and Planets V, 165

\bibitem[{{Beuther} {et~al.}(2015{\natexlab{a}}){Beuther}, {Ragan}, {Johnston},
  {Henning}, {Hacar}, \& {Kainulainen}}]{beuther15b}
{Beuther}, H., {Ragan}, S.~E., {Johnston}, K., {et~al.} 2015{\natexlab{a}},
  \aap, 584, A67, \dodoi{10.1051/0004-6361/201527108}

\bibitem[{{Beuther} {et~al.}(2015{\natexlab{b}}){Beuther}, {Henning}, {Linz},
  {Feng}, {Ragan}, {Smith}, {Bihr}, {Sakai}, \& {Kuiper}}]{beuther15a}
{Beuther}, H., {Henning}, T., {Linz}, H., {et~al.} 2015{\natexlab{b}}, \aap,
  581, A119, \dodoi{10.1051/0004-6361/201526759}

\bibitem[{{Beuther} {et~al.}(2018){Beuther}, {Mottram}, {Ahmadi}, {Bosco},
  {Linz}, {Henning}, {Klaassen}, {Winters}, {Maud}, {Kuiper}, {Semenov},
  {Gieser}, {Peters}, {Urquhart}, {Pudritz}, {Ragan}, {Feng}, {Keto},
  {Leurini}, {Cesaroni}, {Beltran}, {Palau}, {Sanchez-Monge}, {Galvan-Madrid},
  {Zhang}, {Schilke}, {Wyrowski}, {Johnston}, {Longmore}, {Lumsden}, {Hoare},
  {Menten}, \& {Csengeri}}]{beuther18}
{Beuther}, H., {Mottram}, J.~C., {Ahmadi}, A., {et~al.} 2018, ArXiv e-prints.
\newblock \doarXiv{1805.01191}

\bibitem[{{Black}(1994)}]{black94}
{Black}, J.~H. 1994, in Astronomical Society of the Pacific Conference Series,
  Vol.~58, The First Symposium on the Infrared Cirrus and Diffuse Interstellar
  Clouds, ed. R.~M. {Cutri} \& W.~B. {Latter}, 355

\bibitem[{{Bonnell} {et~al.}(2001){Bonnell}, {Clarke}, {Bate}, \&
  {Pringle}}]{bonnell01}
{Bonnell}, I.~A., {Clarke}, C.~J., {Bate}, M.~R., \& {Pringle}, J.~E. 2001,
  \mnras, 324, 573, \dodoi{10.1046/j.1365-8711.2001.04311.x}

\bibitem[{{Bonnor}(1956)}]{bonnor56}
{Bonnor}, W.~B. 1956, \mnras, 116, 351, \dodoi{10.1093/mnras/116.3.351}

\bibitem[{{Busquet} {et~al.}(2016){Busquet}, {Estalella}, {Palau}, {Liu},
  {Zhang}, {Girart}, {de Gregorio-Monsalvo}, {Pillai}, {Anglada}, \&
  {Ho}}]{busquet16}
{Busquet}, G., {Estalella}, R., {Palau}, A., {et~al.} 2016, \apj, 819, 139,
  \dodoi{10.3847/0004-637X/819/2/139}

\bibitem[{{Cardelli} {et~al.}(1989){Cardelli}, {Clayton}, \&
  {Mathis}}]{cardelli89}
{Cardelli}, J.~A., {Clayton}, G.~C., \& {Mathis}, J.~S. 1989, \apj, 345, 245,
  \dodoi{10.1086/167900}

\bibitem[{{Carniani} {et~al.}(2015){Carniani}, {Maiolino}, {De Zotti},
  {Negrello}, {Marconi}, {Bothwell}, {Capak}, {Carilli}, {Castellano},
  {Cristiani}, {Ferrara}, {Fontana}, {Gallerani}, {Jones}, {Ohta}, {Ota},
  {Pentericci}, {Santini}, {Sheth}, {Vallini}, {Vanzella}, {Wagg}, \&
  {Williams}}]{carniani15}
{Carniani}, S., {Maiolino}, R., {De Zotti}, G., {et~al.} 2015, \aap, 584, A78,
  \dodoi{10.1051/0004-6361/201525780}

\bibitem[{{Chambers} {et~al.}(2009){Chambers}, {Jackson}, {Rathborne}, \&
  {Simon}}]{chambers09}
{Chambers}, E.~T., {Jackson}, J.~M., {Rathborne}, J.~M., \& {Simon}, R. 2009,
  \apjs, 181, 360, \dodoi{10.1088/0067-0049/181/2/360}

\bibitem[{{Chandrasekhar} \& {Fermi}(1953)}]{chandrasekhar53}
{Chandrasekhar}, S., \& {Fermi}, E. 1953, \apj, 118, 116,
  \dodoi{10.1086/145732}

\bibitem[{{Churchwell} {et~al.}(2009){Churchwell}, {Babler}, {Meade},
  {Whitney}, {Benjamin}, {Indebetouw}, {Cyganowski}, {Robitaille}, {Povich},
  {Watson}, \& {Bracker}}]{churchwell09}
{Churchwell}, E., {Babler}, B.~L., {Meade}, M.~R., {et~al.} 2009, \pasp, 121,
  213, \dodoi{10.1086/597811}

\bibitem[{{Contreras} {et~al.}(2013){Contreras}, {Schuller}, {Urquhart},
  {Csengeri}, {Wyrowski}, {Beuther}, {Bontemps}, {Bronfman}, {Henning},
  {Menten}, {Schilke}, {Walmsley}, {Wienen}, {Tackenberg}, \&
  {Linz}}]{contreras13}
{Contreras}, Y., {Schuller}, F., {Urquhart}, J.~S., {et~al.} 2013, \aap, 549,
  A45, \dodoi{10.1051/0004-6361/201220155}

\bibitem[{{Csengeri} {et~al.}(2014){Csengeri}, {Urquhart}, {Schuller}, {Motte},
  {Bontemps}, {Wyrowski}, {Menten}, {Bronfman}, {Beuther}, {Henning}, {Testi},
  {Zavagno}, \& {Walmsley}}]{csengeri14}
{Csengeri}, T., {Urquhart}, J.~S., {Schuller}, F., {et~al.} 2014, \aap, 565,
  A75, \dodoi{10.1051/0004-6361/201322434}

\bibitem[{{Cyganowski} {et~al.}(2017){Cyganowski}, {Brogan}, {Hunter}, {Smith},
  {Kruijssen}, {Bonnell}, \& {Zhang}}]{cyganowski17}
{Cyganowski}, C.~J., {Brogan}, C.~L., {Hunter}, T.~R., {et~al.} 2017, \mnras,
  468, 3694, \dodoi{10.1093/mnras/stx043}

\bibitem[{{Cyganowski} {et~al.}(2008){Cyganowski}, {Whitney}, {Holden},
  {Braden}, {Brogan}, {Churchwell}, {Indebetouw}, {Watson}, {Babler},
  {Benjamin}, {Gomez}, {Meade}, {Povich}, {Robitaille}, \&
  {Watson}}]{cyganowski08}
{Cyganowski}, C.~J., {Whitney}, B.~A., {Holden}, E., {et~al.} 2008, \aj, 136,
  2391, \dodoi{10.1088/0004-6256/136/6/2391}

\bibitem[{{Draine}(1978)}]{draine78}
{Draine}, B.~T. 1978, \apjs, 36, 595, \dodoi{10.1086/190513}

\bibitem[{{Dullemond} {et~al.}(2012){Dullemond}, {Juhasz}, {Pohl}, {Sereshti},
  {Shetty}, {Peters}, {Commercon}, \& {Flock}}]{dullemond12}
{Dullemond}, C.~P., {Juhasz}, A., {Pohl}, A., {et~al.} 2012, {RADMC-3D: A
  multi-purpose radiative transfer tool}, Astrophysics Source Code Library.
\newblock \doeprint{1202.015}

\bibitem[{{Dunham} {et~al.}(2014){Dunham}, {Stutz}, {Allen}, {Evans},
  {Fischer}, {Megeath}, {Myers}, {Offner}, {Poteet}, {Tobin}, \&
  {Vorobyov}}]{dunham14}
{Dunham}, M.~M., {Stutz}, A.~M., {Allen}, L.~E., {et~al.} 2014, Protostars and
  Planets VI, 195, \dodoi{10.2458/azu_uapress_9780816531240-ch009}

\bibitem[{{Dzib} {et~al.}(2010){Dzib}, {Loinard}, {Mioduszewski}, {Boden},
  {Rodr{\'{\i}}guez}, \& {Torres}}]{dzib10}
{Dzib}, S., {Loinard}, L., {Mioduszewski}, A.~J., {et~al.} 2010, \apj, 718,
  610, \dodoi{10.1088/0004-637X/718/2/610}

\bibitem[{{Ebert}(1955)}]{ebert55}
{Ebert}, R. 1955, \zap, 37, 217

\bibitem[{{Eden} {et~al.}(2017){Eden}, {Moore}, {Plume}, {Urquhart},
  {Thompson}, {Parsons}, {Dempsey}, {Rigby}, {Morgan}, {Thomas}, {Berry},
  {Buckle}, {Brunt}, {Butner}, {Carretero}, {Chrysostomou}, {Currie},
  {deVilliers}, {Fich}, {Gibb}, {Hoare}, {Jenness}, {Manser}, {Mottram},
  {Natario}, {Olguin}, {Peretto}, {Pestalozzi}, {Polychroni}, {Redman},
  {Salji}, {Summers}, {Tahani}, {Traficante}, {diFrancesco}, {Evans}, {Fuller},
  {Johnstone}, {Joncas}, {Longmore}, {Martin}, {Richer}, {Weferling}, {White},
  \& {Zhu}}]{eden17}
{Eden}, D.~J., {Moore}, T.~J.~T., {Plume}, R., {et~al.} 2017, \mnras, 469,
  2163, \dodoi{10.1093/mnras/stx874}

\bibitem[{{Elia} {et~al.}(2017){Elia}, {Molinari}, {Schisano}, {Pestalozzi},
  {Pezzuto}, {Merello}, {Noriega-Crespo}, {Moore}, {Russeil}, {Mottram},
  {Paladini}, {Strafella}, {Benedettini}, {Bernard}, {Di Giorgio}, {Eden},
  {Fukui}, {Plume}, {Bally}, {Martin}, {Ragan}, {Jaffa}, {Motte}, {Olmi},
  {Schneider}, {Testi}, {Wyrowski}, {Zavagno}, {Calzoletti}, {Faustini},
  {Natoli}, {Palmeirim}, {Piacentini}, {Piazzo}, {Pilbratt}, {Polychroni},
  {Baldeschi}, {Beltr{\'a}n}, {Billot}, {Cambr{\'e}sy}, {Cesaroni},
  {Garc{\'{\i}}a-Lario}, {Hoare}, {Huang}, {Joncas}, {Liu}, {Maiolo}, {Marsh},
  {Maruccia}, {M{\`e}ge}, {Peretto}, {Rygl}, {Schilke}, {Thompson},
  {Traficante}, {Umana}, {Veneziani}, {Ward-Thompson}, {Whitworth}, {Arab},
  {Bandieramonte}, {Becciani}, {Brescia}, {Buemi}, {Bufano}, {Butora},
  {Cavuoti}, {Costa}, {Fiorellino}, {Hajnal}, {Hayakawa}, {Kacsuk}, {Leto}, {Li
  Causi}, {Marchili}, {Martinavarro-Armengol}, {Mercurio}, {Molinaro},
  {Riccio}, {Sano}, {Sciacca}, {Tachihara}, {Torii}, {Trigilio}, {Vitello}, \&
  {Yamamoto}}]{elia17}
{Elia}, D., {Molinari}, S., {Schisano}, E., {et~al.} 2017, \mnras, 471, 100,
  \dodoi{10.1093/mnras/stx1357}

\bibitem[{{Ellsworth-Bowers} {et~al.}(2015){Ellsworth-Bowers}, {Rosolowsky},
  {Glenn}, {Ginsburg}, {Evans}, {Battersby}, {Shirley}, \&
  {Svoboda}}]{ellsworthbowers15a}
{Ellsworth-Bowers}, T.~P., {Rosolowsky}, E., {Glenn}, J., {et~al.} 2015, \apj,
  799, 29, \dodoi{10.1088/0004-637X/799/1/29}

\bibitem[{{Enoch} {et~al.}(2011){Enoch}, {Corder}, {Duch{\^e}ne}, {Bock},
  {Bolatto}, {Culverhouse}, {Kwon}, {Lamb}, {Leitch}, {Marrone}, {Muchovej},
  {P{\'e}rez}, {Scott}, {Teuben}, {Wright}, \& {Zauderer}}]{enoch11}
{Enoch}, M.~L., {Corder}, S., {Duch{\^e}ne}, G., {et~al.} 2011, \apjs, 195, 21,
  \dodoi{10.1088/0067-0049/195/2/21}

\bibitem[{{Feng} {et~al.}(2016{\natexlab{a}}){Feng}, {Beuther}, {Zhang}, {Liu},
  {Zhang}, {Wang}, \& {Qiu}}]{feng16}
{Feng}, S., {Beuther}, H., {Zhang}, Q., {et~al.} 2016{\natexlab{a}}, \apj, 828,
  100, \dodoi{10.3847/0004-637X/828/2/100}

\bibitem[{{Feng} {et~al.}(2016{\natexlab{b}}){Feng}, {Beuther}, {Zhang}, {Liu},
  {Zhang}, {Wang}, \& {Qiu}}]{feng16b}
---. 2016{\natexlab{b}}, \apj, 828, 100, \dodoi{10.3847/0004-637X/828/2/100}

\bibitem[{{Frank} {et~al.}(2014){Frank}, {Ray}, {Cabrit}, {Hartigan}, {Arce},
  {Bacciotti}, {Bally}, {Benisty}, {Eisl{\"o}ffel}, {G{\"u}del}, {Lebedev},
  {Nisini}, \& {Raga}}]{frank14}
{Frank}, A., {Ray}, T.~P., {Cabrit}, S., {et~al.} 2014, Protostars and Planets
  VI, 451, \dodoi{10.2458/azu_uapress_9780816531240-ch020}

\bibitem[{{Friesen} {et~al.}(2016){Friesen}, {Bourke}, {Di Francesco},
  {Gutermuth}, \& {Myers}}]{friesen16}
{Friesen}, R.~K., {Bourke}, T.~L., {Di Francesco}, J., {Gutermuth}, R., \&
  {Myers}, P.~C. 2016, \apj, 833, 204, \dodoi{10.3847/1538-4357/833/2/204}

\bibitem[{{Friesen} {et~al.}(2017){Friesen}, {Pineda}, {co-PIs}, {Rosolowsky},
  {Alves}, {Chac{\'o}n-Tanarro}, {How-Huan Chen}, {Chun-Yuan Chen}, {Di
  Francesco}, {Keown}, {Kirk}, {Punanova}, {Seo}, {Shirley}, {Ginsburg},
  {Hall}, {Offner}, {Singh}, {Arce}, {Caselli}, {Goodman}, {Martin}, {Matzner},
  {Myers}, {Redaelli}, \& {GAS Collaboration}}]{friesen17}
{Friesen}, R.~K., {Pineda}, J.~E., {co-PIs}, {et~al.} 2017, \apj, 843, 63,
  \dodoi{10.3847/1538-4357/aa6d58}

\bibitem[{{Ginsburg} {et~al.}(2013){Ginsburg}, {Glenn}, {Rosolowsky},
  {Ellsworth-Bowers}, {Battersby}, {Dunham}, {Merello}, {Shirley}, {Bally},
  {Evans}, {Stringfellow}, \& {Aguirre}}]{ginsburg13}
{Ginsburg}, A., {Glenn}, J., {Rosolowsky}, E., {et~al.} 2013, \apjs, 208, 14,
  \dodoi{10.1088/0067-0049/208/2/14}

\bibitem[{{Hartmann} {et~al.}(2012){Hartmann}, {Ballesteros-Paredes}, \&
  {Heitsch}}]{hartmann12}
{Hartmann}, L., {Ballesteros-Paredes}, J., \& {Heitsch}, F. 2012, \mnras, 420,
  1457, \dodoi{10.1111/j.1365-2966.2011.20131.x}

\bibitem[{{Hatsukade} {et~al.}(2013){Hatsukade}, {Ohta}, {Seko}, {Yabe}, \&
  {Akiyama}}]{hatsukade13}
{Hatsukade}, B., {Ohta}, K., {Seko}, A., {Yabe}, K., \& {Akiyama}, M. 2013,
  \apjl, 769, L27, \dodoi{10.1088/2041-8205/769/2/L27}

\bibitem[{{Hocuk} {et~al.}(2017){Hocuk}, {Szucs}, {Caselli}, {Cazaux},
  {Spaans}, \& {Esplugues}}]{hocuk17}
{Hocuk}, S., {Szucs}, L., {Caselli}, P., {et~al.} 2017, ArXiv e-prints.
\newblock \doarXiv{1704.02763}

\bibitem[{{Hosokawa} \& {Omukai}(2009)}]{hosokawa09}
{Hosokawa}, T., \& {Omukai}, K. 2009, \apj, 691, 823,
  \dodoi{10.1088/0004-637X/691/1/823}

\bibitem[{Hunter(2007)}]{matplotlib07}
Hunter, J.~D. 2007, Computing In Science \& Engineering, 9, 90,
  \dodoi{10.1109/MCSE.2007.55}

\bibitem[{{Ilee} {et~al.}(2016){Ilee}, {Cyganowski}, {Nazari}, {Hunter},
  {Brogan}, {Forgan}, \& {Zhang}}]{ilee16}
{Ilee}, J.~D., {Cyganowski}, C.~J., {Nazari}, P., {et~al.} 2016, \mnras, 462,
  4386, \dodoi{10.1093/mnras/stw1912}

\bibitem[{{Inutsuka} \& {Miyama}(1992)}]{inutsuka92}
{Inutsuka}, S.-I., \& {Miyama}, S.~M. 1992, \apj, 388, 392,
  \dodoi{10.1086/171162}

\bibitem[{{Jackson} {et~al.}(2010){Jackson}, {Finn}, {Chambers}, {Rathborne},
  \& {Simon}}]{jackson10}
{Jackson}, J.~M., {Finn}, S.~C., {Chambers}, E.~T., {Rathborne}, J.~M., \&
  {Simon}, R. 2010, \apjl, 719, L185, \dodoi{10.1088/2041-8205/719/2/L185}

\bibitem[{{Jim{\'e}nez-Serra} {et~al.}(2010){Jim{\'e}nez-Serra}, {Caselli},
  {Tan}, {Hernandez}, {Fontani}, {Butler}, \& {van Loo}}]{jimenezserra10}
{Jim{\'e}nez-Serra}, I., {Caselli}, P., {Tan}, J.~C., {et~al.} 2010, \mnras,
  406, 187, \dodoi{10.1111/j.1365-2966.2010.16698.x}

\bibitem[{{Kainulainen} {et~al.}(2013){Kainulainen}, {Ragan}, {Henning}, \&
  {Stutz}}]{kainulainen13b}
{Kainulainen}, J., {Ragan}, S.~E., {Henning}, T., \& {Stutz}, A. 2013, \aap,
  557, A120, \dodoi{10.1051/0004-6361/201321760}

\bibitem[{{Kainulainen} {et~al.}(2017){Kainulainen}, {Stutz}, {Stanke},
  {Abreu-Vicente}, {Beuther}, {Henning}, {Johnston}, \&
  {Megeath}}]{kainulainen17a}
{Kainulainen}, J., {Stutz}, A.~M., {Stanke}, T., {et~al.} 2017, \aap, 600,
  A141, \dodoi{10.1051/0004-6361/201628481}

\bibitem[{{Kauffmann} \& {Pillai}(2010)}]{kauffmann10}
{Kauffmann}, J., \& {Pillai}, T. 2010, \apj, 723, L7,
  \dodoi{10.1088/2041-8205/723/1/L7}

\bibitem[{{Kong} {et~al.}(2017){Kong}, {Tan}, {Caselli}, {Fontani}, {Liu}, \&
  {Butler}}]{kong17}
{Kong}, S., {Tan}, J.~C., {Caselli}, P., {et~al.} 2017, \apj, 834, 193,
  \dodoi{10.3847/1538-4357/834/2/193}

\bibitem[{{Kroupa}(2001)}]{kroupa01}
{Kroupa}, P. 2001, \mnras, 322, 231, \dodoi{10.1046/j.1365-8711.2001.04022.x}

\bibitem[{{Larson}(1985)}]{larson85}
{Larson}, R.~B. 1985, \mnras, 214, 379, \dodoi{10.1093/mnras/214.3.379}

\bibitem[{{Larson}(2005)}]{larson05}
---. 2005, \mnras, 359, 211, \dodoi{10.1111/j.1365-2966.2005.08881.x}

\bibitem[{{Lippok} {et~al.}(2016){Lippok}, {Launhardt}, {Henning}, {Balog},
  {Beuther}, {Kainulainen}, {Krause}, {Linz}, {Nielbock}, {Ragan},
  {Robitaille}, {Sadavoy}, \& {Schmiedeke}}]{lippok16}
{Lippok}, N., {Launhardt}, R., {Henning}, T., {et~al.} 2016, \aap, 592, A61,
  \dodoi{10.1051/0004-6361/201525792}

\bibitem[{{Louvet} {et~al.}(2016){Louvet}, {Motte}, {Gusdorf}, {Nguy{\^e}n
  Luong}, {Lesaffre}, {Duarte-Cabral}, {Maury}, {Schneider}, {Hill}, {Schilke},
  \& {Gueth}}]{louvet16}
{Louvet}, F., {Motte}, F., {Gusdorf}, A., {et~al.} 2016, \aap, 595, A122,
  \dodoi{10.1051/0004-6361/201629077}

\bibitem[{{Lu} {et~al.}(2015){Lu}, {Zhang}, {Wang}, \& {Gu}}]{lu15b}
{Lu}, X., {Zhang}, Q., {Wang}, K., \& {Gu}, Q. 2015, \apj, 805, 171,
  \dodoi{10.1088/0004-637X/805/2/171}

\bibitem[{{McGuire} {et~al.}(2016){McGuire}, {Fuller}, {Peretto}, {Zhang},
  {Traficante}, {Avison}, \& {Jimenez-Serra}}]{mcguire16}
{McGuire}, C., {Fuller}, G.~A., {Peretto}, N., {et~al.} 2016, \aap, 594, A118,
  \dodoi{10.1051/0004-6361/201527062}

\bibitem[{{McKee} \& {Ostriker}(2007)}]{mckee07}
{McKee}, C.~F., \& {Ostriker}, E.~C. 2007, \araa, 45, 565,
  \dodoi{10.1146/annurev.astro.45.051806.110602}

\bibitem[{{McKee} \& {Tan}(2002)}]{mckee02}
{McKee}, C.~F., \& {Tan}, J.~C. 2002, \nat, 416, 59, \dodoi{10.1038/416059a}

\bibitem[{{McKee} \& {Tan}(2003)}]{mckee03}
---. 2003, \apj, 585, 850, \dodoi{10.1086/346149}

\bibitem[{McKinney(2010)}]{pandas10}
McKinney, W. 2010, in Proceedings of the 9th Python in Science Conference, ed.
  S.~van~der Walt \& J.~Millman, 51 -- 56

\bibitem[{{McMullin} {et~al.}(2007){McMullin}, {Waters}, {Schiebel}, {Young},
  \& {Golap}}]{casa07}
{McMullin}, J.~P., {Waters}, B., {Schiebel}, D., {Young}, W., \& {Golap}, K.
  2007, in Astronomical Society of the Pacific Conference Series, Vol. 376,
  Astronomical Data Analysis Software and Systems XVI, ed. R.~A. {Shaw},
  F.~{Hill}, \& D.~J. {Bell}, 127

\bibitem[{{Molinari} {et~al.}(2010){Molinari}, {Swinyard}, {Bally}, {Barlow},
  {Bernard}, {Martin}, {Moore}, {Noriega-Crespo}, {Plume}, {Testi}, {Zavagno},
  {Abergel}, {Ali}, {Andr{\'e}}, {Baluteau}, {Benedettini}, {Bern{\'e}},
  {Billot}, {Blommaert}, {Bontemps}, {Boulanger}, {Brand}, {Brunt}, {Burton},
  {Campeggio}, {Carey}, {Caselli}, {Cesaroni}, {Cernicharo}, {Chakrabarti},
  {Chrysostomou}, {Codella}, {Cohen}, {Compiegne}, {Davis}, {de Bernardis}, {de
  Gasperis}, {Di Francesco}, {di Giorgio}, {Elia}, {Faustini}, {Fischera},
  {Fukui}, {Fuller}, {Ganga}, {Garcia-Lario}, {Giard}, {Giardino}, {Glenn},
  {Goldsmith}, {Griffin}, {Hoare}, {Huang}, {Jiang}, {Joblin}, {Joncas},
  {Juvela}, {Kirk}, {Lagache}, {Li}, {Lim}, {Lord}, {Lucas}, {Maiolo},
  {Marengo}, {Marshall}, {Masi}, {Massi}, {Matsuura}, {Meny}, {Minier},
  {Miville-Desch{\^e}nes}, {Montier}, {Motte}, {M{\"u}ller}, {Natoli}, {Neves},
  {Olmi}, {Paladini}, {Paradis}, {Pestalozzi}, {Pezzuto}, {Piacentini},
  {Pomar{\`e}s}, {Popescu}, {Reach}, {Richer}, {Ristorcelli}, {Roy}, {Royer},
  {Russeil}, {Saraceno}, {Sauvage}, {Schilke}, {Schneider-Bontemps},
  {Schuller}, {Schultz}, {Shepherd}, {Sibthorpe}, {Smith}, {Smith},
  {Spinoglio}, {Stamatellos}, {Strafella}, {Stringfellow}, {Sturm}, {Taylor},
  {Thompson}, {Tuffs}, {Umana}, {Valenziano}, {Vavrek}, {Viti}, {Waelkens},
  {Ward-Thompson}, {White}, {Wyrowski}, {Yorke}, \& {Zhang}}]{molinari10}
{Molinari}, S., {Swinyard}, B., {Bally}, J., {et~al.} 2010, \pasp, 122, 314,
  \dodoi{10.1086/651314}

\bibitem[{{Molinari} {et~al.}(2016){Molinari}, {Schisano}, {Elia},
  {Pestalozzi}, {Traficante}, {Pezzuto}, {Swinyard}, {Noriega-Crespo}, {Bally},
  {Moore}, {Plume}, {Zavagno}, {di Giorgio}, {Liu}, {Pilbratt}, {Mottram},
  {Russeil}, {Piazzo}, {Veneziani}, {Benedettini}, {Calzoletti}, {Faustini},
  {Natoli}, {Piacentini}, {Merello}, {Palmese}, {Del Grande}, {Polychroni},
  {Rygl}, {Polenta}, {Barlow}, {Bernard}, {Martin}, {Testi}, {Ali},
  {Andr{\'e}}, {Beltr{\'a}n}, {Billot}, {Carey}, {Cesaroni}, {Compi{\`e}gne},
  {Eden}, {Fukui}, {Garcia-Lario}, {Hoare}, {Huang}, {Joncas}, {Lim}, {Lord},
  {Martinavarro-Armengol}, {Motte}, {Paladini}, {Paradis}, {Peretto},
  {Robitaille}, {Schilke}, {Schneider}, {Schulz}, {Sibthorpe}, {Strafella},
  {Thompson}, {Umana}, {Ward-Thompson}, \& {Wyrowski}}]{molinari16a}
{Molinari}, S., {Schisano}, E., {Elia}, D., {et~al.} 2016, \aap, 591, A149,
  \dodoi{10.1051/0004-6361/201526380}

\bibitem[{{Motte} {et~al.}(2017){Motte}, {Bontemps}, \& {Louvet}}]{motte17}
{Motte}, F., {Bontemps}, S., \& {Louvet}, F. 2017, ArXiv e-prints.
\newblock \doarXiv{1706.00118}

\bibitem[{{M{\"u}ller} {et~al.}(2005){M{\"u}ller}, {Schl{\"o}der}, {Stutzki},
  \& {Winnewisser}}]{cdms}
{M{\"u}ller}, H. S.~P., {Schl{\"o}der}, F., {Stutzki}, J., \& {Winnewisser}, G.
  2005, Journal of Molecular Structure, 742, 215,
  \dodoi{10.1016/j.molstruc.2005.01.027}

\bibitem[{{Myers}(2017)}]{myers17}
{Myers}, P.~C. 2017, \apj, 838, 10, \dodoi{10.3847/1538-4357/aa5fa8}

\bibitem[{{Nagasawa}(1987)}]{nagasawa87}
{Nagasawa}, M. 1987, Progress of Theoretical Physics, 77, 635,
  \dodoi{10.1143/PTP.77.635}

\bibitem[{{Nguyen-Lu'o'ng} {et~al.}(2013){Nguyen-Lu'o'ng}, {Motte}, {Carlhoff},
  {Louvet}, {Lesaffre}, {Schilke}, {Hill}, {Hennemann}, {Gusdorf}, {Didelon},
  {Schneider}, {Bontemps}, {Duarte-Cabral}, {Menten}, {Martin}, {Wyrowski},
  {Bendo}, {Roussel}, {Bernard}, {Bronfman}, {Henning}, {Kramer}, \&
  {Heitsch}}]{luong13}
{Nguyen-Lu'o'ng}, Q., {Motte}, F., {Carlhoff}, P., {et~al.} 2013, \apj, 775,
  88, \dodoi{10.1088/0004-637X/775/2/88}

\bibitem[{Oliphant(2007)}]{numpy07}
Oliphant, T.~E. 2007, Computing in Science \& Engineering, 9

\bibitem[{{Ossenkopf} \& {Henning}(1994)}]{ossenkopf94}
{Ossenkopf}, V., \& {Henning}, T. 1994, \aap, 291, 943

\bibitem[{{Ostriker}(1964)}]{ostriker64}
{Ostriker}, J. 1964, \apj, 140, 1056, \dodoi{10.1086/148005}

\bibitem[{{Palau} {et~al.}(2015){Palau}, {Ballesteros-Paredes},
  {V{\'a}zquez-Semadeni}, {S{\'a}nchez-Monge}, {Estalella}, {Fall}, {Zapata},
  {Camacho}, {G{\'o}mez}, {Naranjo-Romero}, {Busquet}, \& {Fontani}}]{palau15}
{Palau}, A., {Ballesteros-Paredes}, J., {V{\'a}zquez-Semadeni}, E., {et~al.}
  2015, \mnras, 453, 3785, \dodoi{10.1093/mnras/stv1834}

\bibitem[{{Palau} {et~al.}(2017){Palau}, {Zapata}, {Roman-Zuniga},
  {Sanchez-Monge}, {Estalella}, {Busquet}, {Girart}, {Fuente}, \&
  {Commercon}}]{palau17}
{Palau}, A., {Zapata}, L.~A., {Roman-Zuniga}, C.~G., {et~al.} 2017, ArXiv
  e-prints.
\newblock \doarXiv{1706.04623}

\bibitem[{{Peretto} \& {Fuller}(2009)}]{peretto09}
{Peretto}, N., \& {Fuller}, G.~A. 2009, \aap, 505, 405,
  \dodoi{10.1051/0004-6361/200912127}

\bibitem[{P{\'e}rez \& Granger(2007)}]{ipython07}
P{\'e}rez, F., \& Granger, B.~E. 2007, Computing in Science \& Engineering, 9

\bibitem[{{Peters} {et~al.}(2011){Peters}, {Banerjee}, {Klessen}, \& {Mac
  Low}}]{peters11}
{Peters}, T., {Banerjee}, R., {Klessen}, R.~S., \& {Mac Low}, M.-M. 2011, \apj,
  729, 72, \dodoi{10.1088/0004-637X/729/1/72}

\bibitem[{{Peters} {et~al.}(2010{\natexlab{a}}){Peters}, {Banerjee}, {Klessen},
  {Mac Low}, {Galv{\'a}n-Madrid}, \& {Keto}}]{peters10a}
{Peters}, T., {Banerjee}, R., {Klessen}, R.~S., {et~al.} 2010{\natexlab{a}},
  \apj, 711, 1017, \dodoi{10.1088/0004-637X/711/2/1017}

\bibitem[{{Peters} {et~al.}(2010{\natexlab{b}}){Peters}, {Klessen}, {Mac Low},
  \& {Banerjee}}]{peters10b}
{Peters}, T., {Klessen}, R.~S., {Mac Low}, M.-M., \& {Banerjee}, R.
  2010{\natexlab{b}}, \apj, 725, 134, \dodoi{10.1088/0004-637X/725/1/134}

\bibitem[{{Plummer}(1911)}]{plummer1911}
{Plummer}, H.~C. 1911, \mnras, 71, 460, \dodoi{10.1093/mnras/71.5.460}

\bibitem[{{Pokhrel} {et~al.}(2018){Pokhrel}, {Myers}, {Dunham}, {Stephens},
  {Sadavoy}, {Zhang}, {Bourke}, {Tobin}, {Lee}, {Gutermuth}, \&
  {Offner}}]{pokhrel18}
{Pokhrel}, R., {Myers}, P.~C., {Dunham}, M.~M., {et~al.} 2018, \apj, 853, 5,
  \dodoi{10.3847/1538-4357/aaa240}

\bibitem[{{Remijan} {et~al.}(2007){Remijan}, {Markwick-Kemper}, \& {ALMA
  Working Group on Spectral Line Frequencies}}]{slaim}
{Remijan}, A.~J., {Markwick-Kemper}, A., \& {ALMA Working Group on Spectral
  Line Frequencies}. 2007, in American Astronomical Society Meeting Abstracts,
  Vol. 211, 132.11

\bibitem[{{Robitaille}(2017)}]{robitaille17}
{Robitaille}, T.~P. 2017, \aap, 600, A11, \dodoi{10.1051/0004-6361/201425486}

\bibitem[{{Robitaille} {et~al.}(2007){Robitaille}, {Whitney}, {Indebetouw}, \&
  {Wood}}]{robitaille07}
{Robitaille}, T.~P., {Whitney}, B.~A., {Indebetouw}, R., \& {Wood}, K. 2007,
  \apjs, 169, 328, \dodoi{10.1086/512039}

\bibitem[{{Rosero} {et~al.}(2016){Rosero}, {Hofner}, {Claussen}, {Kurtz},
  {Cesaroni}, {Araya}, {Carrasco-Gonz{\'a}lez}, {Rodr{\'{\i}}guez}, {Menten},
  {Wyrowski}, {Loinard}, \& {Ellingsen}}]{rosero16}
{Rosero}, V., {Hofner}, P., {Claussen}, M., {et~al.} 2016, ArXiv e-prints.
\newblock \doarXiv{1609.03269}

\bibitem[{{Rosolowsky} {et~al.}(2010){Rosolowsky}, {Dunham}, {Ginsburg},
  {Bradley}, {Aguirre}, {Bally}, {Battersby}, {Cyganowski}, {Dowell},
  {Drosback}, {Evans}, {Glenn}, {Harvey}, {Stringfellow}, {Walawender}, \&
  {Williams}}]{rosolowsky10}
{Rosolowsky}, E., {Dunham}, M.~K., {Ginsburg}, A., {et~al.} 2010, \apjs, 188,
  123, \dodoi{10.1088/0067-0049/188/1/123}

\bibitem[{{Rosolowsky} {et~al.}(2008){Rosolowsky}, {Pineda}, {Kauffmann}, \&
  {Goodman}}]{rosolowsky08b}
{Rosolowsky}, E.~W., {Pineda}, J.~E., {Kauffmann}, J., \& {Goodman}, A.~A.
  2008, \apj, 679, 1338, \dodoi{10.1086/587685}

\bibitem[{{Sanhueza} {et~al.}(2013){Sanhueza}, {Jackson}, {Foster},
  {Jimenez-Serra}, {Dirienzo}, \& {Pillai}}]{sanhueza13}
{Sanhueza}, P., {Jackson}, J.~M., {Foster}, J.~B., {et~al.} 2013, \apj, 773,
  123, \dodoi{10.1088/0004-637X/773/2/123}

\bibitem[{{Sanhueza} {et~al.}(2017){Sanhueza}, {Jackson}, {Zhang},
  {Guzm{\'a}n}, {Lu}, {Stephens}, {Wang}, \& {Tatematsu}}]{sanhueza17}
{Sanhueza}, P., {Jackson}, J.~M., {Zhang}, Q., {et~al.} 2017, \apj, 841, 97,
  \dodoi{10.3847/1538-4357/aa6ff8}

\bibitem[{{Sato} {et~al.}(2014){Sato}, {Wu}, {Immer}, {Zhang}, {Sanna}, {Reid},
  {Dame}, {Brunthaler}, \& {Menten}}]{sato14}
{Sato}, M., {Wu}, Y.~W., {Immer}, K., {et~al.} 2014, \apj, 793, 72,
  \dodoi{10.1088/0004-637X/793/2/72}

\bibitem[{{Schilke} {et~al.}(1997){Schilke}, {Walmsley}, {Pineau des Forets},
  \& {Flower}}]{schilke97}
{Schilke}, P., {Walmsley}, C.~M., {Pineau des Forets}, G., \& {Flower}, D.~R.
  1997, \aap, 321, 293

\bibitem[{{Schuller} {et~al.}(2009){Schuller}, {Menten}, {Contreras},
  {Wyrowski}, {Schilke}, {Bronfman}, {Henning}, {Walmsley}, {Beuther},
  {Bontemps}, {Cesaroni}, {Deharveng}, {Garay}, {Herpin}, {Lefloch}, {Linz},
  {Mardones}, {Minier}, {Molinari}, {Motte}, {Nyman}, {Reveret}, {Risacher},
  {Russeil}, {Schneider}, {Testi}, {Troost}, {Vasyunina}, {Wienen}, {Zavagno},
  {Kovacs}, {Kreysa}, {Siringo}, \& {Wei{\ss}}}]{schuller09}
{Schuller}, F., {Menten}, K.~M., {Contreras}, Y., {et~al.} 2009, \aap, 504,
  415, \dodoi{10.1051/0004-6361/200811568}

\bibitem[{{Shirley} {et~al.}(2007){Shirley}, {Claussen}, {Bourke}, {Young}, \&
  {Blake}}]{shirley07}
{Shirley}, Y.~L., {Claussen}, M.~J., {Bourke}, T.~L., {Young}, C.~H., \&
  {Blake}, G.~A. 2007, \apj, 667, 329, \dodoi{10.1086/520570}

\bibitem[{{Shirley} {et~al.}(2005){Shirley}, {Nordhaus}, {Grcevich}, {Evans},
  {Rawlings}, \& {Tatematsu}}]{shirley05}
{Shirley}, Y.~L., {Nordhaus}, M.~K., {Grcevich}, J.~M., {et~al.} 2005, \apj,
  632, 982, \dodoi{10.1086/431963}

\bibitem[{{Smith} {et~al.}(2016){Smith}, {Glover}, {Klessen}, \&
  {Fuller}}]{smithr16}
{Smith}, R.~J., {Glover}, S.~C.~O., {Klessen}, R.~S., \& {Fuller}, G.~A. 2016,
  \mnras, 455, 3640, \dodoi{10.1093/mnras/stv2559}

\bibitem[{{Smith} {et~al.}(2009){Smith}, {Longmore}, \& {Bonnell}}]{smithr09}
{Smith}, R.~J., {Longmore}, S., \& {Bonnell}, I. 2009, \mnras, 400, 1775,
  \dodoi{10.1111/j.1365-2966.2009.15621.x}

\bibitem[{{Svoboda} {et~al.}(2016){Svoboda}, {Shirley}, {Battersby},
  {Rosolowsky}, {Ginsburg}, {Ellsworth-Bowers}, {Pestalozzi}, {Dunham},
  {Evans}, {Bally}, \& {Glenn}}]{svoboda16}
{Svoboda}, B.~E., {Shirley}, Y.~L., {Battersby}, C., {et~al.} 2016, \apj, 822,
  59, \dodoi{10.3847/0004-637X/822/2/59}

\bibitem[{{Tan} {et~al.}(2014){Tan}, {Beltr{\'a}n}, {Caselli}, {Fontani},
  {Fuente}, {Krumholz}, {McKee}, \& {Stolte}}]{tan14}
{Tan}, J.~C., {Beltr{\'a}n}, M.~T., {Caselli}, P., {et~al.} 2014, Protostars
  and Planets VI, 149, \dodoi{10.2458/azu_uapress_9780816531240-ch007}

\bibitem[{{Tan} {et~al.}(2013){Tan}, {Kong}, {Butler}, {Caselli}, \&
  {Fontani}}]{tan13}
{Tan}, J.~C., {Kong}, S., {Butler}, M.~J., {Caselli}, P., \& {Fontani}, F.
  2013, \apj, 779, 96, \dodoi{10.1088/0004-637X/779/2/96}

\bibitem[{{Tan} {et~al.}(2016){Tan}, {Kong}, {Zhang}, {Fontani}, {Caselli}, \&
  {Butler}}]{tan16}
{Tan}, J.~C., {Kong}, S., {Zhang}, Y., {et~al.} 2016, \apjl, 821, L3,
  \dodoi{10.3847/2041-8205/821/1/L3}

\bibitem[{{Teixeira} {et~al.}(2016){Teixeira}, {Takahashi}, {Zapata}, \&
  {Ho}}]{teixeira16}
{Teixeira}, P.~S., {Takahashi}, S., {Zapata}, L.~A., \& {Ho}, P.~T.~P. 2016,
  \aap, 587, A47, \dodoi{10.1051/0004-6361/201526807}

\bibitem[{{The Astropy Collaboration} {et~al.}(2018){The Astropy
  Collaboration}, {Price-Whelan}, {Sip{\H o}cz}, {G{\"u}nther}, {Lim},
  {Crawford}, {Conseil}, {Shupe}, {Craig}, {Dencheva}, {Ginsburg},
  {VanderPlas}, {Bradley}, {P{\'e}rez-Su{\'a}rez}, {de Val-Borro}, {Paper
  Contributors}, {Aldcroft}, {Cruz}, {Robitaille}, {Tollerud}, {Coordination
  Committee}, {Ardelean}, {Babej}, {Bach}, {Bachetti}, {Bakanov}, {Bamford},
  {Barentsen}, {Barmby}, {Baumbach}, {Berry}, {Biscani}, {Boquien}, {Bostroem},
  {Bouma}, {Brammer}, {Bray}, {Breytenbach}, {Buddelmeijer}, {Burke},
  {Calderone}, {Cano Rodr{\'{\i}}guez}, {Cara}, {Cardoso}, {Cheedella},
  {Copin}, {Corrales}, {Crichton}, {D'Avella}, {Deil}, {Depagne}, {Dietrich},
  {Donath}, {Droettboom}, {Earl}, {Erben}, {Fabbro}, {Ferreira}, {Finethy},
  {Fox}, {Garrison}, {Gibbons}, {Goldstein}, {Gommers}, {Greco}, {Greenfield},
  {Groener}, {Grollier}, {Hagen}, {Hirst}, {Homeier}, {Horton}, {Hosseinzadeh},
  {Hu}, {Hunkeler}, {Ivezi{\'c}}, {Jain}, {Jenness}, {Kanarek}, {Kendrew},
  {Kern}, {Kerzendorf}, {Khvalko}, {King}, {Kirkby}, {Kulkarni}, {Kumar},
  {Lee}, {Lenz}, {Littlefair}, {Ma}, {Macleod}, {Mastropietro}, {McCully},
  {Montagnac}, {Morris}, {Mueller}, {Mumford}, {Muna}, {Murphy}, {Nelson},
  {Nguyen}, {Ninan}, {N{\"o}the}, {Ogaz}, {Oh}, {Parejko}, {Parley}, {Pascual},
  {Patil}, {Patil}, {Plunkett}, {Prochaska}, {Rastogi}, {Reddy Janga},
  {Sabater}, {Sakurikar}, {Seifert}, {Sherbert}, {Sherwood-Taylor}, {Shih},
  {Sick}, {Silbiger}, {Singanamalla}, {Singer}, {Sladen}, {Sooley},
  {Sornarajah}, {Streicher}, {Teuben}, {Thomas}, {Tremblay}, {Turner},
  {Terr{\'o}n}, {van Kerkwijk}, {de la Vega}, {Watkins}, {Weaver}, {Whitmore},
  {Woillez}, {Zabalza}, \& {Contributors}}]{astropy18}
{The Astropy Collaboration}, {Price-Whelan}, A.~M., {Sip{\H o}cz}, B.~M.,
  {et~al.} 2018, \aj, 156, 123, \dodoi{10.3847/1538-3881/aabc4f}

\bibitem[{{Traficante} {et~al.}(2017){Traficante}, {Fuller}, {Billot},
  {Duarte-Cabral}, {Merello}, {Molinari}, {Peretto}, \&
  {Schisano}}]{traficante17}
{Traficante}, A., {Fuller}, G.~A., {Billot}, N., {et~al.} 2017, \mnras, 470,
  3882, \dodoi{10.1093/mnras/stx1375}

\bibitem[{{Traficante} {et~al.}(2015){Traficante}, {Fuller}, {Peretto},
  {Pineda}, \& {Molinari}}]{traficante15}
{Traficante}, A., {Fuller}, G.~A., {Peretto}, N., {Pineda}, J.~E., \&
  {Molinari}, S. 2015, \mnras, 451, 3089, \dodoi{10.1093/mnras/stv1158}

\bibitem[{{Veneziani} {et~al.}(2013){Veneziani}, {Elia}, {Noriega-Crespo},
  {Paladini}, {Carey}, {Faimali}, {Molinari}, {Pestalozzi}, {Piacentini},
  {Schisano}, \& {Tibbs}}]{veneziani13}
{Veneziani}, M., {Elia}, D., {Noriega-Crespo}, A., {et~al.} 2013, \aap, 549,
  A130, \dodoi{10.1051/0004-6361/201219570}

\bibitem[{{Wang} {et~al.}(2012){Wang}, {Zhang}, {Wu}, {Li}, \&
  {Zhang}}]{wangk12}
{Wang}, K., {Zhang}, Q., {Wu}, Y., {Li}, H.-b., \& {Zhang}, H. 2012, \apjl,
  745, L30, \dodoi{10.1088/2041-8205/745/2/L30}

\bibitem[{{Wang} {et~al.}(2011){Wang}, {Zhang}, {Wu}, \& {Zhang}}]{wangk11}
{Wang}, K., {Zhang}, Q., {Wu}, Y., \& {Zhang}, H. 2011, \apj, 735, 64,
  \dodoi{10.1088/0004-637X/735/1/64}

\bibitem[{{Wang} {et~al.}(2014){Wang}, {Zhang}, {Testi}, {van der Tak}, {Wu},
  {Zhang}, {Pillai}, {Wyrowski}, {Carey}, {Ragan}, \& {Henning}}]{wangk14}
{Wang}, K., {Zhang}, Q., {Testi}, L., {et~al.} 2014, \mnras, 439, 3275,
  \dodoi{10.1093/mnras/stu127}

\bibitem[{{Wang} {et~al.}(2010){Wang}, {Li}, {Abel}, \& {Nakamura}}]{wangp10}
{Wang}, P., {Li}, Z.-Y., {Abel}, T., \& {Nakamura}, F. 2010, \apj, 709, 27,
  \dodoi{10.1088/0004-637X/709/1/27}

\bibitem[{{Wang} {et~al.}(2006){Wang}, {Zhang}, {Rathborne}, {Jackson}, \&
  {Wu}}]{wangy06}
{Wang}, Y., {Zhang}, Q., {Rathborne}, J.~M., {Jackson}, J., \& {Wu}, Y. 2006,
  \apjl, 651, L125, \dodoi{10.1086/508939}

\bibitem[{{Weingartner} \& {Draine}(2001)}]{weingartner01}
{Weingartner}, J.~C., \& {Draine}, B.~T. 2001, \apj, 548, 296,
  \dodoi{10.1086/318651}

\bibitem[{{Whitworth} \& {Ward-Thompson}(2001)}]{whitworth01}
{Whitworth}, A.~P., \& {Ward-Thompson}, D. 2001, \apj, 547, 317,
  \dodoi{10.1086/318373}

\bibitem[{{Wienen} {et~al.}(2012){Wienen}, {Wyrowski}, {Schuller}, {Menten},
  {Walmsley}, {Bronfman}, \& {Motte}}]{wienen12}
{Wienen}, M., {Wyrowski}, F., {Schuller}, F., {et~al.} 2012, \aap, 544, A146,
  \dodoi{10.1051/0004-6361/201118107}

\bibitem[{{Williams} {et~al.}(1994){Williams}, {de Geus}, \&
  {Blitz}}]{williams94}
{Williams}, J.~P., {de Geus}, E.~J., \& {Blitz}, L. 1994, \apj, 428, 693,
  \dodoi{10.1086/174279}

\bibitem[{{Young} \& {Evans}(2005)}]{young05}
{Young}, C.~H., \& {Evans}, II, N.~J. 2005, \apj, 627, 293,
  \dodoi{10.1086/430436}

\bibitem[{{Yuan} {et~al.}(2017){Yuan}, {Wu}, {Ellingsen}, {Evans}, {Henkel},
  {Wang}, {Liu}, {Liu}, {Li}, \& {Zavagno}}]{yuan17}
{Yuan}, J., {Wu}, Y., {Ellingsen}, S.~P., {et~al.} 2017, \apjs, 231, 11,
  \dodoi{10.3847/1538-4365/aa7204}

\bibitem[{{Zhang} \& {Wang}(2011)}]{zhang11}
{Zhang}, Q., \& {Wang}, K. 2011, \apj, 733, 26,
  \dodoi{10.1088/0004-637X/733/1/26}

\bibitem[{{Zhang} {et~al.}(2015){Zhang}, {Wang}, {Lu}, \&
  {Jim{\'e}nez-Serra}}]{zhang15}
{Zhang}, Q., {Wang}, K., {Lu}, X., \& {Jim{\'e}nez-Serra}, I. 2015, \apj, 804,
  141, \dodoi{10.1088/0004-637X/804/2/141}

\bibitem[{{Zhang} {et~al.}(2009){Zhang}, {Wang}, {Pillai}, \&
  {Rathborne}}]{zhang09}
{Zhang}, Q., {Wang}, Y., {Pillai}, T., \& {Rathborne}, J. 2009, \apj, 696, 268,
  \dodoi{10.1088/0004-637X/696/1/268}

\bibitem[{{Zhang} {et~al.}(2014){Zhang}, {Tan}, \& {Hosokawa}}]{zhangy14}
{Zhang}, Y., {Tan}, J.~C., \& {Hosokawa}, T. 2014, \apj, 788, 166,
  \dodoi{10.1088/0004-637X/788/2/166}

\end{thebibliography}

%%%%%%%%%%%%%%%%%%%%%%%%%%%%%%%%%%%%%%%%%%%%%%%%%%%%%%%%%%%%%%%%%%%%%%%%%%%%%%%%
%				  Appendices
%%%%%%%%%%%%%%%%%%%%%%%%%%%%%%%%%%%%%%%%%%%%%%%%%%%%%%%%%%%%%%%%%%%%%%%%%%%%%%%%

\appendix
\section{Example CO outflow analysis}\label{apx:PvDiagram}
\begin{figure*}
    \centering
    \includegraphics[width=0.80\textwidth]{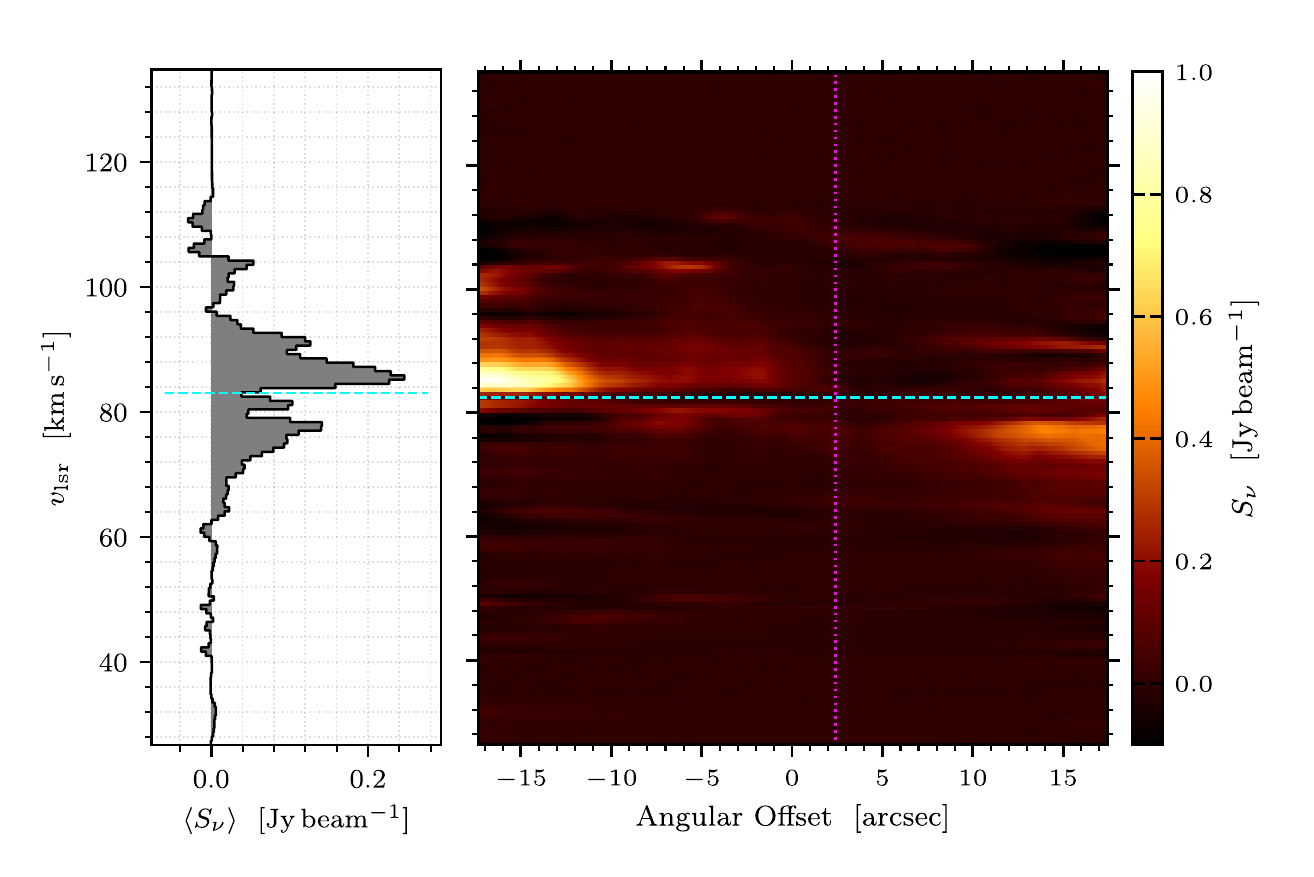}
\caption{
Average spectrum and position-velocity diagram for a $6\farcs0$ wide rectangular aperture lying along the outflow axis.
\textit{Left:} Spatially averaged spectrum.
The center LSR velocity $v_\mathrm{lsr} = \SI{83.0}{\kms}$ traced by \ce{H2CO} $3_{0,3} \rightarrow 2_{0,2}$ is shown in both panels (cyan dashed line).
\textit{Right:} Position-velocity diagram.
The peak position of continuum source G24051 S4 is visualized (magenta dotted line).
}
\label{fig:PvDiagramSpectrum}
\end{figure*}

The \ce{CO} $J = 2 \rightarrow 1$ image cubes show complex emission structures that complicate the identification of coherent velocity structures such as outflows.
Effects may be observed from spatial filtering, foreground and background clouds, and strong self-absorption at the clump systemic velocities.
Bipolar outflows with red- and blue-shifted velocity components may still be easily observed in the data however because they are bright and are coherent in velocity over many independent channels.
To illustrate these effects, we present a spatially averaged spectrum and position-velocity diagram (PV; Fig.~\ref{fig:PvDiagramSpectrum}) for the prominent NW-SE outflow originating from G24051 S4 (see Fig.~\ref{fig:CoOutflowMosaic}).
The spectrum and PV diagram are extracted from a $6\farcs0$ diameter rectangular aperture centered along the outflow axis.
Figure \ref{fig:PvDiagramSpectrum} shows bright, extended emission spanning up to $\sim\! \SI{20}{\kms}$ from the center LSR velocity of $v_\mathrm{lsr} = \SI{83}{\kms}$ determined from the dense gas tracer \ce{H2CO} $3_{0,3} \rightarrow 2_{0,2}$.
The red-shifted lobe (SE) and blue-shifted lobe (NW) are clearly observed in the PV diagram at negative and positive angular offsets along the rectangular aperture axis.

\section{\ce{SiO} $J=5\rightarrow4$ maps}\label{apx:SiOOutflows}
\begin{figure*}
    \centering
    \includegraphics[width=0.99\textwidth,trim={12mm 3mm 19mm 7mm},clip]{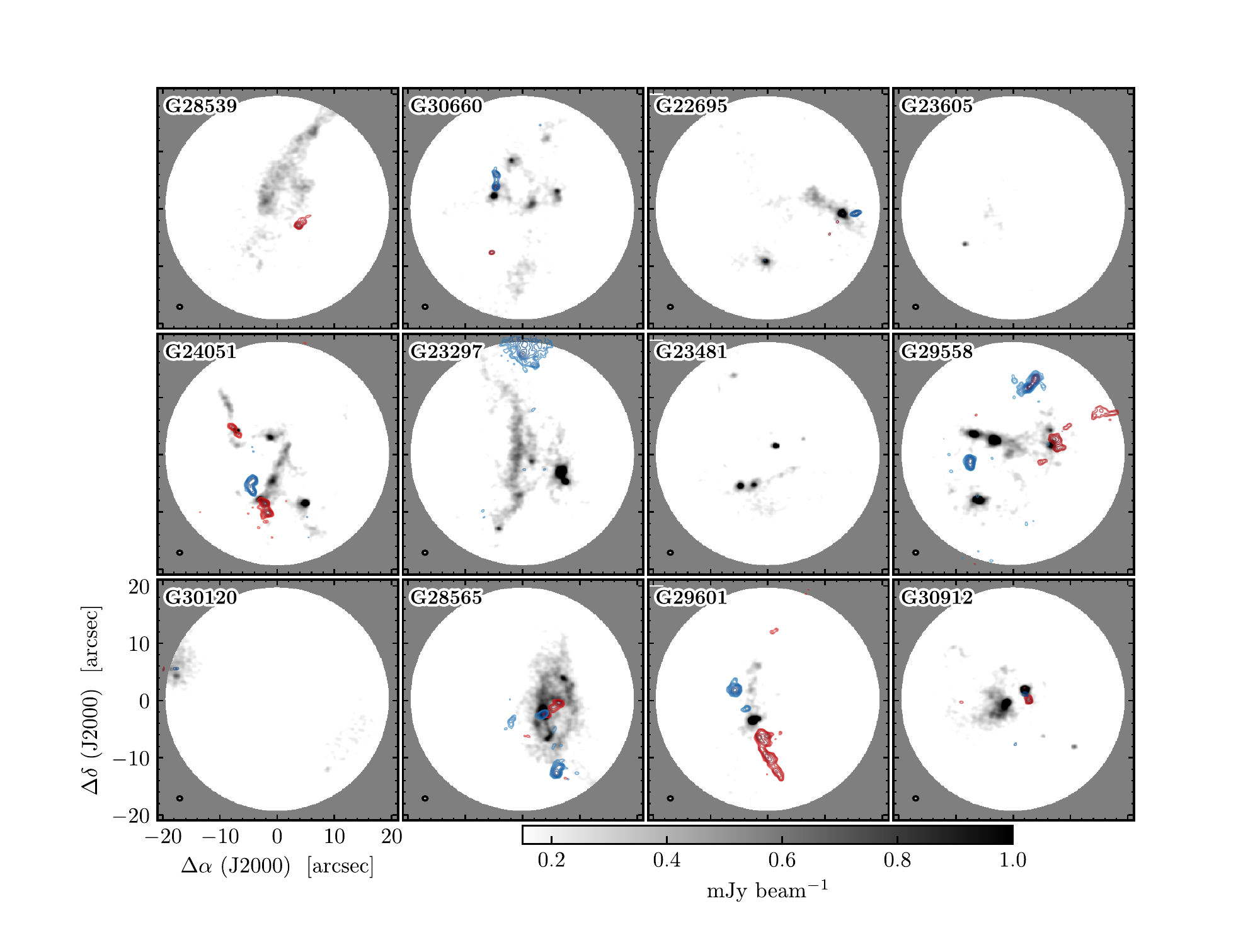}
\caption{
ALMA joint $12+\SI{7}{m}$ array \ce{SiO} $J=5\rightarrow4$ intensity of velocity components integrated between offsets \SIrange{+2}{+15}{\kms} (red contours) and between offsets \SIrange{-2}{-15}{\kms} (blue contours).
Bi-polar outflows are observed in 3 out of 12 clumps.
Contours are shown at logarithmically spaced steps of
0.050, 0.062, 0.075, 0.093, 0.114, 0.139, 0.171, 0.210, 0.258, and \SI{0.316}{\jy\kilo\meter\per\second}.
The inverted grayscale image shows the \SI{230}{GHz} continuum.
The image extends down to the 20\% power point ($40\arcsec$).
The maps are made from the dirty image cubes and have not been deconvolved with \textsc{clean}.
}
\label{fig:SiOOutflowMosaic}
\end{figure*}

Maps of the \ce{SiO} $J=5\rightarrow4$ red- and blue-shifted integrated intensities are shown in Figure \ref{fig:SiOOutflowMosaic}.
Three clumps have clear bipolar outflows: G24051 S5, G28565 S1, and G29601 S1.
All three outflows have CO $J=2\rightarrow1$ counterparts at similar positions and velocities.

\section{Core Model Properties}\label{apx:Models}
We follow a similar approach to modeling starless cores as found in \cite{shirley05} and \cite{lippok16}.
A similar approach is also used in \cite{mcguire16}.
We assume dust opacities $\kappa$ for coagulated grains and thin ice mantles in \citeauthor{ossenkopf94} (\citeyear{ossenkopf94}, hereafter \citetalias{ossenkopf94}) for moderately processed grains with a coagulation timescale of \SI{e5}{yr} at densities between \SIrange{e4}{e8}{cm^{-3}} (i.e.\ ``OH4'' through ``OH6'').
The coagulation density $n_\mathrm{cg}$ from \citetalias{ossenkopf94} is selected for each model core based on whether the mean density (weighted by mass) is in the range $0.5 \times n_\mathrm{cg} - 5 \times n_\mathrm{cg}$.
The value of the dust opacity when interpolated at $\lambda = \SI{1.3}{mm}$ for \SI{e5}{cm^{-3}} (``OH5a'') is $\kappa = \SI{0.90}{cm^2.g^{-1}}$ and varies between \SIrange{0.51}{1.11}{cm^2.g^{-1}} over the full range of densities.
We calculate total the gas mass using a dust-to-gas mass ratio of $f_\mathrm{d} \equiv m_\mathrm{d} / m_\mathrm{g} = 1/110$ and an ISM mean molecular weight of $\mu = 2.33$.
To fully sample the spectral range of the ISRF we extrapolate the the dust opacities from \SI{1}{\um} to \SI{90}{nm} using the prescription of \cite{cardelli89} and from \SI{1.3}{mm} to \SI{10}{mm} using the power law $\kappa_\nu \propto \nu^\beta$ with $\beta = 1.75$.
In addition, scattering efficiencies for the the \citetalias{ossenkopf94} models are added following \cite{young05} and albedos from the \cite{weingartner01} WD3.1 model.

\begin{figure}
    \centering
    \includegraphics[width=0.47\textwidth]{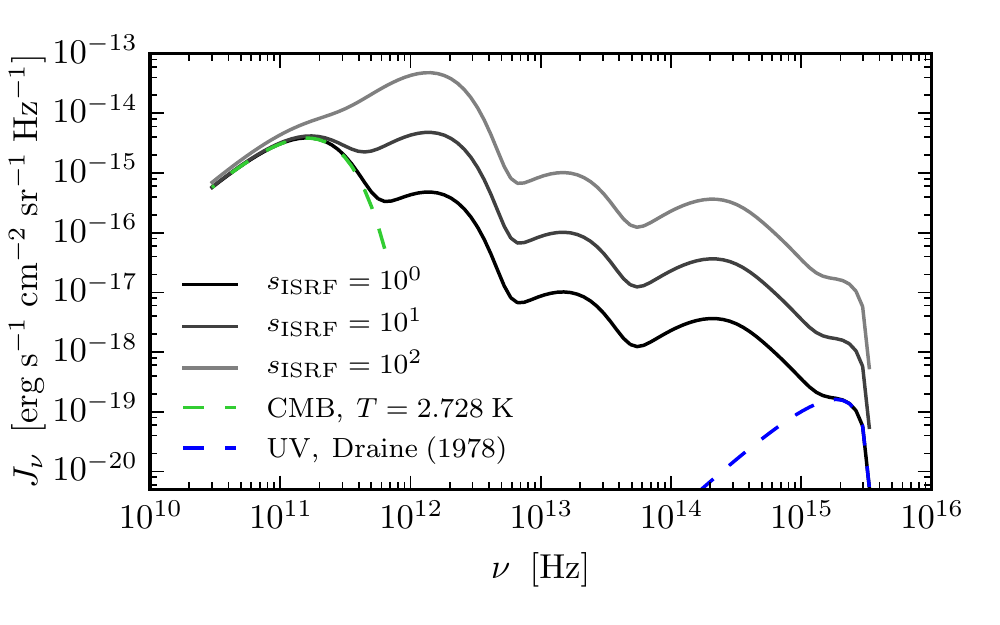} \\
    \includegraphics[width=0.47\textwidth]{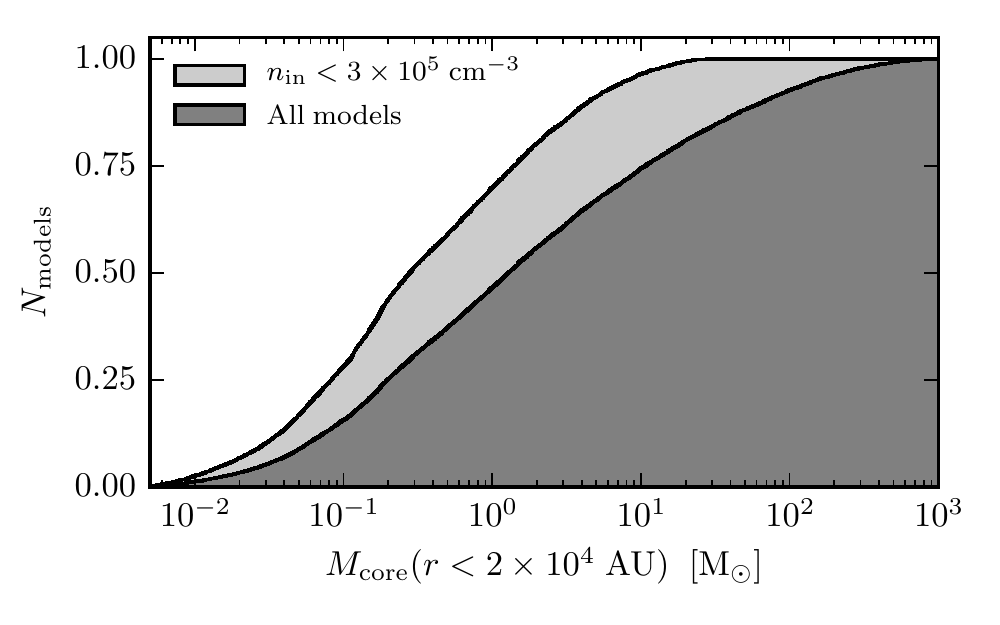}
\caption{
{\it Top:} ISRF parametrization used to self-consistently calculate the temperature profiles of starless core radiative transfer models.
Flux densities are scaled by factors of \num{e0} (black), \num{e1} (grey), and \num{e2} (light grey), excluding the contribution from the CMB.
{\it Bottom:} CDF of the gas mass enclosed within a radius of $r < \SI{2e4}{au}$ for all models (dark gray) and those with central densities $n_\mathrm{in} < \SI{3e5}{cm^{-3}}$ (light gray).
The typical core mass is between $0.2-\SI{20}{\msun}$.
}
\label{fig:ModelMassIsrf}
\end{figure}

\begin{figure}
    \centering
    \includegraphics[width=0.47\textwidth]{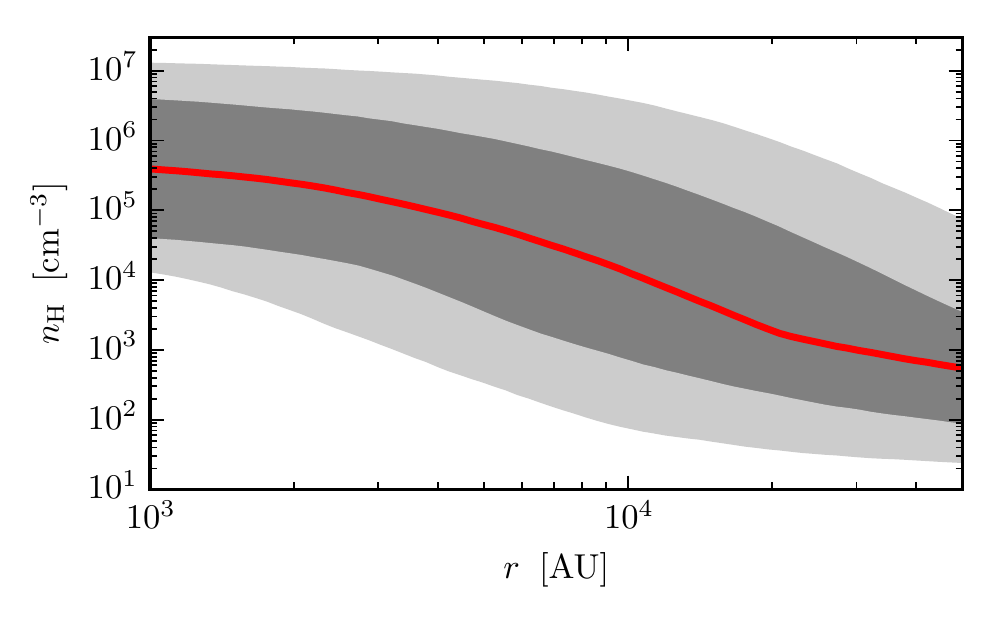} \\
    \includegraphics[width=0.47\textwidth]{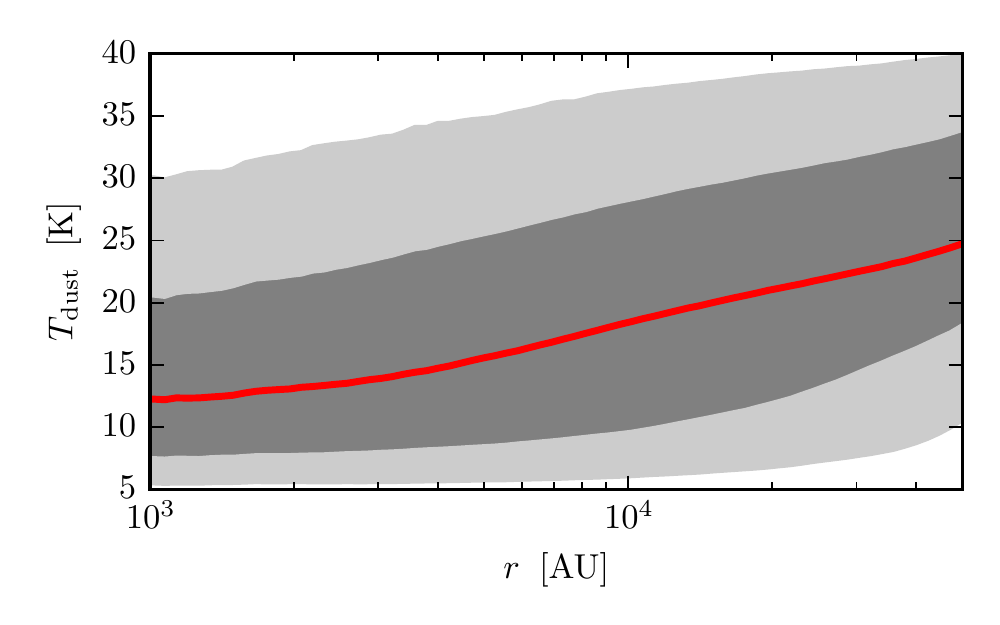} \\
    \includegraphics[width=0.47\textwidth]{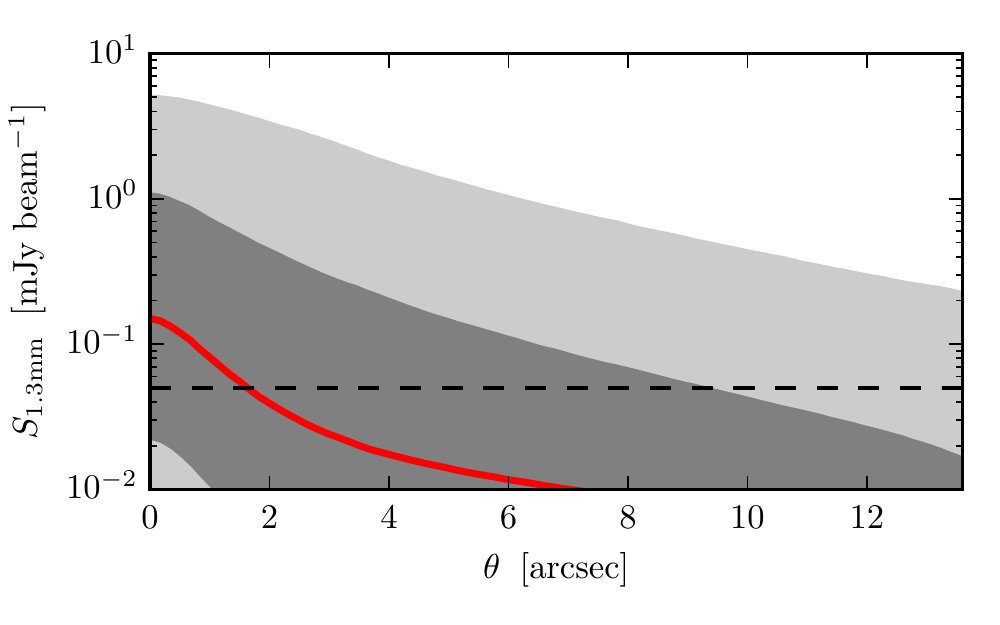}
\caption{
Parameter profiles for the suite of \num{e4} models computed with \radmc.
For each radii bin, the median value (red line), $16-84$ percentile interval (dark gray region), and $2.5-97.5$ percentile interval (light gray region) are shown.
{\it Top:} Input radial gas volume density profiles.
{\it Middle:} Output radial dust temperature profiles varying the ISRF and extinction.
Typical central temperatures range from $8-\SI{20}{K}$.
{\it Bottom:} Output radial surface brightness profiles produced at a fiducial distance of $d_\odot = \SI{4}{kpc}$.
The dashed horizontal line indicates the observed image \sigrms.
}
\label{fig:AllModelProfiles}
\end{figure}

The Plummer-like density profile in Eq.~(\ref{eq:Plummer}) is then irradiated in \radmc\ with an external source input using the spectral energy distribution of the ISRF for a self-consistent calculation of the dust temperature distribution.
We use the \cite{black94} ISRF spectrum as parametrized by \citet[][see Appendix B]{hocuk17} with the UV portion of the spectrum adopted from \cite{draine78}.
The ISRF is then varied in relative strength from the local value of the solar neighborhood by a multiplicative factor \sisrf, excluding the contribution from the CMB.
Figure~\ref{fig:ModelMassIsrf}a shows the ISRF specific intensity $J_\nu$ for $\sisrf = 10^0$, \num{e1}, and \num{e2} with the five parametrized components clearly visible.
Models are computed on a 1D radial grid from \SIrange{2.5e2}{6.0e4}{au} with 100 zones with \num{2e6} photons to ensure convergence in the output $T_\mathrm{dust}$ profiles over the tested range in $n_\mathrm{H}$.
The median core mass $M_\mathrm{core}$ integrated out a radius of \SI{2e4}{au} is $\sim \! \SI{1}{\msun}$ with the $(25,75)$ percentile interval ranging between \SIrange{0.2}{10}{\msun}, extending to $> \SI{100}{\msun}$ at the 92 percentile.
Figure \ref{fig:AllModelProfiles} shows the distributions of radial profiles in $n_\mathrm{H}(r)$, $T_\mathrm{dust}(r)$, and $S_\mathrm{1.3mm}(\theta)$ at a fiducial distance of \SI{4}{kpc}.
The typical $n_\mathrm{H}$ at $r = \SI{10}{kau}$ range from $n_\mathrm{H} = \num{8e2}-\SI{3e5}{cm^{-3}}$ and have typical central $T_\mathrm{dust} = 7-\SI{20}{K}$, with the maximum central $T_\mathrm{dust} = \SI{35}{K}$.

%%%%%%%%%%%%%%%%%%%%%%%%%%%%%%%%%%%%%%%%%%%%%%%%%%%%%%%%%%%%%%%%%%%%%%%%%%%%%%%%
%				  Tables
%%%%%%%%%%%%%%%%%%%%%%%%%%%%%%%%%%%%%%%%%%%%%%%%%%%%%%%%%%%%%%%%%%%%%%%%%%%%%%%%

\begin{deluxetable*}{lrrllrr}
\tabletypesize{\footnotesize}
\tablewidth{0pt}
\tablecaption{\label{tab:TargetPositions}Target Positions}
\tablehead{
% heading names
  \colhead{Name}
& \colhead{$\ell$}
& \colhead{$b$}
& \colhead{$\alpha$ (ICRS)}
& \colhead{$\delta$ (ICRS)}
& \colhead{$v_{\rm lsr}$}
& \colhead{\textsc{bgps id}$^{a}$} \vspace{-2mm} \\
% heading units
  \colhead{}
& \colhead{(deg)}
& \colhead{(deg)}
& \colhead{(h:m:s)}
& \colhead{(d:m:s)}
& \colhead{(\si{km.s^{-1}})}
& \colhead{}
}
\startdata
 G22695 &  22.695381 & -0.454657 &  18:34:14.58 &  -09:18:35.84 &  77.80 &  3686 \\
 G23297 &  23.297388 &  0.055330 &  18:33:32.06 &  -08:32:26.27 &  55.00 &  3822 \\
 G23481 &  23.479544 & -0.534764 &  18:35:59.56 &  -08:39:02.53 &  63.80 &  3892 \\
 G23605 &  23.605390 &  0.181325 &  18:33:39.40 &  -08:12:33.24 &  87.00 &  3929 \\
 G24051 &  24.051381 & -0.214655 &  18:35:54.40 &  -07:59:44.60 &  81.10 &  4029 \\
 G28539 &  28.538652 & -0.270358 &  18:44:22.60 &  -04:01:57.70 &  88.60 &  4732 \\
 G28565 &  28.527846 & -0.252172 &  18:44:17.52 &  -04:02:02.40 &  87.46 &  4729 \\
 G29558 &  29.557855 &  0.185321 &  18:44:37.07 &  -02:55:04.40 &  79.72 &  5021 \\
 G29601 &  29.604891 & -0.576768 &  18:47:25.20 &  -03:13:26.04 &  75.78 &  5030 \\
 G30120 &  30.119855 & -1.146674 &  18:50:23.54 &  -03:01:31.58 &  65.31 &  5114 \\
 G30660 &  30.657875 &  0.044680 &  18:47:07.76 &  -02:00:12.17 &  80.20 &  5265 \\
 G30912 &  30.913113 &  0.720803 &  18:45:11.28 &  -01:28:03.72 &  50.74 &  5360
\enddata 
\tablecomments{
($^{a}$) Catalog ID number in the BGPS v2.1.0 \citep{ginsburg13}.
}
\end{deluxetable*}
  % tab:TargetPositions

\begin{deluxetable}{lrrrr}
\tabletypesize{\footnotesize}
\tablewidth{0pt}
\tablecaption{\label{tab:PhysicalProperties}Clump Physical Properties}
\tablehead{
% heading names
  \colhead{Name}
& \colhead{$d_\odot$}
& \colhead{$\Sigma_\mathrm{pk}$}
& \colhead{$M_\mathrm{cl}$}
& \colhead{$T_\mathrm{K}$} \vspace{-2mm} \\
% heading units
  \colhead{}
& \colhead{(\si{kpc})}
& \colhead{(\si{g.cm^{-2}})}
& \colhead{(\si{\msun})}
& \colhead{(\si{K})}
}
\startdata
 G22695 &  4450 (190) &  0.0580 (0.013) &   930  (110) &  14.70   (0.42)     \\
 G23297 &  3480 (281) &  0.0760 (0.019) &   420 (\ 85) &  11.73   (0.41)     \\
 G23481 &  3780 (220) &  0.1100 (0.024) &   760  (120) &  11.29   (0.14)     \\
 G23605 &  4800 (240) &  0.0370 (0.015) &   880  (260) &  \nodata (\nodata)  \\
 G24051 &  4490 (210) &  0.0790 (0.015) &   760  (110) &  11.87   (0.37)     \\
 G28539 &  4780 (220) &  0.1280 (0.011) &  3610  (360) &  12.38   (0.14)     \\
 G28565 &  4680 (200) &  0.0830 (0.019) &   910  (220) &  \nodata (\nodata)  \\
 G29558 &  4370 (240) &  0.0690 (0.014) &   590 (\ 86) &  12.11   (0.17)     \\
 G29601 &  4270 (280) &  0.0900 (0.018) &   660  (130) &  15.98   (0.27)     \\
 G30120 &  3680 (260) &  0.0750 (0.031) &   820  (160) &  14.12   (0.15)     \\
 G30660 &  4410 (240) &  0.0770 (0.019) &  1380  (360) &  \nodata (\nodata)  \\
 G30912 &  2980 (250) &  0.0990 (0.019) &   450 (\ 88) &  11.67   (0.12)
\enddata 
\tablecomments{
Uncertainties are reported as the MAD in parentheses.
Properties are taken from \cite{svoboda16}, except for mass measurements of G29601 and G30912 which are taken from \cite{traficante15}.
}
\end{deluxetable}
  % tab:PhysicalProperties

\begin{deluxetable}{lrrrrrr}
\tabletypesize{\footnotesize}
\tablewidth{0pt}
\tablecaption{\label{tab:Correlator}ALMA Correlator Configuration}
\tablehead{
% heading names
  \colhead{\textsc{spw}}
& \colhead{Cen.\ Freq.}
& \colhead{$N$}
& \colhead{Bandwidth}
& \colhead{Bandwidth}
& \colhead{$\Delta f$}
& \colhead{$\Delta v$} \vspace{-2mm} \\
% heading units
  \colhead{}
& \colhead{(GHz)}
& \colhead{}
& \colhead{(kHz)}
& \colhead{($\mathrm{km/s}$)}
& \colhead{(kHz)}
& \colhead{($\mathrm{km/s}$)}
}
\startdata
1      & 216.112580 & 960               & 468750.0  & 650.252  & 488.28     & 0.677      \\
2      & 217.104980 & 960               & 468750.0  & 647.280  & 488.28     & 0.674      \\
3      & 218.222192 & 480               & 117187.2  & 160.991  & 244.14     & 0.335      \\
4      & 218.475632 & 480               & 117187.2  & 160.804  & 244.14     & 0.335      \\
5      & 218.760066 & 480               & 117187.2  & 160.595  & 244.14     & 0.335      \\
6      & 219.560358 & 240               & 117187.2  & 160.010  & 488.28     & 0.667      \\
7      & 230.538000 & 960               & 468750.0  & 609.564  & 488.28     & 0.635      \\
8      & 231.321828 & 960               & 468750.0  & 607.499  & 488.28     & 0.632      \\
9      & 233.820000 & 128               & 2000000.0 & 2564.301 & 15625.00   & 20.033     \\
\enddata 
\tablecomments{
Column descriptions:
(1)   Spectral window (SPW) ID number,
(2)   Center frequency of SPW in the rest frame,
(3)   Number of channels, 
(4,5) SPW total bandwidth,
(6,7) SPW channel resolution.
Uncertainties are given in parentheses.
}
\end{deluxetable}  % tab:Correlator

\begin{deluxetable}{lcrrccr}
\tabletypesize{\footnotesize}
\tablewidth{0pt}
\tablecaption{\label{tab:LineList}Spectral Line Transition Properties}
\tablehead{
% heading names
  \colhead{Specie}
& \colhead{Transition}
& \colhead{Rest Freq.}
& \colhead{$E_\mathrm{u}/k$}
& \colhead{Ref.}
& \colhead{\textsc{spw}}
& \colhead{$\Delta v$} \vspace{-2mm} \\
% heading units
  \colhead{} 
& \colhead{}
& \colhead{(GHz)}
& \colhead{(K)}
& \colhead{}
& \colhead{}
& \colhead{($\mathrm{km/s}$)}
}
\startdata
\ce{DCO+}   & $3 \rightarrow 2$              & 216.1125800 & 20.74 & (1)   & 1   & 0.68      \\
c-\ce{HC3H} & $3_{3,0} \rightarrow 2_{2,1}$  & 216.2787560 & 19.47 & (1)   & 1   & 0.68      \\
\ce{SiO}    & $5 \rightarrow 4$              & 217.1049800 & 31.26 & (1)   & 2   & 0.68      \\
\ce{DCN}    & $3 \rightarrow 2$              & 217.2385378 & 20.85 & (2)   & 2   & 0.68      \\
p-\ce{H2CO} & $3_{0,3} \rightarrow 2_{0,2}$  & 218.2221920 & 20.96 & (1)   & 3   & 0.34      \\
p-\ce{H2CO} & $3_{2,2} \rightarrow 2_{2,1}$  & 218.4756320 & 68.09 & (1)   & 4   & 0.34      \\
\ce{CH3OH}  & $4_{2,2} \rightarrow 3_{1,2}$  & 218.4400500 & 45.46 & (1)   & 4   & 0.34      \\
p-\ce{H2CO} & $3_{2,1} \rightarrow 2_{2,0}$  & 218.7600660 & 68.11 & (1)   & 5   & 0.34      \\
\ce{C^18O}  & $2 \rightarrow 1$              & 219.5603580 & 15.81 & (1)   & 6   & 0.68      \\
\ce{CO}     & $2 \rightarrow 1$              & 230.5380000 & 16.60 & (1)   & 7   & 0.68      \\
\ce{N2D+}   & $3 \rightarrow 2$              & 231.3218283 & 22.20 & (2)   & 8   & 0.68     
\enddata 
\tablecomments{
Transition property reference key: (1) SLAIM, (2) CDMS.
}
\end{deluxetable}  % tab:LineList

\clearpage 

\startlongtable
\begin{deluxetable*}{llllrrrrrrrcc}
\tabletypesize{\footnotesize}
\tablewidth{0pt}
\tablecaption{\label{tab:CoreProperties}Core Observed Properties}
\tablehead{
% heading names
  \colhead{Name}
& \colhead{ID}
& \colhead{$\alpha$ (ICRS)}
& \colhead{$\delta$ (ICRS)}
& \colhead{$\Omega_\mathrm{c}$}
& \colhead{$a$}
& \colhead{$b$}
& \colhead{PA}
& \colhead{$S_\nu$}
& \colhead{$\delta S_\nu$}
& \colhead{$S_{\nu,\mathrm{pk}}$}
& \colhead{CO}
& \colhead{SiO} \vspace{-2mm} \\
% heading units
  \colhead{}
& \colhead{}
& \colhead{(h:m:s)}
& \colhead{(d:m:s)}
& \colhead{(\si{as^2})}
& \colhead{(\si{as})}
& \colhead{(\si{as})}
& \colhead{(\si{deg})}
& \colhead{(\si{mJy})}
& \colhead{(\si{mJy})}
& \colhead{(\si{mJy/bm})}
& \colhead{}
& \colhead{}
}
\startdata
% Target &      ID &                           J2000 &        $\Omega$ &          $a$ &          $b$ &         PA & $S_\nu$ & $\delta S_\nu$ & $S_{pk}$ \\
 G22695 &        1 &  18:34:13.7336 & -09:18:36.6910 &            9.30 &        2.597 &        1.213 &      146.0 &  17.284 &      0.109 &        6.414 & 1 &   \\
 G22695 &        2 &  18:34:14.6345 & -09:18:44.6915 &           16.54 &        3.022 &        1.976 &      126.1 &   9.083 &      0.196 &        1.518 &   &   \\
 G22695 &        3 &  18:34:14.0405 & -09:18:33.7904 &            7.17 &        2.256 &        1.444 &      135.6 &   5.348 &      0.133 &        0.737 &   &   \\
\midrule
 G23297 &        1 &  18:33:32.1419 & -08:32:25.9306 &           22.70 &        7.853 &        1.461 &       85.5 &  16.431 &      0.309 &        0.727 &   &   \\
 G23297 &        2 &  18:33:31.6164 & -08:32:29.2777 &            2.79 &        1.055 &        0.870 &       72.4 &  11.515 &      0.088 &        6.381 & 2 &   \\
 G23297 &        3 &  18:33:32.1439 & -08:32:34.3684 &            6.86 &        3.124 &        1.180 &       61.9 &   5.699 &      0.130 &        0.925 &   &   \\
 G23297 &        4 &  18:33:32.3527 & -08:32:38.5247 &            3.77 &        1.880 &        0.911 &       94.0 &   4.139 &      0.066 &        1.906 & 1 &   \\
 G23297 &        5 &  18:33:31.5694 & -08:32:30.9485 &            0.86 &        0.692 &        0.488 &      148.8 &   2.542 &      0.045 &        3.026 &   &   \\
 G23297 &        6 &  18:33:31.9659 & -08:32:27.1673 &            2.29 &        1.161 &        0.784 &      107.4 &   1.740 &      0.097 &        0.893 &   &   \\
 G23297 &        7 &  18:33:31.6280 & -08:32:22.9146 &            3.53 &        1.982 &        0.798 &       88.6 &   1.645 &      0.099 &        0.472 &   &   \\
\midrule
 G23481 &        1 &  18:35:59.8961 & -08:39:08.0500 &            2.68 &        0.998 &        0.853 &     -164.0 &   3.514 &      0.085 &        2.311 &   &   \\
 G23481 &        2 &  18:35:59.7404 & -08:39:07.8724 &            3.04 &        1.464 &        0.758 &     -137.0 &   2.647 &      0.098 &        1.282 & 1 &   \\
 G23481 &        3 &  18:35:59.9838 & -08:38:48.7847 &            3.02 &        1.513 &        0.806 &      176.9 &   2.529 &      0.046 &        1.263 &   &   \\
 G23481 &        4 &  18:35:59.4861 & -08:39:01.0450 &            2.56 &        0.962 &        0.702 &      176.2 &   2.265 &      0.103 &        2.030 &   &   \\
 G23481 &        5 &  18:35:59.3916 & -08:39:05.8982 &            5.46 &        4.296 &        0.681 &     -153.8 &   1.920 &      0.142 &        0.385 &   &   \\
\midrule
 G23605 &        1 &  18:33:39.9726 & -08:12:39.4169 &            1.88 &        0.938 &        0.651 &      177.4 &   1.371 &      0.058 &        1.197 &   &   \\
\midrule
 G24051 &        1 &  18:35:54.1219 & -07:59:53.4130 &           13.49 &        2.213 &        1.980 &     -168.1 &  12.535 &      0.164 &        5.085 &   &   \\
 G24051 &        2 &  18:35:55.0045 & -07:59:35.7741 &            8.14 &        3.347 &        0.969 &      112.6 &   6.905 &      0.100 &        1.108 &   &   \\
 G24051 &        3 &  18:35:53.9805 & -07:59:58.0198 &            7.95 &        2.563 &        1.491 &      155.0 &   6.397 &      0.077 &        0.924 &   &   \\
 G24051 &        4 &  18:35:54.4771 & -07:59:41.2910 &           11.24 &        3.038 &        1.379 &     -176.6 &   5.844 &      0.207 &        1.192 & 1 &   \\
 G24051 &        5 &  18:35:54.5839 & -07:59:52.2422 &            4.87 &        1.920 &        1.103 &     -148.3 &   5.500 &      0.111 &        1.810 & 1 & 1 \\
 G24051 &        6 &  18:35:54.9096 & -07:59:40.4092 &            4.36 &        1.417 &        1.011 &      150.3 &   3.820 &      0.100 &        2.154 &   & 1 \\
 G24051 &        7 &  18:35:54.4693 & -07:59:49.2862 &            4.14 &        1.641 &        0.987 &       62.3 &   3.810 &      0.120 &        1.044 &   &   \\
 G24051 &        8 &  18:35:54.8945 & -07:59:42.6850 &            3.87 &        1.813 &        0.959 &      154.2 &   1.911 &      0.102 &        0.612 &   &   \\
 G24051 &        9 &  18:35:54.3071 & -07:59:43.7882 &            1.42 &        1.354 &        0.487 &       68.1 &   1.057 &      0.076 &        0.643 &   &   \\
 G24051 &       10 &  18:35:54.3750 & -07:59:45.8972 &            1.44 &        1.396 &        0.454 &       55.4 &   1.045 &      0.077 &        0.645 &   &   \\
 G24051 &       11 &  18:35:54.8697 & -07:59:51.3991 &            1.47 &        1.060 &        0.555 &      135.5 &   0.659 &      0.054 &        0.506 &   &   \\
\midrule
 G28539 &        1 &  18:44:22.2420 & -04:01:44.7142 &           16.35 &        6.175 &        1.334 &       45.3 &  19.114 &      0.121 &        1.842 &   &   \\
 G28539 &        2 &  18:44:22.7348 & -04:01:56.1668 &            8.58 &        2.585 &        1.671 &       73.6 &   5.411 &      0.183 &        0.711 &   &   \\
 G28539 &        3 &  18:44:22.8536 & -04:02:03.2640 &           12.63 &        4.346 &        2.086 &       46.9 &   4.587 &      0.191 &        0.439 &   &   \\
 G28539 &        4 &  18:44:22.3397 & -04:01:54.0246 &            7.39 &        2.372 &        1.682 &      130.3 &   4.055 &      0.156 &        0.603 &   &   \\
 G28539 &        5 &  18:44:22.8195 & -04:02:07.5292 &            2.36 &        1.403 &        0.682 &       80.2 &   1.249 &      0.064 &        0.526 &   &   \\
 G28539 &        6 &  18:44:22.7434 & -04:01:53.8880 &            1.41 &        0.930 &        0.656 &      169.0 &   0.842 &      0.070 &        0.551 &   &   \\
\midrule
 G28565 &        1 &  18:44:17.2674 & -04:02:03.5257 &           14.00 &        3.641 &        2.023 &       79.3 &  17.943 &      0.226 &        2.499 & 1 & 1 \\
 G28565 &        2 &  18:44:16.9912 & -04:02:01.1285 &            3.52 &        2.395 &        0.776 &       89.0 &   3.894 &      0.093 &        0.946 &   &   \\
 G28565 &        3 &  18:44:17.2241 & -04:02:08.5328 &            2.60 &        1.532 &        0.830 &       78.1 &   3.604 &      0.083 &        1.443 &   &   \\
 G28565 &        4 &  18:44:17.1101 & -04:02:09.7546 &            3.33 &        1.914 &        0.810 &       62.9 &   3.087 &      0.082 &        0.827 &   &   \\
 G28565 &        5 &  18:44:17.0529 & -04:01:58.7236 &            1.87 &        1.082 &        0.690 &      143.2 &   2.401 &      0.069 &        1.347 &   &   \\
 G28565 &        6 &  18:44:17.3596 & -04:02:06.5025 &            1.47 &        0.935 &        0.670 &      109.7 &   1.561 &      0.071 &        0.903 &   &   \\
\midrule
 G29558 &        1 &  18:44:37.5015 & -02:55:12.4812 &           20.09 &        2.432 &        2.174 &     -146.5 &  20.697 &      0.184 &        6.613 &   &   \\
 G29558 &        2 &  18:44:37.3029 & -02:55:01.9117 &            4.83 &        1.372 &        1.029 &      161.9 &  10.569 &      0.130 &        4.941 & 1 &   \\
 G29558 &        3 &  18:44:37.5338 & -02:55:00.8673 &            3.41 &        1.524 &        0.742 &      171.2 &   6.624 &      0.093 &        3.588 & 1 &   \\
 G29558 &        4 &  18:44:36.6483 & -02:55:02.6587 &            4.40 &        1.548 &        1.109 &      134.4 &   4.624 &      0.114 &        1.666 &   &   \\
 G29558 &        5 &  18:44:37.0267 & -02:55:08.7202 &            6.09 &        2.893 &        1.085 &      125.8 &   1.881 &      0.146 &        0.332 &   &   \\
 G29558 &        6 &  18:44:37.8020 & -02:55:10.4118 &            3.46 &        1.516 &        1.090 &      125.5 &   1.810 &      0.065 &        0.681 &   &   \\
 G29558 &        7 &  18:44:36.6640 & -02:55:00.1446 &            1.87 &        0.937 &        0.766 &      176.3 &   1.688 &      0.070 &        1.116 &   &   \\
 G29558 &        8 &  18:44:36.6775 & -02:54:57.2062 &            3.91 &        1.871 &        1.128 &      171.4 &   1.626 &      0.090 &        0.472 &   &   \\
 G29558 &        9 &  18:44:37.1252 & -02:55:04.1785 &            2.00 &        1.141 &        0.739 &       45.8 &   1.357 &      0.090 &        0.584 &   &   \\
\midrule
 G29601 &        1 &  18:47:25.3865 & -03:13:29.3698 &           16.05 &        2.547 &        1.728 &       79.6 &  15.771 &      0.240 &        6.686 & 2 & 1 \\
 G29601 &        2 &  18:47:25.3644 & -03:13:20.3497 &            4.94 &        1.908 &        1.077 &     -140.3 &   2.192 &      0.123 &        0.567 &   &   \\
 G29601 &        3 &  18:47:25.3951 & -03:13:23.8223 &            4.15 &        2.104 &        0.976 &       78.6 &   1.751 &      0.124 &        0.501 &   &   \\
 G29601 &        4 &  18:47:25.5623 & -03:13:24.6212 &            1.69 &        0.999 &        0.716 &      151.3 &   0.541 &      0.074 &        0.341 &   &   \\
\midrule
 G30120 &        1 &  18:50:24.7282 & -03:01:27.2884 &            1.38 &        0.820 &        0.683 &      147.5 &   3.924 &      0.020 &        2.709 & 1 &   \\
 G30120 &        2 &  18:50:24.7785 & -03:01:26.1411 &            0.93 &        0.737 &        0.548 &       53.8 &   2.868 &      0.014 &        2.807 &   &   \\
 G30120 &        3 &  18:50:22.9654 & -03:01:43.6061 &            1.61 &        0.912 &        0.680 &     -137.5 &   1.232 &      0.034 &        0.854 &   &   \\
\midrule
 G30660 &        1 &  18:47:07.7985 & -02:00:24.1287 &           27.58 &        5.270 &        2.513 &       66.2 &  15.430 &      0.191 &        0.902 &   &   \\
 G30660 &        2 &  18:47:08.0553 & -02:00:09.6124 &           14.88 &        2.912 &        2.347 &      178.5 &   8.489 &      0.224 &        2.140 &   &   \\
 G30660 &        3 &  18:47:07.8433 & -02:00:04.5482 &           18.40 &        4.397 &        2.139 &      142.3 &   8.257 &      0.219 &        1.318 &   &   \\
 G30660 &        4 &  18:47:07.4677 & -01:59:58.7398 &           11.09 &        3.618 &        1.832 &       52.6 &   7.588 &      0.095 &        0.984 &   &   \\
 G30660 &        5 &  18:47:07.6647 & -02:00:11.1585 &           11.19 &        2.859 &        1.779 &       53.0 &   5.183 &      0.210 &        0.812 &   &   \\
 G30660 &        6 &  18:47:07.3910 & -02:00:09.7055 &            8.59 &        2.296 &        1.622 &      119.3 &   4.896 &      0.162 &        1.128 &   &   \\
\midrule
 G30912 &        1 &  18:45:11.4745 & -01:28:04.9508 &           44.25 &        4.523 &        3.651 &       51.0 &  27.344 &      0.403 &        2.728 & 1 &   \\
 G30912 &        2 &  18:45:11.1447 & -01:28:02.2048 &           12.95 &        2.595 &        1.372 &      124.9 &   9.536 &      0.224 &        4.324 & 2 &   \\
 G30912 &        3 &  18:45:11.9211 & -01:27:55.2127 &            4.29 &        2.778 &        0.976 &      151.1 &   2.249 &      0.069 &        0.682 &   &   \\
 G30912 &        4 &  18:45:11.4095 & -01:27:58.6881 &            4.05 &        2.365 &        1.096 &     -161.7 &   1.498 &      0.114 &        0.430 &   &   \\
 G30912 &        5 &  18:45:10.5824 & -01:28:11.8467 &            1.53 &        0.820 &        0.629 &      179.5 &   1.399 &      0.039 &        1.353 &   &   \\
 G30912 &        6 &  18:45:10.9468 & -01:28:09.9788 &            0.83 &        0.600 &        0.516 &      168.0 &   0.360 &      0.045 &        0.518 &   &   \\
\enddata 
\tablecomments{
Column descriptions:
 (1) Target clump name,
 (2) sub-structure ID number,
 (3) centroid right ascension coordinate,
 (4) centroid declination coordinate,
 (5) total dendrogram area,
 (6) Gaussian major FWHM,
 (7) Gaussian minor FWHM,
 (8) Gaussian position angle,
 (9) source integrated \SI{1.3}{mm} flux density,
(10) uncertainty in source integrated flux density,
(11) source peak flux density,
(12) number of bipolar CO outflows,
(13) number of bipolar SiO outflows.
}
\end{deluxetable*}
  % tab:CoreProperties

\newcommand{\tabdet}{\textcolor{blue}{\bf D}}
\newcommand{\tabnon}{\textcolor{red}{\bf N}}
\newcommand{\tabwek}{\textcolor{cyan}{\bf W}}
\newcommand{\tabofl}{\textcolor{violet}{\bf B}}
\newcommand{\tabemm}{\tabdet}

\begin{deluxetable*}{*{17}{c}}
\tabletypesize{\scriptsize}
\tablewidth{0pt}
\tablecaption{\label{tab:Detections}Band 6 Detections$^{a}$}
\tablehead{
% heading groups
  \colhead{Name}
& \multicolumn{1}{c}{Cont.} & $\,$
& \multicolumn{3}{c}{Deuteration} & $\,$
& \multicolumn{2}{c}{Kinematic} & $\,$
& \multicolumn{4}{c}{High-Excitation} & $\,$
& \multicolumn{2}{c}{Outflow} \vspace{-1mm} \\
  \cmidrule{4-6}
  \cmidrule{8-9}
  \cmidrule{11-14}
  \cmidrule{16-17}
% heading names
  \colhead{}
& \colhead{1.3 mm} &
& \colhead{$\mathrm{DCO^+}$}
& \colhead{$\mathrm{DCN}$}
& \colhead{$\mathrm{N_2D^+}$} &
& \colhead{$\mathrm{C^{18}O}$}
& \colhead{$\mathrm{H_2CO}^b$} &
& \colhead{$\mathrm{H_2CO}$}
& \colhead{$\mathrm{H_2CO}$}
& \colhead{c-$\mathrm{C_3H_2}$}
& \colhead{$\mathrm{CH_3OH}$} &
& \colhead{$\mathrm{CO}$}
& \colhead{$\mathrm{SiO}$}
%% heading units
%  \colhead{}
%& \colhead{(cont.)}
%& \colhead{$3-2$}
%& \colhead{$3-2$}
%& \colhead{$3-2$}
%& \colhead{$2-1$}
%& \colhead{$3_{0,3}-2_{0,2}$}
%& \colhead{$3_{2,1}-2_{2,0}$}
%& \colhead{$3_{2,2}-2_{2,1}$}
%& \colhead{$3_{3,0}-2_{2,1}$}
%& \colhead{$4_2-3_1$}
%& \colhead{$2-1$}
%& \colhead{$5-4$}
}
\startdata
% name & continuum && dcop & dcn & n2dp && c18o & p-h2co\_303\_202 && p-h2co\_321\_220 & p-h2co\_322\_221 & c-hc3h\_330\_221 & ch3oh\_42\_31 && co & sio
G28539 &\tabdet    &&\tabwek   &\tabwek  &\tabnon   &&\tabdet   &\tabdet               &&\tabnon               &\tabwek               &\tabnon               &\tabwek            &&\tabdet  &\tabdet   \\
G30660 &\tabdet        &&\tabdet   &\tabnon  &\tabnon   &&\tabdet   &\tabdet               &&\tabwek               &\tabwek               &\tabdet               &\tabdet            &&\tabdet  &\tabdet   \\
G22695 &\tabdet        &&\tabwek   &\tabwek  &\tabnon   &&\tabdet   &\tabdet               &&\tabdet               &\tabdet               &\tabnon               &\tabdet            &&\tabofl  &\tabdet   \\
G23605 &\tabdet        &&\tabnon   &\tabnon  &\tabnon   &&\tabdet   &\tabdet               &&\tabnon               &\tabwek               &\tabnon               &\tabwek            &&\tabdet  &\tabnon   \\
G24051 &\tabdet        &&\tabdet   &\tabdet  &\tabdet   &&\tabdet   &\tabdet               &&\tabdet               &\tabwek               &\tabdet               &\tabdet            &&\tabofl  &\tabofl   \\
G23297 &\tabdet        &&\tabdet   &\tabdet  &\tabwek   &&\tabdet   &\tabdet               &&\tabdet               &\tabdet               &\tabwek               &\tabdet            &&\tabofl  &\tabdet   \\
G23481 &\tabdet        &&\tabnon   &\tabnon  &\tabnon   &&\tabdet   &\tabdet               &&\tabdet               &\tabdet               &\tabnon               &\tabdet            &&\tabofl  &\tabwek   \\
G29558 &\tabdet        &&\tabdet   &\tabwek  &\tabdet   &&\tabdet   &\tabdet               &&\tabdet               &\tabdet               &\tabdet               &\tabdet            &&\tabofl  &\tabdet   \\
G30120 &\tabwek        &&\tabnon   &\tabnon  &\tabnon   &&\tabdet   &\tabwek               &&\tabnon               &\tabnon               &\tabnon               &\tabwek            &&\tabofl  &\tabdet   \\
G28565 &\tabdet        &&\tabdet   &\tabwek  &\tabdet   &&\tabdet   &\tabdet               &&\tabdet               &\tabdet               &\tabdet               &\tabdet            &&\tabofl  &\tabofl   \\
G29601 &\tabdet        &&\tabdet   &\tabwek  &\tabnon   &&\tabdet   &\tabdet               &&\tabdet               &\tabdet               &\tabwek               &\tabdet            &&\tabofl  &\tabofl   \\
G30912 &\tabdet        &&\tabdet   &\tabdet  &\tabwek   &&\tabdet   &\tabdet               &&\tabdet               &\tabdet               &\tabwek               &\tabdet            &&\tabofl  &\tabdet  
\enddata
\tablecomments{
({\it a}) Detection flags:
\tabdet\ detection with $\mathrm{SNR} \geq 7\sigma$,
\tabwek\ weak detection with $5 \sigma \leq \mathrm{SNR} < 7\sigma$,
\tabnon\ non-detection with $\mathrm{SNR} < 5 \sigma$,
and \tabofl\ detection of bipolar outflow.
({\it b}) $\mathrm{H_2CO}$ transitions listed in order of $3_{0,3}-2_{0,2}$, $3_{2,1}-2_{2,0}$, and $3_{2,2}-2_{2,1}$.
}
\end{deluxetable*}  % tab:Detections

\end{document}